\documentclass[a4paper,11pt,english]{article}

\pdfoutput=1
\usepackage[utf8x]{inputenc}	
\usepackage[english]{babel}
\usepackage[bindingoffset=0.3cm,textheight=22.5cm,hdivide={2.7cm,*,2.7cm}, vdivide={*,22cm,*}]{geometry}
\usepackage[svgnames]{xcolor}
\usepackage[pdftex,breaklinks,colorlinks=true,urlcolor=RoyalBlue,%
linkcolor=blue,citecolor=blue]{hyperref}
\usepackage{amsmath,amsfonts,amssymb,slashed,graphicx,color,empheq}
\usepackage{paralist}

\usepackage{cite}

\makeatletter

\newcommand{\dd}{\mathrm{d}}

\newcommand{\del}{\partial}

\def\nn{\nonumber} 
\def\obar{\overline}
\numberwithin{equation}{section}

\def\a{\alpha}  \def\b{\beta}
 \def\g{\gamma} 
 \def\d{\delta} 
    \def\k{\kappa}
 \def\L{\Lambda}  \def\m{\mu}
\def\n{\nu}    \def\r{\rho}
\def\s{\sigma}  \def\t{\tau}

\def\cA{{\cal A}} \def\cB{{\cal B}} \def\cC{{\cal C}} 
\def\cD{{\cal D}}   
\def\cG{{\cal G}} \def\cH{{\cal H}} \def\cI{{\cal I}} 
\def\cJ{{\cal J}} \def\cK{{\cal K}} \def\cL{{\cal L}} 
\def\cM{{\cal M}}   
 \def\cQ{{\cal Q}} \def\cR{{\cal R}} 
\def\cS{{\cal S}}

\def\R{{\mathbb R}} \def\C{{\mathbb C}} 

 \def\one{\mbox{1 \kern-.59em {\rm l}}}

\def\msu{\mathfrak{su}}
\def\mso{\mathfrak{so}}

\newcommand{\Tr}{\mathrm{Tr}}
\newcommand{\tr}{\mathrm{tr}}
\newcommand{\End}{\mathrm{End}}

\def\Tr{\mbox{Tr}}

\newcommand{\im}{\mathrm{i}}
\newcommand{\ad}{\mathrm{ad}}

\newcommand{\diag}{\rm diag}

 \allowdisplaybreaks[3]

\begin{document}


\parindent=0cm

\renewcommand{\title}[1]{\vspace{10mm}\noindent{\Large{\bf

#1}}\vspace{8mm}} \newcommand{\authors}[1]{\noindent{\large

#1}\vspace{5mm}} \newcommand{\address}[1]{{\itshape #1\vspace{2mm}}}


\begin{titlepage}
\begin{flushright}
 UWThPh-2019-02 
\end{flushright}
\begin{center}
\title{ {\Large  Covariant cosmological quantum space-time, \\[1ex]  higher-spin 
and gravity in the IKKT matrix model}  }

\vskip 3mm

\authors{Marcus Sperling${}^\dagger$, Harold C.\ Steinacker${}^\ddagger$}

\vskip 3mm

 \address{ 
 ${}^\dagger$\textit{Yau Mathematical Sciences Center, Tsinghua University\\
Haidian District, Beijing, 100084, China } \\
Email: {\tt marcus.sperling@univie.ac.at}  
 \\
 \vskip 3mm
${}^\ddagger${\it Faculty of Physics, University of Vienna\\
Boltzmanngasse 5, A-1090 Vienna, Austria  }  \\
Email: {\tt harold.steinacker@univie.ac.at}  
  }

\bigskip

\vskip 1.4cm

\textbf{Abstract}
\vskip 3mm

\begin{minipage}{14cm}%
We discuss a $(3{+}1)$-dimensional covariant quantum space-time describing a 
FLRW cosmology 
with Big Bounce, obtained by a projection of the fuzzy hyperboloid $H^4_n$. 
This provides a background solution of the IKKT matrix model with mass term. 
We characterize the bosonic fluctuation spectrum, which consists of a tower 
of higher-spin modes.
The modes are organized in terms of an underlying $SO(4,2)$ structure group,
which is broken to the $SO(3,1)$ isometry of the background.
The resulting higher-spin gauge theory includes all degrees of freedom required 
for gravity, and should be well suited for quantization.  
All modes propagate with the same speed of light,
even though local 
boost invariance is not manifest. 
The propagating metric perturbation modes comprise those of a massless graviton,
as well as a scalar mode.
Gauge invariance allows to obtain the analog of the linearized  
Einstein-Hilbert action, 
which is expected to be induced upon quantization.

\end{minipage}

\end{center}

\end{titlepage}

\tableofcontents
%
%
\section{Introduction} 
The proper formulation of gravity as a quantum theory is a long-standing and controversial problem. 
Much of the difficulty is due to the fact that the gravitational constant is dimensionful, 
which means that gravity becomes 
strong at short scales, and arguably requires a quantum notion of space-time itself.
It is thus plausible that the starting point of such a quantum theory could be very different from general relativity (GR),
but GR should be approximately recovered at macroscopic scales, in the sense of an effective theory.

With this in mind, we choose the framework of matrix models of Yang-Mills type 
as a starting point.
These models naturally describe dynamical noncommutative (``quantum'') geometries,
and they automatically lead to a gauge theory, 
which is crucial in  Minkowski signature. 
Moreover, the kinetic term arises from a universal commutator structure,
which encodes a universal metric \cite{Steinacker:2010rh}.
To avoid the dangerous  UV/IR mixing
which typically arises on non-commutative spaces \cite{Minwalla:1999px},
one is led to the maximally supersymmetric IKKT or IIB model \cite{Ishibashi:1996xs},
where UV/IR mixing is sufficiently mild\footnote{UV/IR mixing 
leads to $10$-dimensional ``bulk'' supergravity in the Type IIB model 
\cite{Ishibashi:1996xs,Chepelev:1997av,Kabat:1997sa,Steinacker:2016nsc}. 
However, this has nothing to do with the mechanism under consideration here,
which arises on a $4$-dimensional brane, where the bulk supergravity only 
induces a short-range $r^{-8}$ interaction.}.
This model can be viewed as a constructive definition of IIB string theory,
and it exhibits a rich set of brane-type solutions, 
such as \cite{Chepelev:1997ug,Aoki:1998vn,Kimura:2001uk,Kimura:2002nq,Steinacker:2011wb,Kim:2012mw} 
to mention only a few. Quite remarkably,
recent numerical simulations of the model 
\cite{Kim:2011cr,Kim:2012mw,Ito:2015mxa,Nishimura:2019qal} 
provide evidence that a 3+1-dimensional structure indeed
arises at the non-perturbative level.

With this motivation,
our objective is to find a suitable background space-time solution of the IKKT matrix model, 
which leads to the desired low-energy physics including gravity.
There has been considerable effort along these lines including
\cite{Steinacker:2010rh,Blaschke:2010rg,Steinacker:2012ra,Steinacker:2012ct,Kawai:2016wfh}, 
and \cite{Rivelles:2002ez,Hanada:2005vr,Yang:2006dk,Yang:2009pm,Chaney:2015ktw,Kaneko:2017zeo,Stern:2018wud} in 
a similar spirit.
One of the problems which arises on simple non-commutative geometries 
is that they typically carry some 
Lorentz-breaking structure, which is well hidden 
classically, but tends to show up in the loop contributions. 
That problem can be avoided  on \emph{covariant quantum spaces}, such as 
fuzzy $S^4_N$, $H^4_n$, and similar spaces 
\cite{Snyder:1946qz,Yang:1947ud,Grosse:1996mz,Ramgoolam:2001zx,Medina:2002pc,Gazeau:2009mi,Heckman:2014xha,Buric:2017yes,Sperling:2017dts}. 
This class of spaces exhibits a rich extra structure,
which typically leads to a higher-spin gauge theory 
\cite{Steinacker:2016vgf,deMedeiros:2004wb,Vasiliev:2004qz}.

In this paper, we focus on an interesting candidate for such a covariant background, 
given by the  quantized or fuzzy cosmological 
space-time $\cM^{3,1}_n$ recently found in \cite{Steinacker:2017bhb}. This 
$(3{+}1)$-dimensional space  is based on the $4$-dimensional 
fuzzy hyperboloid $H^4_n$ \cite{Sperling:2018xrm},
which upon a projection acquires a semi-classical  
Friedmann-Lema\^{\i}tre-Robertson-Walker (FLRW) geometry describing an 
expanding universe with initial singularity, 
or rather a Big Bounce. 
In particular,
we study the linearized fluctuation spectrum and the effective gauge theory arising on this  $\cM^{3,1}_n$, which is a  solution
of the matrix model with a mass term.
For previous work on FLRW solutions in the context of higher spin gravity we refer to \cite{Sezgin:2005hf,Aros:2017ror}.

One important feature of $\cM^{3,1}_n$ is that it admits a global 
$SO(3,1)$ symmetry, while local Lorentz invariance is not manifest. 
This $SO(3,1)$ symmetry is  the isometry group of the space-like hyperbolas 
$H^3_t$ of the cosmological space-time  with $k=-1$.  
It comprises space-like translations and local $SO(3)$ rotations, 
and the latter can be viewed as a manifest part of Lorentz invariance.
However, there is no boost invariance, because
the cosmic background defines an invariant  time-like vector field. 
Moreover, the model enjoys a large gauge invariance, which can be 
viewed as a 
higher-spin algebra associated to $\mso(4,1)$ 
\cite{Vasiliev:2004qz,Joung:2014qya}, 
acting in an intricate way. 
These symmetries protect the theory from becoming pathological,  
 and almost enforce local Lorentz invariance.
For example, it turns out that the propagation of all physical higher-spin modes 
is governed by a single effective metric, as it should be,
and we expect that the resulting physics will 
largely respect local Lorentz invariance.
Thus the breaking of (local) Lorentz-invariance seems to be well hidden 
in the unphysical fluctuations, as far as we can see. 
Nevertheless, some breaking of Lorentz invariance is bound to show up 
somewhere. 

The present $\cM^{3,1}_n$ solution is somewhat different from
(and in a sense dual to) the solution recently found in 
\cite{Steinacker:2017bhb,Steinacker:2017vqw}. 
The present realization is preferred, because it allows to systematically  
organize and study the (linearized) fluctuation spectrum.
This involves organizing the fluctuations into towers of higher-spin
modes, finding the on-shell modes and their spectrum, and  
identifying among them the metric fluctuation modes. 
All these are non-trivial tasks which requires some technology. Fortunately,
due to the close relation with the Euclidean case of $H^4_n$  and 
its enhanced $SO(4,1)$ symmetry, we can use many of the 
results obtained in \cite{Sperling:2018xrm}, and largely solve these problems. 
That $SO(4,1)$ is in fact part of an $SO(4,2)$ structure group, which is underlying the construction of 
$H^4_n$ as a coadjoint orbit. In particular 
(and in contrast to previous attempts),  
all physical metric degrees of freedom are obtained from the linearized fluctuation
modes, as well as three of the usual four diffeomorphism modes, which arise as 
pure gauge modes
of the higher-spin algebra. The reduction to three diffeomorphism modes 
reflects the presence of an 
invariant volume form on the noncommutative space.

Since GR is not a Yang-Mills theory, one would not expect to
obtain the 
Einstein equations directly from the matrix model. It is therefore very remarkable that 
we do indeed obtain the 
linearized vacuum Einstein equations directly from the matrix model. 
More precisely, the classical action leads to the  
two propagating Ricci-flat graviton modes as in linearized GR\footnote{This also 
works for $\R^4_\theta$  \cite{Rivelles:2002ez,Yang:2006dk,Steinacker:2010rh}, but 
its extension to full gravity seems problematic, notably due to more serious 
Lorentz violations. There is in fact a 
way to formally obtain the full vacuum Einstein equations \cite{Hanada:2005vr},  which however 
leads to other issues.}, as well as a scalar metric mode
whose significance is not yet clear. 
However to obtain the (linearized) Einstein equations  in the presence of matter,
quantum effects in the matrix model are 
presumably required.
Here the higher-spin symmetry is extremely useful, because it strongly restricts 
the possible terms in the quantum effective action. Using this gauge invariance, 
we identify\footnote{This would not be 
possible on simple noncommutative spaces such as $\R^4_n$,
where extra terms are possible which are forbidden here \cite{Klammer:2009dj}. 
This is
very important for obtaining a realistic physics.} the  analog of the 
linearized Einstein-Hilbert action,
which is expected to be induced upon quantization  \cite{Sakharov:1967pk}.
Moreover, there seems to be no analog of the cosmological constant term,
which is replaced by the Yang-Mills action.
This suggests that the cosmological constant problem may not arise in this approach.

Due to the Yang-Mills structure of the underlying (matrix) model, 
the gauge-fixing procedure is straightforward, and
the model should provide a well-defined quantum theory.
There are only finitely many degrees of freedom per volume due to the 
intrinsic UV cutoff of the underlying noncommutative space-time, and
the maximal supersymmetry of the model (broken by the soft mass term)
will strongly restrict possible non-local quantum effects due to UV/IR mixing.
Since Ricci-flat deformations are already solutions of the bare action and all 
metric modes of gravity are present, the model is an excellent candidate for realizing 
 gravity in a quantum theory of space-time and matter.

As an extra bonus, the background under consideration provides a rather attractive
cosmological scenario featuring a Big Bounce and an asymptotically coasting late-time
evolution. 
It is fascinating that a reasonable cosmological evolution
is obtained  without any fine-tuning, and even without requiring 
the presence of matter.
If the resulting gravity at intermediate length scales turns out to be viable, 
this would be a significant advantage over more conventional approaches, 
which typically require a  delicate balance of 
various matter and energy constituents of the universe.

The  outline of the paper is as follows: 
We start in section \ref{sec:class_geometry} by reviewing the classical geometry of $\C 
P^{1,2}$, which is understood either as  
coadjoint orbit or as total space of an 
$S^2$-bundle over $H^4$ or $H^{2,2}$. Its quantization leading to 
fuzzy $H_n^{4}$ as well as the projection to fuzzy $\cM_n^{3,1}$ are then  detailed.
We study the algebra of functions in section \ref{sec:wavefunct-higherspin} by 
means of spin Casimirs and various intertwiners. It will turn out that two 
descriptions are available: either divergence-free symmetric traceless tensors 
on $H^4$ or space-like symmetric traceless tensors on $\cM^{3,1}$. As shown in 
section \ref{sec:matrix-model}, the IKKT matrix model allows three kinds of 
solutions from various embeddings of the $\mso(4,2)$ generators.  
Focusing on the \emph{momentum} solution as classical background, we 
investigate the fluctuation modes and the resulting higher-spin gauge theory in 
section \ref{sec:fluctuations}. We pay attention to gauge-fixing questions as well as the inner products and kinetic 
terms. All these considerations prepare the stage for section \ref{sec:graviton},
in which we address the definition of gravition modes as well as the 
construction of an gauge-invariant quadratic action. We elaborate the properties 
of this linearized Einstein-Hilbert-like action in detail.
Finally we conclude and summarize in section \ref{sec:conclusion}.
The appendix \ref{app:proofs} provides details and proofs of some elaborate 
calculations that were omitted in the main text.
To make the paper self-contained and to facilitate follow-up work,
we include all the required background and details.
%
%
\section{Geometric preliminaries}
\label{sec:class_geometry}
\subsection{Classical hyperboloid \texorpdfstring{$H^4$}{H4} from
 \texorpdfstring{$\C P^{1,2}$}{H4}}
 We start with a discussion of the  classical geometry underlying fuzzy $H^4_n$ 
which is $\C P^{1,2}$,
viewed as an $S^2$ bundle over the 4-hyperboloid $H^4$. 
More precisely, $\C P^{1,2}$  is an 
 $SO(4,1)$-equivariant bundle over $H^4$, meaning that there exists an $SO(4,1)$ 
action on both the total space and the 
base space that are compatible with the bundle projection \cite[Def.\ 
1.5]{Berline:2004}.
The local stabilizer group $SO(4)$ acts non-trivially on the 
$S^2$ fiber such that the internal excitations become higher-spin fields on 
$H^4$.
In addition, $\C P^{1,2}$ is a coadjoint orbit of $SO(4,2)$, which provides
extra geometrical structure and, naturally, leads to a quantization in terms of 
fuzzy $H^4_n$.
The construction shares some similarities with twistor constructions for Minkowski space.
\paragraph{Conventions.} 
Latin letters $a,b,c,\ldots$ will typically imply the natural representation of 
$SO(4,1)$, and Greek letters $\mu, \nu, \gamma,\ldots$ the $SO(3,1)$ version.
We will generally raise and lower indices with the $SO(4,1)$-invariant tensor 
$\eta^{ab}$ or 
its $SO(3,1)$ version $\eta^{\mu\nu}$, unless otherwise stated.
Moreover, Einstein summation convention is adopted, i.e.\ repeated indices are 
summed over.
\paragraph{\texorpdfstring{$\C P^{1,2}$}{CP12} as 
\texorpdfstring{$SO(4,1)$}{SO(4,1)}-equivariant bundle over the hyperboloid 
\texorpdfstring{$H^4$}{H4}.}
Let $\psi \in \C^4$ be a spinor of  $\mso(4,1)$  with normalization $\bar \psi 
\psi = 1$, and consider the following Hopf map:
 \begin{align}
  \label{Hopf}
 \begin{aligned}
  x^a: \quad \C P^{1,2}  &\to H^4 \ \subset \R^{1,4}   \\
 \psi\ &\mapsto  x^a = \frac r2\bar{\psi} \g^{a} \psi , \qquad a = 0,1,2,3,4 \, 
 \end{aligned}
\end{align}
noting that the  phase of $\psi$ drops out. 
Here $\C P^{1,2}$ is defined as space of 
unit spinors $\bar\psi \psi = 1$ modulo $U(1)$, and $r$ introduces a length 
scale. One can  verify that 
\begin{align}
 \sum_{a,b=0}^4 \eta_{ab} x^a x^b = -\frac {r^2}4 \eqqcolon -  R^2
 \label{xx-class-bundle}
\end{align}
implying that the right-hand-side is indeed in $H^4$.
The map \eqref{Hopf} is a non-compact version of the Hopf map $\C P^3 \to S^4$,
and, moreover, it
is an $SO(4,1)$-intertwiner provided the $x^a$ transform as $SO(4,1)$ vectors.
The fiber can be seen to be $S^2$ such that  $\C P^{1,2}$ is an
equivariant $S^2$-bundle over $H^4$; for more details, we
refer to \cite{Sperling:2018xrm}.
Note that the  metric on the $H^4 \subset \R^{1,4}$  
is  Euclidean, despite the $SO(4,1)$ metric on target space. 
This is obvious at the point 
$\xi=(R,0,0,0,0)$, where the tangent space is $\R^4_{1234}$.
\paragraph{$SO(4,2)$ formulation and embedding functions.}
As mentioned above, $\C P^{1,2}$ is a 6-dimensional 
coadjoint orbit of $SU(2,2)$,
given by
\begin{align}
 \C P^{1,2} \  \cong \ \{U^{-1} Z U, \quad U\in SU(2,2) \} \ \hookrightarrow \ \msu(2,2) 
 \label{SU22-orbit}
\end{align}
where $Z$ is a rank one $4\times 4$ matrix defined by
\begin{align}
 Z = \psi\bar\psi, \qquad Z^2 = Z, \qquad \tr(Z) = 1 , \qquad
 Z^\dagger = \g^0 Z {\g^0}^{-1} \ . 
 \label{Z-relations}
\end{align}
The embedding \eqref{SU22-orbit} is  described by 
the  embedding functions 
\begin{align}
 \label{Mab-trace-Z}
 \begin{aligned}
 m^{ab} &= \tr(Z \Sigma^{ab}) = \obar \psi\Sigma^{ab}\psi = (m^{ab})^*, \\
 x^a &=  r \, m^{a5}, \qquad a,b=0,\ldots,4
\end{aligned}
\end{align}
noting that $\frac{1}{2}\g^a =  \Sigma^{a5}$.
Upon restricting to $\mso(4,1)\subset\mso(4,2) \cong \msu(2,2)$, we recover 
\eqref{Hopf}, which reflects the transitivity of the $SO(4,1)$ action on $\C 
P^{1,2}$.
The last equation in \eqref{Mab-trace-Z} encodes the Hopf map, 
which will generalize to the non-commutative case.
The $SO(4,2)$ structure is often useful, but it does not respect 
the projection to $H^4$. 

We briefly recall some of the resulting algebraic relations, which are 
derived and discussed  
in \cite{Sperling:2018xrm}.
One can  compute the invariant functions
\begin{align}
\sum_{0\leq a<b\leq 4} m^{ab} m_{ab} 
  &= \frac{1}{2} \; ,
 \label{mm-id-so41} \\
\sum_{0\leq a<b\leq 5} m^{ab} m_{ab} 
  &= \frac{3}{4} \; .
  \label{mm-id-so42}
\end{align}
 Here, the indices are raised and lowered with $\eta_{ab} = 
\diag(-1,1,1,1,1,-1)$. Recalling \eqref{Mab-trace-Z}, we recover
\begin{align}
  x_a x^a  \equiv \sum_{a=0}^4  x_a x^a = -\frac {r^2}4 = - R^2 \,.
  \label{eq:radial_constraint_H4}
\end{align}
It is remarkable that the $SO(4,1)$-invariant $x^a x_a$ is constant on 
the $SO(4,2)$ orbit $\C P^{1,2}$.
Similarly, the relation $Z^2 = Z$ of \eqref{Z-relations} entails the $SO(4,2)$ 
identity
\begin{align}
 \sum_{c,d=0}^5 \eta_{cd}m^{ac} m^{bd} = \frac{1}{4} \eta^{ab}  , \qquad 
a,b=0,\ldots,5 \,,
   \label{so42-MM-rel-class}
\end{align}
which reduces to the $SO(4,1)$ relation
\begin{align}
\eta_{cd}m^{ac} m^{bd}  - r^{-2} x^a x^b   &= \frac{1}{4} \eta^{ab}  , \qquad 
a,b=0,\ldots,4 \ .
   \label{so41-MM-rel-class}
\end{align}  
 In particular, \eqref{so41-MM-rel-class} implies that $m^{ab}$ is orthogonal to 
$x^a$, i.e.\
\begin{align}
 x_a m^{ab} = 0 , \qquad 
b=0,\ldots,4\ .
 \label{m-x-orthogonal}
\end{align}
Furthermore, the following $SO(4,2)$ identity holds
\begin{align}
\epsilon_{abcdef}m^{ab} m^{cd} &=  4 m_{ef} \,,
  \label{epsilonMM-id-1}
 \end{align}
 which reduced to $SO(4,1)$ implies
 \begin{align}
  \epsilon_{abcde} m^{ab} m^{cd} = -\frac 4r\, x_e \ , \qquad e =0,\ldots,4  .
   \label{epsilonMM-id-2}
 \end{align}
Finally, there exists a self-duality relation 
\begin{align}
 \epsilon_{abcde}m^{ab} x^{c} &= m_{de} \ .
 \label{selfdual-class}
\end{align}
Thus $m^{ab}$ is a tangential self-dual rank 2 
tensor on $H^{4}$,
 in complete analogy to the fuzzy 4-sphere $S^4_N$ discussed in  
\cite{Sperling:2017dts}.
One can thus express $m^{ab}$ 
in terms of the $SO(4)$ t'Hooft symbols
\begin{align}
  m^{\mu\nu} = \eta_{\mu\nu}^i\, J_i ,  \qquad\quad   J_i J^i = 1
\label{tHooft-classical}
\end{align}
where $ J_i,\ i=1,2,3$ describes the internal $S^2$. 
This exhibits the structure of $\C P^{1,2}$ is an $SO(4,1)$-equivariant bundle over $H^4$. 
The $S^2$ fiber is generated by the local $SU(2)_L$, while $SU(2)_R$ acts 
trivially.
\paragraph{Projection to  \texorpdfstring{$H^{2,2}$}{H22}.}
Alternatively, the $SO(4,2)$ homogeneous space $\C P^{1,2}$ can be 
viewed as $S^2$-bundle over $H^{2,2}$, which arises from  
the $SO(3,2)$-equivariant Hopf map
\begin{align}
\label{eq:Hopf_H22}
\begin{aligned}
 t^a:\quad \C P^{1,2} &\to H^{2,2} \subset \R^{2,3}  \\
   \psi &\mapsto \frac 1R  \obar\psi \Sigma^{a4}\psi   =  \frac{1}{R}  m^{a4}, \qquad 
a=0,1,2,3,5 \,. \ 
\end{aligned}
\end{align}
Now, the coordinate functions $t^a$ define a hyperboloid $H^{2,2}\subset 
\R^{3,2}$ with intrinsic signature $(+,+,-,-)$. 
Using analogous identities as before, we obtain the constraints 
\begin{align}
 \label{p-normalization-class}
\begin{aligned}
 \tilde\eta_{ab} t^a t^b &= r^{-2} , \qquad \tilde\eta_{ab} = 
\diag(-1,1,1,1,-1)  \; ,\\
 t_a x^a &= 0 \equiv t_\mu x^\mu \; , 
 \qquad \mu = 0,1,2,3 \,.
 \end{aligned}
\end{align}
The last relation in \eqref{p-normalization-class} follows from 
\eqref{so42-MM-rel-class}, noting that $t^4 \equiv 0$.
%
%
\paragraph{\texorpdfstring{$SO(3,1)$}{SO(3,1)}-invariant projection and 
Minkowski signature.}
Neither of the  spaces $H^4$ or $H^{2,2}$   considered so far has
Minkowski signature. Fortunately, space-times with Minkowski signature can be 
obtained by $SO(3,1)$-covariant 
projections of the above hyperboloids. 
Explicitly, consider the  projections \cite{Sperling:2018xrm,Steinacker:2017bhb}
\begin{alignat}{2}
  \label{proj-class}
\begin{aligned}
 \Pi_x: \quad \C P^{1,2} \ &\to \R^{1,3}_x & & \\
               m &\mapsto x^\mu = r m^{\mu 5} \,,&\qquad \mu &= 0,1,2,3 \,, \\
\Pi_t: \quad \C P^{1,2} \ &\to \R^{1,3}_t  & &\\
               m &\mapsto t^\mu = R^{-1} m^{\mu 4}
               \,,&\qquad \mu &= 0,1,2,3 \,,
 \end{aligned}
\end{alignat}
which respect $SO(3,1)$.
In section  \ref{sec:projection-Lorentzian},
the image $\cM^{3,1}\subset \R_x^{3,1}$ of $\Pi_x$ will serve as cosmological FLRW 
space-time with $k=-1$ and initial singularity due to $x^0 > 0$. 
The local stabilizer $SO(3)$ acts on the $S^2$ fiber.
In contrast, the image of  $\Pi_t$ covers the space-like region of $\R^{3,1}_t$
with norm $t_\mu t^\mu \geq r^2$,
and the local stabilizer $SO(2,1)$ acts on the $H^{1,1}$ fiber.
This  does not seem to give an interesting space-time, 
but it will be related to momentum space.
%
%
\subsection{The fuzzy hyperboloid \texorpdfstring{$H^4_n$}{H4n}}
Now we consider the quantization of the geometry discussed above. This is based on
the fuzzy 4-hyperboloid $H^4_n$, 
which is a quantization of the $S^2$-bundle $\C P^{1,2}\to H^4$ with the 
canonical Poisson structure on $\C P^{1,2}$.
Fuzzy $H^4_n$ was introduced in \cite{Hasebe:2012mz} and further developed in  
\cite{Steinacker:2017vqw}, and we briefly recapitulate the main results. 
As for any coadjoint orbit, the canonical quantization of $\C P^{1,2}$ proceeds 
in terms of the operator algebra $\End(\cH_n)$
where $\cH_n$ is a suitable unitary irreducible representation (irrep) of 
$SU(2,2)\cong SO(4,2)$.
The representation is chosen such that 
the Lie algebra generators 
$\cM^{ab} \in \End(\cH_n)$ generate a quantized 
algebra of functions, interpreted as fuzzy $\C P^{1,2}_n$.
The $\cM^{ab}$ are naturally viewed as quantized coordinate 
functions $m^{ab}$ on $\mso(4,2)$.
Analogously 
to the Hopf map \eqref{Hopf}, fuzzy $H^4_n$ is defined via a 
projection of $\C P^{1,2}_n$ that is generated by Hermitian operators $X^a \sim x^a$, 
transforming as vectors under $SO(4,1)\subset SO(4,2)$.

To define fuzzy $H^4_n$ explicitly, let 
$\eta_{ab} = \diag(-1,1,1,1,1,-1)$ be the invariant metric of $SO(4,2)$, and
$\cM_{ab}$ be Hermitian generators of $\mso(4,2)$ which satisfy
\begin{align}
  [\cM_{ab},\cM_{cd}] &=\im \left(\eta_{ac}\cM_{bd} - \eta_{ad}\cM_{bc} - 
\eta_{bc}\cM_{ad} + \eta_{bd}\cM_{ac}\right) \ .
 \label{M-M-relations-noncompact}
\end{align}
We choose a particular type of (discrete series) 
positive-energy unitary irreps $\cH_{n}$ 
known as \emph{minireps} or \emph{doubletons} 
\cite{Mack:1975je,Fernando:2009fq}, which, remarkably, remain irreducible under 
the restriction to $SO(4,1) \subset SO(4,2)$.
This follows from the minimal oscillator construction of $\cH_n$, 
where all $SO(4,2)$ weight multiplicities 
are at most one, cf.\ \cite{Mack:1969dg,Mack:1975je,Heidenreich:1980xi}. 
Strictly speaking there are two versions 
$\cH_{n}^L$ or $\cH_{n}^R$ with opposite chirality, 
but this distinction is irrelevant in the present paper and can therefore 
be dropped.

The doubletons $\cH_n$ have positive discrete spectrum 
\begin{align}
 {\rm spec}(\cM^{05}) = \{E_0, E_0+1, \ldots \}, \qquad E_0 = 1+\frac{n}2 
\end{align}
where the 
eigenspace with lowest eigenvalue of $\cM^{05}$ is an $(n {+} 1)$-dimensional 
irreducible
representation of either $SU (2)_L$ or $SU (2)_R$. Then the Hermitian generators
\begin{alignat}{2}
 X^a &\coloneqq r\cM^{a 5}, & \qquad a &= 0,1,2,3,4  \,,\nn\\
 T^a &\coloneqq R^{-1} \cM^{a 4}, & \qquad a &=0,1,2,3,5  
\end{alignat}  
 satisfy 
 \begin{subequations}
 \label{basic-CR-H4}
 \begin{align}
  [X^a,X^b] &= -  \im\, r^2\cM^{ab}  \eqqcolon \im \Theta^{ab} \,, \\
   [T^\mu,X^\nu] &= \frac{r}{R} [\cM^{\mu 4},\cM^{\nu 5}] = \im  
\frac{1}{R}\eta^{\mu\nu} X^4 \,, \\
  [T^\mu,X^4] &= \frac{r}{R} [\cM^{\mu 4},\cM^{4 5}] = - \im \frac{1}{R}  
X^\mu \,,\\
[T^5,X^\nu] &= \frac{r}{R} [\cM^{5 4},\cM^{\nu 5}] = -\im \,r T^\nu \,, \\
[T^\mu, T^\nu] &= \frac{\im}{R^2} \cM^{\mu \nu} = -\frac{\im}{r^2 R^2} 
\Theta^{\mu  \nu} \,.
\end{align}
\end{subequations}
The $X^a$ transform as $SO(4,1)$ vectors, i.e.\
\begin{align}
    [\cM_{ab},\cM_{cd}] &=\im \left(\eta_{ac}\cM_{bd} - \eta_{ad}\cM_{bc} - 
\eta_{bc}\cM_{ad} + \eta_{bd}\cM_{ac}\right) 
 \label{M-M-relations} \\
  [\cM_{ab},X_c] &=  \im\left(\eta_{ac} X_b - \eta_{bc} X_a\right) ,
 \label{M-X-relations}
\end{align}
for $a,b,c,d\in \{0,\ldots,4\}$.
Since the restriction $SO(4,1) \subset SO(4,2)$ is irreducible, it 
follows that the $X^a$ live on a hyperboloid, 
\begin{align} 
\sum_{a,b= 0,1,2,3,4}  \eta_{ab} X^a X^b &= \sum_{a=1}^4 X^a X^a - X^0 X^0 = - 
R^2 \one \   
  \label{X-constraint}
\end{align}
with radius \cite{Sperling:2018xrm}
\begin{align}
 R^2 = \frac{r^2}{4}(n^2-4) \label{eq:radius}
\end{align}
and 
\begin{align}
 \sum_{a,b= 0,1,2,3,4,5}  \eta_{ab} \cM^{ac} \cM^{bd} + (c\leftrightarrow d) = 
\frac{1}{2}(n^2-4) \eta^{cd} = 2 \frac{R^2}{r^2} \eta^{cd}  \,.
 \label{MM-contraction-general}
\end{align}
Since $X^0 = r \cM^{05} > 0$ has positive spectrum, this describes a one-sided hyperboloid in $\R^{1,4}$,
denoted as $H^4_n$.
Similarly, the $T^{a}$, for $a=0,1,2,3,5$ transform as $SO(3,2)$ vectors and 
satisfy
\begin{align} 
 \sum_{a,b= 0,1,2,3,5} \eta_{ab} T^a T^b &= \sum_{i=1}^3 T^i T^i  - T^5 
T^5  - T^0 T^0 = \frac 1{r^2} \one \  
 \label{hyperboloid-T-constraint}
\end{align}
 and the $T^a$ generate a hyperboloid $H^{2,2}$ with signature $(--++)$.
 This reflects the fact that the reduction to $SO(3,2)\subset SO(4,2)$  of 
$\cH_n$ is also irreducible.
Further identities include 
\begin{align}
 \label{X-T-M}
\begin{aligned}
\sum_{b=0}^4 \cM^{ab}X_b + X_b \cM^{ab} &= 0 , \qquad  a=0,1,2,3,4 \:,\\
\sum_{b=0,1,2,3,5} \cM^{ab}T_b + T_b \cM^{ab} &= 0 ,  \qquad 
 a=0,1,2,3,5 \,.
 \end{aligned}
\end{align}
Analogous to fuzzy $S^4_N$, the semi-classical geometry underlying  
$H^4_n$ and $H^{2,2}$ is $\C P^{1,2}$ \cite{Hasebe:2012mz},
which is an $S^2$-bundle over $H^4$ or $H^{2,2}$ carrying a canonical symplectic 
structure, respectively. 
In the fuzzy case, the typical fiber becomes a fuzzy 2-sphere $S^2_n$.
However, for most parts of this paper, we work again in the semi-classical 
limit.
It is important to note that the 
induced metric on the hyperboloid $ H^4\subset \R^{1,4}$ is  Euclidean, despite 
the $SO(4,1)$ isometry. 
This can be seen at the point 
$x=(R,0,0,0,0)$, where the tangent space is $\R^4_{1234}$.
\paragraph{Coherent states and quantization map.}
Since $\cH_n$ are lowest weight representations,  there is a natural definition of coherent states on $\C P^{1,2}_n$,
which are defined to be the $SO(4,2)$ orbits of the lowest weight state in $\cH_n$.
The ambiguity in the choice of the  group element
$g\in SO(4,2)$ leads to a $U(1)$ phase ambiguity, so that the coherent states 
$\{|z\rangle\}$ furnish a $U(1)$-bundle over $\C P^{1,2}$,
or a $U(2)$-bundle over $H^4$. For more details on coherent states we refer to
\cite{Perelomov:1986tf,Grosse:1993uq,Hawkins:1997gj,Ishiki:2015saa,Ishiki:2018aja}.

Given these coherent states, there is a natural $SO(4,2)$-equivariant
quantization map from the classical space of functions on $\C P^{1,2}$
to the fuzzy functions $\End(\cH_n)$:
\begin{align}
 \label{quantization-map}
\begin{aligned}
 \cQ: \quad \cC(\C P^{1,2}) &\to \End(\cH_n)  \\
 \phi(z)  &\mapsto \hat \phi = \int\limits_{\C P^{1,2}} d\mu\, \phi(z) 
\left|z\right\rangle \left\langle z\right| \ .
\end{aligned}
\end{align}
Up to a cutoff, this map is one-to-one\footnote{To prove this, 
one can show that the decomposition 
into irreps is the same, cf.\ \cite{Pawelczyk:2002kd}. }, and
the inverse is given by the symbol 
\begin{align}
 \hat{\phi} \in \End(\cH_n) \mapsto 
\langle z | \hat{\phi} |z\rangle \ 
= \phi(z) \in  \cC( \C P^{1,2} ) \,,
 \label{quantization-inverse}
\end{align}
up to mode-dependent normalization.
Since $\cQ$ respects $SO(4,2)$, the generators act as
\begin{align}
  [\cM^{ab},\cQ(\phi)] &= \cQ(\{m^{ab},\phi(z)\})
  \label{Q-intertwiner}
\end{align}
since $\{m^{ab},\cdot\}$ implements the $SO(4,2)$ action on  $\cC(\C P^{1,2})$.
In particular,
the Laplacian operator(s) are respected too, e.g.
\begin{align}
 \Box \cQ(\phi) 
 = \cQ(\Box_{cl} \phi)
 \label{Laplacian-intertwiner}
 \qquad 
 \text{with} 
 \qquad 
 \Box \ = \  [T^\mu,[T_\mu,\cdot]]
 \,, \quad 
\Box_{cl}  \ =  \ -\{t^\mu,\{t_\mu,\cdot\}\} \,,
\end{align}
and similarly for other operators based on $SO(4,2)$.
%
%
\subsection{Projected quantum space-time   \texorpdfstring{$\cM^{3,1}_n$}{M(3,1)} from \texorpdfstring{$H^4_n$}{H4n}}
\label{sec:projection-Lorentzian}
In this paper, we study the $SO(3,1)$-invariant fuzzy space-time 
generated by the coordinates $X^\mu$ for $\mu=0,1,2,3$.
As introduced in \cite{Steinacker:2017bhb},
this space-time can be viewed as 
projection  $\Pi_x$ of $H^4$ onto the $0123$ plane, and 
describes a homogeneous and isotropic quantized cosmological 
space-time\footnote{We change notation from \cite{Steinacker:2017bhb}, 
where $Y^1$ was dropped instead of $Y^4$.} $\cM_n^{1,3}$.
The relations \eqref{X-constraint} and  \eqref{hyperboloid-T-constraint} lead to
\begin{align}
 \eta_{\mu\nu} X^{\mu} X^{\nu}  &= -R^2 -  X^4 X^4, \nn\\
 \eta_{\mu\nu} T^{\mu} T^{\nu}  &= \frac 1{r^2} + \frac 1{r^2 R^{2}}\, X^4 X^4, \qquad \mu,\nu = 0,\ldots,3 
 \label{P-constraint}
\end{align} 
recalling that $r R\, T^5=-X^4$.
Finally, the constraint \eqref{MM-contraction-general}  yields 
\begin{align}
0 =  X_\mu T^\mu + T^\mu X_\mu  \ .
\label{X-X-orthogonal}
\end{align}
The cosmological $\cM^{3,1}$ background embedded by  $X^\mu \sim x^\mu$ is 
covariant under $SO(3,1)$, which is the symmetry respected by  $\Box$. 
The $T^\mu$ will serve as momentum generators on $\cM_n^{1,3}$, even though they 
could considered 
as defining a quantum space in their own right \cite{Steinacker:2017bhb}.
%
%
\subsection{Semi-classical structure of \texorpdfstring{$\cM^{3,1}$}{M(3,1)}}
Now we discuss the semi-classical limit  $\cM^{3,1}$ 
 and its associated bundle structure in more detail.
The basic object is 
the space $\cC$ of functions on $\C P^{2,1}$, which can be viewed as equivariant bundle 
over $\cM^{3,1}$ or $H^4$, i.e.\ as module  over the algebra $\cC^0\subset \cC$ of 
functions on the base. 
The $SO(4,1)$-covariant functions $\theta^{ab}$ and $x^a$ on 
$\C P^{2,1}$ over $\cM^{3,1}$ separate as follows: the $x^\mu$  
serve as coordinates on $\cM^{3,1}$ and generate the algebra $\cC^0 \subset \cC$ 
of functions\footnote{Strictly speaking the $x^\mu$ cannot distinguish the two 
sheets of $\cM^{3,1}$, which requires e.g. $x^4$.}
on $\cM^{3,1}$, 
whereas the $\theta^{\mu\nu}$ and $t^\mu$ generate the module $\cC$
over $\cC^0$.
The constraints \eqref{MM-contraction-general} imply the following relations on 
the  generators:
\begin{subequations}
\label{geometry-H-M}
\begin{align}
 x_\mu x^\mu &= -R^2 - x_4^2 = -R^2 \cosh^2(\eta) \, ,  \\
 t_{\mu} t^{\mu}  &=  r^{-2}\, \cosh^2(\eta) \,, \\
 t_\mu x^\mu &= 0, \ \\
 t_\mu \theta^{\mu\a} &= - \sinh(\eta) x^\a , \\
 x_\mu \theta^{\mu\a} &= - r^2 R^2 \sinh(\eta) t^\a , \\
 \eta_{\mu\nu}\theta^{\mu\a} \theta^{\nu\b} &= R^2 r^2 \eta^{\a\b} - R^2 r^4 
t^\a t^\b + r^2 x^\a x^\b  \\
 \theta^{\mu\nu}\theta_{\mu\nu} &= 2R^2r^2\big(2-\cosh^2(\eta)\big)
\end{align}
\end{subequations}
where $\mu,\a = 0,\ldots ,3$.
Here $\eta$ serves as a global time coordinate, and $x_4$ is a measure for the 
current 
size i.e.\ the curvature scale of the universe.
From the radial constraint \eqref{eq:radial_constraint_H4} on 
$H^4$  one deduces $\{x_a x^a, 
x^\mu\} = 0$, which further implies  
\begin{align}
 0 &= \sum_{a=0}^4 x_a m^{a\mu} = \sum_{\nu =0}^3 x_\nu m^{\nu \mu} + x_4 
m^{4\mu} \ .
 \label{radial-constraint}
\end{align}
This establishes a relation between the momenta and the  $t^{\mu}$,
\begin{align}
 t^\mu &=  \frac 1R\, m^{\mu4} = -\frac{1}{R r^2 x^4}\, x_\nu\theta^{\nu\mu} 
  \ \stackrel{\xi}{=} \ \frac{1}{R r^2}\, \frac{1}{\tanh(\eta)}\theta^{0\mu} 
\,,
\end{align}
where $\xi$ is a reference point  on $H^4$ which can be chosen as
\begin{align}
 \xi = (x^0,0,0,0,x^4) \stackrel{\Pi}{\to} \ (x^0,0,0,0) , \qquad x^0 = 
R\cosh(\eta),  \ \ x^4 = R\sinh(\eta) \ 
\end{align}
via $SO(3,1)$-invariance.
Furthermore, the self-duality constraint in $H^4_n$ \cite{Sperling:2018xrm}
\begin{align}
 \epsilon_{abcde} \theta^{ab} x^{c} &= n r\theta_{de}  
 \label{SD-H-class}
\end{align}
 reduces to 
\begin{align}
\label{SD-H-ref_point}
\begin{aligned}
t^i &= \frac 1R m_{i 4} = \frac 1{nRr^3} \epsilon_{abc i 4} \theta^{ab} x^{c} 
\ \stackrel{\xi}{=} \ \frac 1{nr^3} \cosh(\eta)\epsilon^{i jk} \theta^{jk}  , \\
t^0  &\stackrel{\xi}{=} \  0 \,, 
\end{aligned}
\end{align}
where the last equation is simply a consequence of $x_\mu t^\mu = 0$.
Therefore $t^\mu$ describes a space-like $S^2$ with radius 
$r^{-2} \cosh^2(\eta)$.
As a remark, the form \eqref{SD-H-ref_point} only applies in the special 
$\mso(3,1)$ adapted frame, and it is not generally covariant; 
of course on Minkowski manifolds, there is no notion of self-duality.
 Such compact extra dimensions which transform covariantly under  $SO(3,1)$
are possible only due to the  partial breaking of
Poincare covariance the present  cosmological background. 

Conversely, the relations \eqref{SD-H-ref_point} allow to express   
$\theta^{\mu\nu}$ 
 in terms of the momenta $t^\mu$ as follows:
\begin{align}
 \label{theta-in-terms-of-p}
\begin{aligned}
 \theta^{ij} &\stackrel{\xi}{=} \frac{n r^3}{2 \cosh(\eta)} \varepsilon^{ijk} \, 
t^k  \;, \\
 \theta^{0i} &\stackrel{\xi}{=}   R r^2\tanh(\eta)\,  t^i \; ,
 \end{aligned}
\end{align}
at the reference point $\xi$.
By means of $R \sim \frac{1}{2} n r$ from \eqref{eq:radius} and using $SO(3,1)$ 
covariance, one can write 
\eqref{theta-in-terms-of-p} globally as follows:
\begin{align}
 \theta^{\mu\nu} &= c(x^\mu t^\nu - x^\nu t^\mu) + b \epsilon^{\mu\nu\a\b} x_\a 
t_\b  \
 \eqqcolon r^2 R\,\eta^{\mu\nu}_\a(x)\, t^\a \  
 \label{theta-P-relation} \\
&\text{with} \qquad 
c =  r^2 \frac{\sinh(\eta)}{\cosh^2(\eta)} 
 \qquad \text{and} \qquad
 b = \frac{n r^3}{2R\cosh^2(\eta)} \,.
\end{align}
Hence 
$\eta^{\mu\nu}_\a(x)$ is a  $SO(3,1)$-invariant tensor field on 
$\cM^{3,1}$, which is analogous of the t'Hooft symbols.
Moreover, note that $\theta^{0i} \gg \theta^{ij}$ for late times $\eta \gg 1$;
this reflects the embedding of $H^4 \subset \R^{4,1}$ which approaches the light cone at late times.
Consequently, space is almost commutative, but space-time is not.
The effects of non-commutativity will be further weakened due to the 
averaging on $S^2$.
\paragraph{Poisson calculus.}
We can  define a canonical  ''Poisson'' connection on $\cC$, viewed as module\footnote{This means that 
$\cC$ describes a vector bundle over $\cM^{3,1}$.  This 
connection respects the refined $\cC^s$ modules, which are introduced 
in section \ref{sec:spin-casimir}. The construction  could  be generalized to the noncommutative case,
but we restrict ourselves to the semi-classical geometry here
for simplicity.}  
over the functions $\cC^0$ on $\cM^{3,1}$, by 
\begin{align} 
 \nabla_{e_\mu} \coloneqq  \{t_\mu,\cdot\}: \quad  \cC \to \cC,
 \quad \qquad  e_\mu  = \frac{x^4}{R}\del_\mu .
\label{tilde-nabla}
\end{align}
Here $e_\mu$ are vector fields on $\cM^{3,1}$ defined in terms of the Cartesian coordinates $x^\mu$.
Explicitly,
\begin{align}
\label{}
\begin{aligned}
\nabla_{e_\mu}  x^\nu &=  \frac{x^4}{R}\d_\mu^{\nu} \,,\\
 \nabla_{e_\mu} \theta^{\a\b} &=r^2 \left( -\d_\mu^\a t^\b +  \d_\mu^\b t^\a 
\right) \,, \\
 \nabla_{e_\mu}  t^{\a} &= -\frac{1}{r^2 R^2} \eta_{\mu\nu}\theta^{\nu\a} \, .
\end{aligned}
\end{align}
Note that $\nabla$ defines an $SO(3,1)$-equivariant connection on each $H^3$.
However $SO(3,1)$-invariance does not uniquely determine a connection on $\cM^{3,1}$,
since any conformal rescaling of $\eta^{\mu\nu}$ defines a different 
FLRW metric and, thus, a different connection on $\cM^{3,1}$.
Nevertheless, $\nabla$ is the unique connection that respects the symplectic 
form, as discussed below. 
This amounts to an extra structure on $\cM^{3,1}$ which is not present in Riemannian geometry.
We also note the following identity:
\begin{align}
 \{\cM^{\mu\nu},\phi\} &= - (x^\mu\del^\nu - x^\nu \del^\mu)\phi , \qquad \phi 
\in \cC^0 \  .
  \label{M-ad-explicit}
 \end{align}
 Furthermore, \eqref{geometry-H-M} implies  
\begin{align}
 \del_{\mu} x_4= -\frac{1}{x^4}  x_\mu 
 \qquad 
 \text{and}
 \qquad
 \del_{\mu} f(\eta) = -\frac{x_\mu}{R^2} 
\frac{1}{\sinh(\eta)\cosh(\eta)}\frac{\del}{\del \eta} f(\eta) \,.
 \label{del-eta-identities}
\end{align}
\paragraph{Integration and measure.}
As for any quantized coadjoint orbit, the trace on $\End(\cH_n)$ corresponds to 
the integral over the underlying symplectic space, i.e.\
\begin{align}
 \Tr \cQ(\phi) =  \int_{\C P^{1,2}} \dd \Omega\, \phi \ = \int_{H^4} \rho_H 
[\phi]_0 \ 
 = \int_{\cM^4} \rho_M [\phi]_0
 \label{integral-trace}
\end{align}
where $\dd\Omega$ 
is the  $SO(4,2)$-invariant volume form on $\C P^{1,2}$
arising from the (Kirillov-Kostant) symplectic form.
We aim to find the explicit form on $H^4$ and $\cM^{3,1}$ in terms of the 
cosmological time $\eta$,
which determines the $SO(3,1)$-invariant space-like slice 
 $H^3 \subset H^4\subset \R^{4,1}$ via $x_\mu x^\mu =   -R^2 \cosh^2(\eta)$  
from \eqref{geometry-H-M}.
At the reference point $\xi$,
 $x_0 = R\cosh(\eta)$ and $\dd x_0 = R\sinh\eta \dd\eta$ such that 
the  $SO(4,1)$-invariant volume form on $H^4$ can be written as
\begin{align}
 \rho_H \  &= \frac 4{L_{NC}^4 } R \  \dd \eta \dd\Omega_3\, 
   = \frac 4{L_{NC}^4 } \frac{1}{\sinh(\eta)}   \dd x_0 \ldots \dd x_3   \ .
 \end{align}
 where $\dd\Omega_3$ is the induced volume form $H^3$. 
 This can be  pulled back to $\cM^{3,1}$ using $\sinh(\eta) = \frac{x_4}{R}$, 
which yields
 \begin{align}
 \rho_M \ &= \ \rho_M(x)\, \dd x_0 \ldots \dd x_3 , 
 \qquad \rho_M(x)  = \frac{4R}{L_{NC}^4 \, x_4}  \,.
 \label{measure-explicit}
\end{align}
Since the symplectic volume form is invariant, we have 
\begin{align}
 \int_\cM \rho_M f (\nabla g)  = - \int_\cM \rho_M (\nabla  f)\, g \,,
\end{align}
which in Cartesian coordinates reduces to 
\begin{align}
 \int_\cM \rho_M f x^4(\del_\mu g) = - \int_\cM \rho_M x^4 (\del_\mu f)\ g  \,,
\end{align} 
in accordance with \eqref{measure-explicit}.
\paragraph{Averaging over $S^2$.}
The above integral \eqref{integral-trace} over the entire bundle space $\C 
P^{1,2}$ 
can be viewed as an integral over the fiber followed by an integral over the base manifold 
$H^4$. Let us discuss the former in some detail:
We define the averaging $[f(t)]_0$ as an integral over the $S^2$ fiber described by the $t$ generators,
\begin{align}
 [f(t)]_0 &= \frac{1}{4\pi r^{-2}\cosh(\eta)^2}\int_{S^2_t}  f(t) 
\end{align}
such that $[1]_0 = 1$.
Explicitly, the $SO(4,1)$ covariant formula of \cite{Sperling:2018xrm}
\begin{align}
  \left[\theta^{ab}  \theta^{cd}\right]_{0} 
   &= \frac{r^2 R^2}{3}  \left(P^{ac} P^{bd} - P^{bc} P^{ad} + 
\varepsilon^{abcde} 
\frac 1{R} x^e \right) \; 
\label{average-H}
\end{align}
with $P^{ac} = \eta^{ab} + R^{-2} x^a x^b\ $ gives, for example,
\begin{align}
 [t^\mu t^\nu]_0 &\eqqcolon \frac{1}{3r^2} \k^{\mu\nu} \,,
   \label{kappa-average} \\
\k^{\mu\nu}&= \cosh^2(\eta)\, \eta^{\mu\nu} + \frac{x^\mu x^\nu}{R^2} , \qquad 
\k_{\mu\nu} x^\nu = 0 \,.
\label{kappa-def}
\end{align}
Here $\kappa^{\m\n}$ 
is the unique $SO(3,1)$-compatible, positive semi-definite metric which 
projects out the time-like directions.
Similarly, one finds
\begin{align}
 \label{average-t-theta}
\begin{aligned}
\left[t^{\a}  \theta^{\mu\nu}\right]_{0} 
  &= \frac{1}{3} \Big(\sinh(\eta) ( \eta^{\a\nu} x^\mu - \eta^{\a\mu} x^\nu) 
  +  x_\b \varepsilon^{\b 4\a\mu \nu}  \Big)\,, \\
   [t^{\mu_1} \ldots t^{\mu_4}]_0 &= \frac 35 
\big([t^{\mu_1}t^{\mu_2}][t^{\mu_3} t^{\mu_4}]_0 
   + [t^{\mu_1}t^{\mu_3}][t^{\mu_2} t^{\mu_4}]_0  + 
[t^{\mu_1}t^{\mu_4}][t^{\mu_2} t^{\mu_3}]_0\big) \,.
\end{aligned}
\end{align}
More generally, one can derive a Wick-type formula
\begin{align}
 [t^{\mu_1} \ldots t^{\mu_{n}}]_0 = c_n \sum_{\rm contractions} [t t]_0\ldots [t 
t]_0 \ .
\end{align}
%
%
\subsection{Metric properties of \texorpdfstring{$\cM^{3,1}$}{M31}}
\label{sec:metric}
The effective metric on the background $Y^\mu = T^\mu$ \eqref{T-solution-4} 
can be extracted from the kinetic term for 
the fluctuations, which 
 in the matrix model has the universal form\footnote{The effective metric on fuzzy spaces is an interesting topic that
 has been discussed from various points of view \cite{Kaneko:2017zeo,Ishiki:2015saa,Ishiki:2018aja}. However on brane solutions in the matrix model, 
 there is no choice but to use 
 the present effective metric, which is encoded in the kinetic term of the action. This is  consistent with the 
 open string metric on D-branes \cite{Seiberg:1999vs,Yang:2006dk}.} 
 \cite{Steinacker:2017vqw,Steinacker:2016vgf,Steinacker:2010rh}
\begin{align}
  S[\phi] &=  R^2\, \Tr [T^\mu,\phi][T_\mu,\phi] 
  \ \sim \ - R^2\int_{\C P^{1,2}} \dd \Omega\,\{t^\mu,\phi\}\{t_\mu,\phi\}  
\nn\\ 
 &= \ -\int _{\cM^{3,1}}\,  \dd x_0 \ldots \dd x_3 \, \rho_M(x) \, 
  \g^{\mu\nu}\del_\mu \phi \del_\nu \phi \ \nn\\
   &= \ -\int_{\cM^{3,1}} \dd^4 x\, \sqrt{|G_{\mu\nu}|}\,
  G^{\mu\nu}\del_\mu \varphi \del_\nu \varphi \ 
  \label{scalar-action-G}
\end{align} 
using \eqref{integral-trace}. Some dimensionful 
constants are absorbed in $\varphi$, and the last form
is manifestly covariant\footnote{The signs are chosen according to the standard form $\cL = T-V$.}.
Here
\begin{align}
 \g^{\a\b} &= \eta_{\mu\nu}e^{\mu\a}e^{\nu\b} 
  = \sinh^2(\eta) \eta^{\a\b} \nn\\[1ex]
  e^\a  &= \{t^\a,\cdot\} = e^{\a\mu} \del_\mu , \qquad  e^{\a\mu} = 
\eta^{\a\mu} \sinh(\eta) \ ,
 \label{gamma-vielbein}
\end{align}
where $e^\a$ plays the role of a (rescaled) vielbein.
The effective metric on $\cM^{3,1}$ is then 
\begin{align}
   G^{\mu\nu} &= \a\, \g^{\mu\nu} \ , 
   \qquad \a = \sqrt{\frac 1{\tilde\rho_M^2|\g^{\mu\nu}|}} 
             =  \sinh^{-3}(\eta) \ 
   \label{eff-metric-G}
\end{align}
which is $SO(3,1)$-invariant with signature $(-+++)$. Some irrelevant 
dimensionful constants have been dropped in $\tilde\rho_M$. Consequently, 
the inverse metric of \eqref{eff-metric-G} is given by
\begin{align}
 G_{\mu\nu} &= \sinh(\eta) \eta_{\mu\nu} \, .
 \label{G-effective}
\end{align}
The metric is conformal to the induced (or closed-string) metric $\eta_{\mu\nu}$, hence 
the causal structures are the same.
\paragraph{FLRW form of the metric.}
Recall the coordinate choice on the 3-hyperboloid of \cite{Steinacker:2017bhb},
\begin{align}
 \begin{pmatrix}
  x^0 \\ x^1 \\ x^2 \\x^3 
 \end{pmatrix}
= R \cosh(\eta) 
\begin{pmatrix}
\cosh(\chi) \\
\sinh(\chi)\sin(\theta) \cos(\varphi) \\
\sinh(\chi)\sin(\theta) \sin(\varphi) \\
\sinh(\chi)\cos(\theta)
\end{pmatrix} \ .
\end{align}
Then  the $SO(1,3)$-invariant 
metric on $H^3_\eta$ has the form
\begin{align}
\dd s^2|_{H^3} = \sum_i \dd x_i^2 = R^2\cosh^2(\eta)^2 \dd \Sigma^2 \,,
\end{align}
where $\dd\Sigma^2$ is the length element on a spatial standard 3-hyperboloid 
$H^3$.
Therefore 
\begin{align}
 \dd s^2_G|_p = G_{\mu\nu} \dd x^\mu \dd x^\nu 
   &= -R^2 \sinh^3(\eta) \dd \eta^2 + R^2\sinh(\eta) \cosh^2(\eta)\, 
\dd \Sigma^2 \ \nn\\
   &= -\dd t^2 + a^2(t)\dd\Sigma^2 \,, 
\end{align}
where the scale parameter $a(t)$ is determined by the following two equations:
\begin{align}
\label{}
\begin{aligned}
a(t)^2 &=  R^2\sinh(\eta) \cosh^2(\eta) , \\
\dd t &=  R \sinh(\eta)^{\frac{3}{2}} \dd\eta \ .
\end{aligned}
\end{align}
The first equation gives
\begin{align}
 2 a \, \dd a 
  &= R^2\cosh(\eta)\left(1+ 3\sinh^2(\eta)\right) \dd \eta 
\end{align}
and combining these we obtain 
\begin{align}
  2\frac{\dd a}{\dd t} &= \sinh^{-2}(\eta)(1+ 3\sinh^2(\eta))
   = 3+\frac{1}{\sinh^2(\eta)} \ ,
\end{align}
which for late times gives
\begin{align}
 a(t)  \approx   \frac{3}{2} t, \qquad t \to \infty \,.
 \label{a-late-coast}
\end{align}
This describes a linear coasting universe as in 
\cite{Steinacker:2017vqw,Steinacker:2017bhb}, cf.\ 
\cite{Kolb:1989bg,John:1999gm,Melia:2011fj}.
For early times, 
we can approximate 
\begin{align}
 R \eta^{\frac{3}{2}}\ d \eta &= dt \ , 
 \qquad \eta\ \propto (t-t_0)^{\frac{2}{5}}
\end{align}
such that
\begin{align}
a(t) \ \propto \ \eta^{\frac 12}  \ \propto \ (t-t_0)^{\frac{1}{5}} \ .
\end{align}
Hence we obtain a reasonable FLRW cosmology that is asymptotically coasting 
at late times and a Big Bang-like initial singularity, or rather a Big Bounce, since
the flow of time is  expected to be the 
opposite\footnote{upon imposing a suitable $\im\varepsilon$ prescription; 
work in progress.} 
on the two sheets of $\cM^{3,1}$.

We note that at late times, one can relate conformal factor $\a$  of 
\eqref{eff-metric-G} to the cosmic scale parameter $a(t)$ as follows
\begin{align}
 \a \sim \sinh^{-3}(\eta) \sim \frac{R^3}{x_0^3} \sim \frac{R^2}{a(t)^2} \ .
 \label{alpha-a-relation}
\end{align}
\paragraph{Effective d'Alembertian.}
 It follows from action \eqref{scalar-action-G} of a scalar field that 
the  Laplacian (or rather d'Alembertian) on  $\cM^{3,1}$  associated to the 
metric $G_{\mu\nu}$ reads
\begin{align}
 \Box &= -\{t^\a,\{t_\a,\cdot\}\} = \a^{-1} \Box_G,
\label{eq:box-def} \\
 \text{with} \qquad 
 \Box_G &= -\frac{1}{\sqrt{|G_{\mu\nu}|}}\del_\mu\big(\sqrt{|G_{\mu\nu}|}\, G^{\mu\nu}\del_\nu\big) \nn\\
  &= -\g^{\a\b}\del_\a \del_\b -  \sinh(\eta)\eta^{\a\b}(\del_\a\sinh(\eta)) 
\del_\b \,.
 \label{G-Box-relation}
\end{align}
For example,  we obtain
\begin{align}
 \Box_G f(\eta) &= -\sinh(\eta)\del^\mu (\sinh(\eta) \del_\mu) f(\eta) 
  =  \frac{1}{R^2}\Big(\del_\eta^2 + 3\frac{\sinh(\eta)}{\cosh(\eta)}\del_\eta 
\Big) f(\eta) \,,
  \label{Box-eta-formula}
\end{align}
which in particular results in 
\begin{align}
 \Box(x^\mu x_\mu) 
  = - 2 -10 \sinh^2(\eta) 
  \qquad \text{and} \qquad  
\Box(x_\mu)  
  = \frac 1{R^2} x_\mu \,.
  \label{Box-x-ids}
\end{align}
%
\paragraph{Geometric structures and Lorentz invariance.}
Let us recapitulate the
basic geometric structures on the semi-classical space-time $\cM^{3,1}$:
\begin{compactenum}[(1)]
\item The metric $\gamma^{\mu\nu}$ or its conformal class, with Minkowski 
signature,
and the corresponding matrix or Poisson d'Alembertian $\Box$ of 
\eqref{eq:box-def}.
This allows to identify uniquely an effective metric $G^{\mu\nu}$ 
\eqref{eff-metric-G},
which defines the meaning of local Lorentz invariance.
\item The space-like metric $\kappa^{\mu\nu}$ \eqref{kappa-def} on the 
space-like sheets $H^3$.
\item The time-like  vector field 
\begin{align}
 \tau = x^\mu \del_\mu   \ = - x^4 \frac{\del}{\del x^4} \,,
 \label{tau-VF-def}
\end{align}
which is the comoving time derivative of the cosmic background, cf.\ 
\eqref{t-x-poisson-id}.
\end{compactenum}
All of these respect  the  $SO(3,1)$ isometry.
Notice that the anti-symmetric tensor $\theta^{\mu\nu}$ is not in this list,
since it vanishes upon averaging $[\cdot]_0$ over the local fiber  and, therefore, is not 
directly observable.
This is a crucial advantage over noncommutative field theory 
on more basic spaces such as \ $\R^4_\theta$ \cite{Douglas:2001ba,Szabo:2001kg}.
Clearly $\kappa_{\mu\nu}$ and $\t$ separate the effective 
metric  $G^{\mu\nu}$ into space-like and time-like components.
This breaks local Lorentz invariance, but in a mild way, respecting the global 
$SO(3,1)$ symmetry
defined by the cosmic background. A crucial question is 
whether 
or not these Lorentz-breaking structures enter significantly into 
the local physics.
We will see  that at least the propagation of all physical 
higher-spin modes is  indeed governed solely by  $G^{\mu\nu}$. 
%
%
\section{Wavefunctions, higher-spin modes and covariance}
\label{sec:wavefunct-higherspin}
Now  consider the full endomorphism algebra $\cC \coloneqq \End(\cH_n)$, which 
is interpreted 
as quantized algebra of functions on $\C P^{1,2}$. This is in 
one-to-one correspondence
with classical functions on $\C P^{1,2}$ via the quantization map 
\eqref{quantization-map}.
Due to the underlying bundle structure, $\cC$ can be viewed as space of 
higher-spin harmonics over $\cM^{3,1}$.
In this section, we show how $\cC$ separates into sub-sectors\footnote{In the 
semi-classical limit, the $\cC^s$ are modules over $\cC^0$. 
In the fuzzy case the $\cC^s$ are no longer modules and hence do not fit into the 
standard scheme of noncommutative geometry,
but nevertheless the interpretation is correct.} $\cC^s$,
which correspond to spin $s$ harmonics.
We will use the same symbols $\cC, \cC^s$ for the semi-classical and the fuzzy case.

We can define an $SO(4,2)$-invariant inner product on $\cC$ via
\begin{align}
 \left\langle\phi,\psi\right\rangle = \Tr \big(\phi^\dagger \psi\big) \  
\sim \ \int\limits_{\C P^{1,2}} \phi^\dagger \psi, 
 \qquad \phi,\psi \in \mathrm{HS}(\cH_n) \ .
 \label{inner-prod-End}
\end{align}
For polynomial functions, this trace diverges; therefore, 
we will mostly consider the analogs of square-integrable functions i.e.\ the 
space of Hilbert-Schmidt operators $\mathrm{HS}(\cH_n)$ in $\End(\cH_n)$.
The space $\mathrm{HS}(\cH_n)$ is itself a Hilbert space and a Banach algebra; 
hence, the inner product is positive definite by construction.
We will nevertheless 
use the somewhat sloppy notation $\phi\in \cC $ while implicitly using $\phi 
\in \mathrm{HS}(\cH_n)$.
\subsection{Spin Casimir}
\label{sec:spin-casimir}
We can now decompose the modes in $\cC$ into unitary representations of  
$SO(3,1) \subset SO(4,1) \subset SO(4,2)$, organized as higher-spin modes.
To achieve that, recall that the following spin Casimir on $H^4_n$ was 
introduced in \cite{Sperling:2018xrm}:
\begin{align}
\cS^2 &\coloneqq \frac 12\sum_{a,b\neq 5} [\cM_{ab},[\cM^{ab},\cdot]] 
  + r^{-2} [X_a,[X^a,\cdot]]  \nn\\
  &= 2 C^2[\mso(4,1)] - C^2[\mso(4,2)]
\label{Spin-casimir}
 \end{align}
 which commutes with $\Box_H$,
 \begin{align}
      [\cS^2,\Box_H ] = 0     \qquad \text{with} \quad 
 \Box_H =  [X_a,[X^a,\cdot]] = r^2(C^2[\mso(4,2)] - C^2[\mso(4,1)])   \, .
 \label{S^2-BoxH-commute} 
\end{align}
This means that $\Box_H$ and $\cS^2$ can be simultaneously diagonalized, defining the spin $s$ modes $\cC^s$
\begin{align}
    \End(\cH_n) =\cC = \cC^0 \oplus \cC^1 \oplus \cC^2 \oplus \ldots  \qquad \text{with} 
\quad 
  \cS^2|_{\cC^s} = 2s(s+1)  \,
  \label{EndH-Cs-decomposition}
\end{align}
as shown\footnote{in an earlier version of this paper, a truncation at $s =n$ was claimed. This truncation only arises in the image of a coherent state quantization map  \cite{Sperling:2018xrm}, but not in the full operator algebra $\End(\cH_n)$.}
in \cite{Sperling:2018xrm}.
Hence $\Box_H$ defines  a Laplacian on the $\cC^s$,
which are modules over the algebra $\cC^0$  of functions on $H^4$.
Moreover, the connection $\nabla$ of \eqref{tilde-nabla} also respects  
 $\cC^s$.
On the other hand, we  observe that $\cS^2$ also commutes with $\Box$ 
which characterizes $\cM^{3,1}_n$:
\begin{align}
    [\cS^2,\Box] = 0  
    \qquad \text{with} 
\quad 
   \Box = \Box_T \equiv [T_\mu,[T^\mu,\cdot]] = R^{-2}\left(C^2[\mso(4,1)] - 
C^2[\mso(3,1)]\right)
 \label{Box-T-SO41} 
\end{align}
because $\cS^2$ respects $SO(4,1)$.
As a consequence, the decomposition \eqref{EndH-Cs-decomposition}
also provides the decomposition of functions on $\cM^{3,1}_n$ into higher-spin 
modes.
Although the space of functions is the same for $H^4_n$ and $\cM^{3,1}_n$,
the different geometry is encoded in the Laplacian or d'Alembertian.
Here $\Box_T$ defines a d'Alembertian on $\cM^{3,1}_n$ and
encodes an effective metric with Minkowski signature, while 
$\Box_H$  defines a Laplacian on $H^4_n$ with Euclidean signature, 
as discussed in \cite{Sperling:2018xrm}.
Both respect the same decomposition into higher spin modes $\cC^s$.
\paragraph{Mode decomposition.}
Next, we can decompose $\cC = \End(\cH_n)$ into $SO(4,2)$ irreps, which 
provides a decomposition into a direct sum or integral
\begin{align}
    \cC = \int \dd \L \, V^{\mso(4,2)}_\L \quad \cong \cC(\C P^{1,2})
\end{align}
of unitary irreps $V^{\mso(4,2)}_\L$  of $\mso(4,2)$. 
Due to \eqref{quantization-map}, this decomposition is
multiplicity-free and uniquely identified
by their quadratic Casimir $C^2(\mso(4,2))$, as in the compact case.
The $V^{\mso(4,2)}_\L$ can be further decomposed into a  sum of $\mso(4,1)$ 
irreps as follows:
\begin{align}
    V^{\mso(4,2)}_\L = \bigoplus\limits_{s \geq 0}\, V^{\mso(4,1)}_{\L,s}
\end{align}
where the spin $s$ eigenvalue of \eqref{Spin-casimir} is restricted to 
integer values  $s \geq 0$, as shown in \cite{Sperling:2018xrm}. 
The elements of $V^{\mso(4,1)}_{\L,s}$ can be viewed as spin $s$ 
fields\footnote{Note that $SO(4,2)$ should \emph{not} be interpreted 
as conformal group on $H^4$, rather it provides the organization of the
higher-spin modes on $H^4$ and $\cM^{3,1}$.} on $H^4$ because  $\C P^{1,2}$ is 
an equivariant $S^2$-bundle over $H^4$, and the harmonics on $S^2$ are captured 
by $s$. 
Finally, these  $V^{\mso(4,1)}_{\L,s}$ can  be further decomposed into 
$\mso(3,1)$ irreps as
\begin{align}
    V^{\mso(4,1)}_{\L,s} &= \bigoplus\limits_{m^2}\, V^{\mso(3,1)}_{\L,s,m^2}   \nn\\
    (\Box + m^2) \phi^{(s)}_{\L,m^2} &= 0, \qquad m^2 =   R^{-2}(C^2[\mso(4,1)] - C^2[\mso(3,1)]) 
 \label{Box-eigenfunctions}
\end{align}
for some spectrum of eigenvalues $m^2$ which could a priori have any sign.
Via the quantization map $\cQ$ of \eqref{quantization-map}, 
these correspond precisely to fields with mass $m^2$ on $\cM^{3,1}$ with
effective  metric $G^{\mu\nu}$.
Quite remarkably, we will see in section \ref{sec:gaugefix} that the on-shell, 
propagating modes in the present
theory are precisely given in terms of eigenmodes of $\Box$ with $m^2=0$.
However they are expected to be non-trivial combinations of different 
$\L$ modes.
\paragraph{Tensor generators and tensor fields.}
Now consider the fluctuation modes on  $\cM^{3,1}$ more explicitly.
They can be expanded in terms of the $\theta^{ab}$ generators 
as follows:
\begin{align}
 \Phi  &= \phi(X) + \phi_{ab}(X) \cM^{ab} + \ldots \quad \in  \End(\cH_n)   \nn\\
  &\sim \phi(x) + \phi_{ab}(x) m^{ab} + \ldots \quad 
   \label{harmonics-M}
\end{align}
This will be refined below in terms of the $\theta^{\mu\nu}$ and/or $t^\mu$
generators. Then the coefficient functions $\phi_{ab}(x)$ etc.\
can  be interpreted as functions (or higher-spin tensor fields) 
on\footnote{We will ignore the dependence on two sheets $\cM^\pm$ for simplicity.}
$\cM^{3,1}$, or  more properly as sections of higher-spin bundles on $H^4$.
Due to the constraints of $\cM^{ab}$, these coefficient functions are not independent, and
there are various possibilities how to parametrize the most general modes. 
For $H^4$, this has been achieved in \cite{Sperling:2018xrm}. 
To make this more transparent, we will focus on the semi-classical limit from now on.
Then $\cC^0$ reduces to the space of functions on $H^4$
resp.\ $\cM^{3,1}$, and the $\cC^s$ are modules over $\cC^0$.
The Casimir $\cS^2$ then measures the spin of the local $SU(2)_L$ acting along the fiber.
%
%
\subsection{Space-like gauge and \texorpdfstring{$SO(3,1)$}{SO(3,1)}-covariant 
tensor fields}
\label{sec:space-like gauge}
Now we establish a  correspondence of the $\cC^s$ with the standard 
representation of higher-spin fields on $\cM^{3,1}$. We can define a map 
\begin{align}
\label{functions-expansion-momentum}
\begin{aligned}
  \Gamma^{(s)} \cM^{3,1} \  &\rightarrow  \    \cC^s  \\
 \phi^{(s)}_{\mu_1 \ldots  \mu_s}(x)   &\mapsto  \ \phi^{(s)}_{\mu_1 \ldots  
\mu_s}(x) t^{\mu_1}\ldots t^{\mu_s}
\end{aligned}
\end{align}
from rank $s$ traceless symmetric tensor field $\phi^{(s)}_{\mu_1\ldots \mu_s}$ 
on $\cM^{3,1}$ to $\cC^s$.
Due to the constraint $x_\mu t^\mu = 0$  of \eqref{geometry-H-M}
the map \eqref{functions-expansion-momentum} has a kernel, which can be viewed 
as an internal gauge invariance 
\begin{align}
 \phi_{\mu_1\ldots\mu_s} \to \phi_{\mu_1\ldots\mu_s} + x_{\mu_1} 
\phi_{\mu_2\ldots\mu_s} \, .
 \label{gauge-t}
\end{align}
We can use this to impose the following gauge-fixing condition 
\begin{align}
 x^{\mu_i} \phi_{\mu_1\ldots\mu_s} = 0 \ 
 \label{inner-gauge-fixing-tau}
\end{align}
denoted as \emph{space-like gauge}.
This gauge-fixing respects $SO(3,1)$, but is not covariant in the standard 
sense 
because it uses the  time-like vector field 
$\t = x^\mu \del_\mu$ of \eqref{tau-VF-def}. Hence, 
\eqref{inner-gauge-fixing-tau} is reminiscent of the Coulomb or radiation gauge.
In particular, the gauge-fixing implies that all time-components of 
$\phi_{\mu_1\ldots\mu_s}$ vanish, such that it
defines a tensor field on the time-slices $H^3$.
Assuming this space-like gauge, we can write down an inverse map to 
\eqref{functions-expansion-momentum}  by
\begin{align}
\begin{aligned}
   \cC^s \  &\rightarrow  \  \Gamma^{(s)} \cM^{3,1}  \\
 \phi^{(s)} &\to  \phi^{(s)}_{\mu_1 \ldots  \mu_s}(x)
 \end{aligned}
  \label{map-Cs-tensorfields}
\end{align}
resulting in a traceless symmetric space-like tensor field.
Note that \eqref{map-Cs-tensorfields} is a $\cC^0$-module map, unlike the 
analogous map 
given in \cite{Sperling:2018xrm} for $H^4$.
To see that the map \eqref{map-Cs-tensorfields} is surjective, it suffices to 
express the $\cM^{\mu\nu}$
in terms of $t^\mu$ generators using \eqref{theta-P-relation}.
This establishes a one-to-one parametrization of $\cC^s$
in terms of traceless symmetric space-like tensor fields 
$\phi_{\mu_1\ldots\mu_s}$,
which is very useful to count degrees of freedom.
More generally, one can also define maps 
\begin{align}
\begin{aligned}
  \cC^s \  &\rightarrow  \  \Gamma^{(1)} \cM^{3,1} \otimes \cC^{s-1}  \\
 \phi^{(s)} &\to   \phi^{(s)}_{\mu} \coloneqq  \phi^{(s)}_{\mu_1 \ldots  
\mu_s}(x)t^{\mu_2}\ldots t^{\mu_s}
\end{aligned}
 \label{map-Cs-tensorfields-1}
\end{align}
etc., but we will avoid such constructions because they 
do not respect the underlying higher symmetry structures.

Note that the internal gauge invariance \eqref{gauge-t} is consistent with the 
positive definite inner product \eqref{inner-prod-End}.
For example,
\begin{align}
 \left\langle\phi_\mu(x) t^\mu,\psi_\nu(x) t^\nu\right\rangle 
  = \frac{1}{3 r^2} \langle\phi_\mu^\dagger(x) \psi_\nu(x)\rangle \kappa^{\mu\nu} \ 
  \label{inner-spin-1}
\end{align}
using \eqref{kappa-average} and the definition \eqref{kappa-def} of 
$\kappa^{\mu\nu}$.
The expression \eqref{inner-spin-1} vanishes on the would-be pure gauge states 
$x_{\mu_1} \phi_{\mu_2\ldots\mu_s}$ in  \eqref{gauge-t}, and
is manifestly positive for fields satisfying the gauge-fixing condition 
\eqref{inner-gauge-fixing-tau}.
Hence the Hilbert space structure inherited from $\cH_n$  takes care of the above 
gauge freedom, and the would-be pure gauge states
are already factored out.
This is a remarkable aspect of the present framework.
\subsection{\texorpdfstring{$SO(3,1)$}{SO(3,1)} intertwiners and sub-structure}
\label{sec:D-substructure}
The above  organization can be refined further by considering the 
$SO(3,1)$  intertwiner  
\begin{align}
\begin{aligned}
 D \coloneqq -\im [X_4,\cdot] \sim \{x^4,\cdot\}: \quad V^{\mso(4,2)}_{\L} &\to 
V^{\mso(4,2)}_{\L}  \\
      V^{\mso(4,1)}_{\L,s} &\mapsto V^{\mso(4,1)}_{\L,s+1} \oplus  
V^{\mso(4,1)}_{\L,s-1}
\end{aligned}
      \label{D-def}
\end{align}
which is  anti-Hermitian in the sense 
\begin{align}
 \int \phi^{(s')} D\phi^{(s)} &= - \int  D\phi^{(s')}\, \phi^{(s)} \ .
 \label{D-adjoint}
\end{align}
Since $D$ is a derivation, it is determined by its action on the generators
\begin{align}
 D(X^\mu) = r^2 R T^\mu  \qquad \text{and} \qquad 
 D(T^\mu) =  R^{-1} X^\mu \,, 
 \label{K-xt-explicit}
\end{align}
which is essentially an exchange of $X^\mu$ and $T^\mu$.
$D$ is a non-compact generator of $SO(4,2)$, 
which transforms as vector under $SO(4,1)$.
Hence  in unitary representations, it should be unbounded in both directions.
Note that $D$ does not commute with $\cS^2$, and it changes\footnote{This can be seen easily by looking at polynomials.} the quantum number $s$ by $\pm 1$, 
cf.\ \eqref{Box-x4-relations}. 
We can therefore separate $D$ into raising and lowering operators
\begin{align}
 D = D^+ + D^-: \quad \cC^{s} &\to \cC^{s+1} \oplus   \cC^{s-1}  
   \label{cQ-def}
\end{align}
 where 
 \begin{align}
 \begin{aligned}
    D^- &= -\im[X_4,\cdot]_-: \quad  \cC^s \to \cC^{s-1}, \qquad  D^-\phi^{(s)} 
\coloneqq -\im[X_4,\phi^{(s)}]_{s-1}  \\
    D^+ &= -\im[X_4,\cdot]_+:\quad  \cC^s \to \cC^{s+1},  \qquad D^+ \phi^{(s)} 
\coloneqq -\im[X_4,\phi^{(s)}]_{s+1} 
\end{aligned}
    \label{Dpm-def}
\end{align}
and similarly in the Poisson case.
 Since modes with different spin are orthogonal, it follows that 
\begin{align}
 D^+ = - (D^-)^\dagger \,.
 \label{D-adjoint-pm}
\end{align}
Hence $D^+ D^-$ respects the spin and can be diagonalized  within 
$V^{\mso(4,2)}_\L $.
Next, define the subspaces
\begin{align}
 \cK^{(s,k)} \coloneqq \left\{\phi^{(s)} \in \cC^s\, \big| \quad (D^-)^{k+1} 
\phi^{(s)} = 0 \right\} \ \subset \cC^{s}
\end{align}
and the quotients
\begin{align}
 \cC^{(s,k)} \coloneqq \cK^{(s,k)} /  \cK^{(s,k-1)}  \,.
 \label{C-sk-def}
\end{align}
As we will see below, $D^-$ is essentially the 
3-divergence on the space-like leaves $H^3 \subset \cM^{3,1}$. 
Thus according to \eqref{map-Cs-tensorfields},
$\cC^{(s,0)}$ is the space of divergence-free space-like rank s symmetric tensor fields,
and the intertwiners $D^\pm$ act as follows
\begin{align}
  D^\pm: \quad \cC^{(s,k)} &\to \cC^{(s-1,k-1)}  \ .
  \label{D-refined}
\end{align}
Now the orthogonal Hilbert space decomposition 
\begin{align}
 \cH = \mathrm{Ker}(A) \oplus \mathrm{Im}(A^\dagger) 
\end{align}
for $A = D^-$ gives
\begin{align}
 \cC^{(s)} = \cC^{(s,0)} \oplus \mathrm{Im}(D^+) \ .
\end{align}
Repeating this argument for $A=(D^-)^k$, 
 we arrive at the orthogonal decomposition
\begin{align}
 \cC^{(s)} = \cC^{(s,0)} \oplus \cC^{(s,1)} \oplus  \ldots   \oplus \cC^{(s,s)}
\end{align}
where $D^-$ resp.\ $D^+$ act as isomorphism
\begin{align}
  D^-: \ \cC^{(s,k)} &\to \cC^{(s-1,k-1)}  , 
  \qquad   D^+: \ \cC^{(s,k)} \to \cC^{(s+1,k+1)} \ .
  \label{div-dagger-explicit}
\end{align}
Therefore
\begin{align}
 \cC^{(s,k)} \cong \cC^{(s-k,0)} 
\end{align}
 correspond to  symmetric traceless space-like divergence-free 
 spin $(s{-}k)$ tensor fields.
 
 Let us emphasize that there are two different notions of \emph{spin} in the 
current context: on the one hand, the quantum number $s$ labels the 
Casimir $\cS^2$ eigenvalue, 
which measures the spin on the internal $S^2$ fiber. On the other hand, as 
tensor fields on $H^3$, the $SO(3,1)$ group theory yields another notion of 
spin,
and in this sense the $\cC^{(s,k)}$ are spin $(s{-}k)$ irreps of $SO(3,1)$.

Furthermore, the inner products between the quotients \eqref{C-sk-def} have the 
following structure:
\begin{align}
 \langle \cC^{(s,k)},\cC^{(s',k')} \rangle = c_{(s,k)} \d^k_{k'} \d^{s}_{s'} \ , \qquad c_{(s,k)} > 0 \ .
 \label{Csk-orthogonality}
\end{align}
In view of this structure, it may be tempting to consider 
the $\cC^{(s-k,0)}$ as analogs of primary fields, and the $\cC^{(s,k)}$ 
for $k>0$ as descendants. However, this notion would be 
quite different from the familiar notion in CFT, where 
primaries are annihilated by special conformal transformations and are lowest weight 
modes for $D = -\im[X^4,\cdot]$. Here, $D$ is  unbounded in both directions 
due to the rigid $SO(4,2)$ signature, and $D^-$  is the lowering operator. 
We will therefore refrain from using such a language, and note that the organization is 
quite different from  CFT even though $SO(4,2)$ plays a central role.

Finally, we  observe that $D$ provides the missing part of the 
$SO(4,1)$-invariant 
Laplace operator on $H^4$ 
\begin{align}
 \Box_H = \quad \sum\limits_{a=0}^4\, [X^a,[X_a,\cdot]] \ = 
[X_\mu,[X^\mu,\cdot]] - D^2 \ .
\end{align}
%
%
\paragraph{\texorpdfstring{$D^\pm$}{D} and divergence.}
Now consider $D$ more explicitly. Due to \eqref{P-constraint},  we have 
\begin{align}
2 x_4 D(\phi )  =   \{x_4^2, \phi\} & = - \{x_\mu x^\mu, \phi\} = - 2x_\mu \{x^\mu, \phi\}  \nn\\
&= r^2 R^2 \{t^\mu t_\mu, \phi\} = 2 r^2 R^2 t^\mu \{t_\mu,\phi\} \ .
\label{D-ids-1}
\end{align}
Assuming space-like gauge, $D$ takes the explicit form 
\begin{align}
 D(\phi) &=  \frac{r^2 R^2}{x_4}  t^\mu \{t_\mu, \phi\} 
   =\frac{r^2 R^2}{x_4}   t^\mu \Big(\sinh(\eta)\del_\mu\phi_{\mu_1\ldots\mu_s}(x) t^{\mu_1}\ldots t^{\mu_s} 
    -\frac{s}{r^2 R^2}\phi_{\mu_1\ldots\mu_s}\theta^{\mu\mu_1} t^{\mu_2}\ldots t^{\mu_s} \Big) \nn\\
 &= r^2 R\, t^\mu \del_\mu\phi_{\mu_1\ldots\mu_s} t^{\mu_1}\ldots t^{\mu_s} 
  \equiv r^2 R\, (\nabla^{(3)}_\mu\phi_{\mu_1\ldots\mu_s}) t^\mu t^{\mu_1}\ldots t^{\mu_s} 
 \label{D-explicit}
\end{align}
using \eqref{geometry-H-M}, where $\nabla^{(3)}$ is the covariant derivative along the $H^3$ sheets. The last form holds because the $t^\mu$ generators 
are space-like.
Then \eqref{D-explicit} splits according to \eqref{Dpm-def} into
\begin{align}
\begin{aligned}
 D = D^- + D^+: \quad \cC^{s} \quad &\to \qquad \cC^{s-1} \quad\oplus  \quad 
\cC^{s+1}  \\
   \phi \quad &\mapsto  r^2 R [t^\mu \nabla^{(3)}_\mu \phi]_- +  r^2 R [t^\mu 
\nabla^{(3)}_\mu \phi]_+ \ ,  
\end{aligned}
\end{align}
where $D^+$  is the symmetrized total derivative along $H^3$,
and $D^-$  is the 3-divergence,
\begin{align}
  D^- \phi^{(s)} &= \b_s r^2 R \nabla_{(3)}^\mu\phi_{\mu\mu_2\ldots\mu_s} t^{\mu_2}\ldots t^{\mu_s}  \
  \eqqcolon \b_s r^2 R\, {\rm div_3}(\phi^{(s)}) \qquad \in \cC^{s-1}
  \label{D-div-general}
\end{align}
for some $\b_s$, and $\phi_{\mu\mu_1\ldots\mu_s}$ is totally symmetric and 
space-like. Then \eqref{D-div-general} is the 
space-like divergence of tensor fields on $H^3 \subset H^4$;
since the tensor fields are viewed as tensor fields on $H^4$ in space-like 
gauge, this can also be viewed as 4-dimensional divergence on $H^4$.
Then $D^+$ is determined via \eqref{D-adjoint-pm}.

Usually,  derivative modes $\cC^{(s,k)}$ with $k>0$ would 
 be considered as pure gauge fields. 
However here, all these modes are space-like and participate in 
the positive definite inner product 
\eqref{Csk-orthogonality}. Such modes will be part of the physical Hilbert space.
On the other hand,  the divergence-free  
fields $D^-\phi^{(s,0)} = 0$ will constitute the massless
spin $s$ fields from the 4-dimensional 
point of view. For example, the $\phi^{(2,0)}$ modes correspond to divergence-free rank 2 traceless 
symmetric tensor fields $\phi_{\mu\nu}$ in space-like gauge, which clearly 
have only 2 independent degrees of freedom, as appropriate for massless gravitons.
This somewhat unusual organization 
reflects the absence of manifest (local) Lorentz invariance.

In the following sections we will encounter the operator $D^- D^+$, which is essentially the 
3-dimensional Laplacian on $H^3$,
\begin{align}
 D^- D^+ \phi^{(s)} \propto (\Delta_{(3)} + c_s) \phi^{(s)} 
\end{align}
where  $\Delta_{(3)} = \nabla_{(3)}^\a \nabla^{}_{(3)\a} $.
For $s=0$, this is explicitly
\begin{align}
  D^- D^+ \phi^{(0)} &= r^2 R D^- (t^\a \del_\a \phi^{(0)})
   = r^4 R^2 [t^\a t^\b  \nabla^{(3)}_\a \del_\b \phi^{(0)}]_0  \nn\\
   &= \frac{r^2 R^2}{3} \k^{\a\b} \nabla^{(3)}_\a \del_\b \phi^{(0)} \
   = \frac{r^2 R^2}{3} \cosh^2(\eta)\Delta^{(3)} \phi^{(0)}
 \label{D-D+-phi0}
\end{align}
using \eqref{kappa-average}.
Similarly for $s=1$, the 3-divergence is
\begin{align}
 D^- \phi^{(1)} &= r^2 R[t^\a \nabla^{(3)}_\a \phi^{(1)}]_0 
  = r^2 R [t^\mu t^\a]_{0} \nabla^{(3)}_\a \phi_\mu  \nn\\
   &=  \frac{R}{3}  \cosh^2(\eta)\nabla_{(3)}^\mu \phi_\mu  
   \label{D--phi1}
\end{align}
which gives $\b_1 = \frac 1{3r^2} \cosh^2(\eta)$ in \eqref{D-div-general}.
If $\phi^{(1)}$ is divergence-free, then
\begin{align}
  D^- D^+ \phi^{(1)} &= r^2 R D^-\left[t^\a \nabla^{(3)}_\a \phi^{(1)}\right]_2 
\nn\\
   &= r^2 R D^-\left(\nabla^{(3)}_\a \phi_\mu \left[t^\mu 
t^\a\right]_{2}\right)  \nn\\
   &= r^4 R^2  \left[t^\b\nabla^{(3)}_\b \nabla^{(3)}_\a \phi_\mu \left(t^\mu 
t^\a  - \frac{1}{3r^2}\k^{\mu\a}\right) \right]_1  \nn\\
   &= \frac{4}{3}  r^2 R^2 \cosh^2(\eta) \left(\nabla_{(3)}^\a \nabla^{(3)}_\a 
\phi_\mu  t^\mu
   + \nabla_{\mu(3)} \nabla^{(3)}_\a \phi^\mu  t^\a \right)  \nn\\
   &=\frac{4}{3} r^2 R^2 \cosh^2(\eta)\left(\Delta^{(3)} \phi_\mu t^\mu
   + (R^{(3)}_{\mu\a})_{\ \nu}^\mu \phi^\nu  t^\a \right)   \nn\\
   &=\frac{4}{3} r^2 \left(R^2 \cosh^2(\eta) \Delta^{(3)} \phi^{(1)} + 
2\phi^{(1)}\right)
    \label{D+D--phi1}
\end{align}
noting that $\nabla^{(3)} \k_{\mu\nu} = 0$ and 
$[\nabla^{(3)}_\mu, \nabla^{(3)}_\b]\phi^\mu = (R^{(3)}_{\mu\b})^\mu_{\ \nu}\phi^\nu =  R^{(3)}_{\b\nu}\phi^\nu$, where 
$R^{(3)}_{\b\mu} = \frac 2{R^2\cosh^2(\eta)} P_{\b\mu}^{(3)}$ is the Ricci tensor on $H^3$  .
Here $P^{(3)}$ is the tangential projector on $H^3$.
\paragraph{Time-like vector field $\t$ as intertwiner.}
Next, we consider
\begin{align}
 t_\mu \{x^\mu,\phi^{(s)}\} &=  \theta^{\mu\nu}\del_\nu \phi_{\mu_1\ldots\mu_s}(x) t^\mu t^{\mu_1}\ldots t^{\mu_s}
  - s \sinh(\eta) \phi_{\mu \mu_2\ldots\mu_s}(x) t^\mu t^{\mu_2}\ldots t^{\mu_s}  \nn\\
  &=    - \sinh(\eta)(s + x^\nu \del_\nu) \phi^{(s)} \, .
 \label{t-x-phi-id}
\end{align}
Therefore 
the time-like vector field $\tau = x^\nu \del_\nu$ of \eqref{tau-VF-def}
on $\cM^{3,1}$
can be extended as an $SO(3,1)$ intertwiner of $\cC^s$  as follows
\begin{align}
 t^\mu \{x_\mu,\phi^{(s)}\} = -\sinh(\eta)(s + \t)  \phi^{(s)} = - x^\mu \{t_\mu,\phi^{(s)}\} 
 \qquad \in \cC^s \, .
 \label{t-x-poisson-id}
\end{align}
\paragraph{Space-like gauge and degrees of freedom.}
The present space-like gauge is also useful to identify the independent physical degrees of freedom contained in $\cC^s$.
Consider e.g.\ $s=1$. Then $\phi^{(1)} = \phi_\mu t^\mu$  is a symmetric 
traceless space-like tensor field 
$\phi_{\mu}$. This separates into $\phi^{(1)} = \phi^{(1,0)} + \phi^{(1,1)}$, where 
$\phi^{(1,1)} = D^+ \phi^{(0)}$ encodes a scalar field, and 
$\phi^{(1,0)} \in \cC^{(1,0)}$  is 3-divergence-free $\nabla_{(3)}^\mu \phi^{(1,0)}_{\mu} = 0$. 
This implies 
 $\nabla^\mu \phi^{(1,0)}_{\mu} = 0$  for any $4$-dimensional 
$SO(3,1)$-invariant connection on 
 $\cM^{3,1}$, because $x^\mu \phi_\mu = 0$. Therefore
 $\phi^{(1,0)}$ provides the 2 degrees of freedom of a massless spin 1 gauge 
field, while  $\phi \in \cC^{(1,1)}$
 contributes the remaining degree of freedom for a massive spin 1 field with 3 degrees of freedom.

Similarly for $s=2$, the  $\phi^{(2)} = \phi_{\mu\nu} t^\mu t^\nu$ are symmetric traceless space-like tensor fields.
The $\phi^{(2,0)}$ are 3-divergence-free $\nabla_{(3)}^\mu \phi^{(2,0)}_{\mu\nu} 
= 0$, which implies 
 $\nabla^\mu \phi^{(2,0)}_{\mu\nu} = 0$  for any 4-dimensional 
$SO(3,1)$-invariant connection on $\cM^{3,1}$.
 Thus the $\phi_{\mu\nu}^{(2,0)}$ are traceless, divergence-free and space-like, hence they contain 2 degrees of freedom,
 as appropriate for gravitons. The remaining sectors $\phi^{(2,1)} = D^+ \phi'^{(1,0)}$ and $\phi^{(2,2)} = D^+D^+ \phi'^{(0)}$ in $\cC^2$ 
 provide the missing degrees of freedom for a general massive spin 2 
 field with 5 degrees of freedom.
 We will see in section \ref{sec:graviton} how this applies to the actual gravitons in the 
 matrix model.
\paragraph{Gauge transformations and diffeomorphisms.} 
Gauge transformations act on 
fields $\phi\in \cC$ via $\phi \mapsto \{\L,\phi\}$, for 
any $\L\in\cC$. For $\L\in\cC^0$, these 
correspond to (noncommutative) $U(1)$ gauge transformations. For
 $\L\in\cC^1$, they  induce volume-preserving diffeomorphisms 
 (and corresponding gauge modes) on $H^4_n$, 
as discussed in \cite{Sperling:2018xrm}.
On $\cM^{3,1}$, the gauge transformations generated by 
a vector field $v^\mu$  (in  space-like gauge) act on functions $\phi\in\cC^0$ as
\begin{align}
 \{v^\mu t_\mu,\phi\}_0 &= \left[t^\mu\theta^{\a\b}\right]_0 \del_\a v_\mu 
\del_\b \phi 
  + v^\mu \{t_\mu,\phi\}_0 \nn\\
  &= -\frac{1}{3} \left(\sinh(\eta)(\eta^{\mu\a}x^\b - \eta^{\mu\b} x^\a) 
    - x_\g \varepsilon^{\g 4\mu\a\b}\right)\del_\a v_\mu \del_\b \phi 
   + \sinh(\eta)v^\mu \del_\mu \phi \nn\\
   &= \frac{1}{3} \left(\sinh(\eta)\left( 3 v^\mu \del_\mu - ({\rm div} v) \t 
   + (\t v^\mu) \del_\mu \right)\phi
    - x_\g \varepsilon^{\g\mu\a\b}\del_\a v_\mu \del_\b \phi\right) 
\end{align}
using \eqref{average-t-theta}. Although
this is not the most general diffeomorphism, 
it does include time-like derivatives on $\cM^{3,1}$ via ${\rm div}v$,
even though $v_\mu$ is in space-like gauge. 
The non-standard form reflects the existence of an invariant volume form on $\C P^{1,2}$.
Nevertheless, this provides a powerful constraint 
for the resulting gravity theory that will be exploited in section 
\ref{sec:lin-EH-sction}.
%
%
\subsection{\texorpdfstring{$SO(4,1)$}{SO(4,1)}-covariant gauge and tensor fields on \texorpdfstring{$H^4$}{H4}}
\label{sec:so41-gauge}
Instead of the above $SO(3,1)$-covariant organization in terms of 
tensor fields in space-like gauge, 
one can alternatively use a $SO(4,1)$-covariant realization of 
$\phi^{(s)} \in \cC^s$ in terms of 
tangential traceless divergence-free tensor fields 
on $H^4$ as in \cite{Sperling:2018xrm}. They are defined by
\begin{align}
 \phi^H_{a_1 \ldots a_s} \propto \{x^{a_1}, \ldots 
\{x^{a_s},\phi^{(s)}\}\ldots \}_0
\label{phi-tensorfield-H}
\end{align}
 normalized such that 
\begin{align}
 \phi^{(s)} =  \{x^{a_1},\ldots\{x^{a_s},\phi^H_{a_1\ldots a_s}\}\ldots\} \, .
\end{align}
The  normalization factor is found recursively 
from\footnote{The notation here is slightly different from 
\cite{Sperling:2018xrm}, where $\phi^{H(s-1)}_a$ is denoted as $\phi^{(s)}_a$.}
\begin{align}
 - \{x_a,\phi^{(s)}\}_{s-1} = \a_s (\Box_H - 2 r^2) \phi^{H(s-1)}_a 
 \label{x-phi-posson-id}
\end{align}
where $\a_1=\frac{1}{3}, \ \a_2 = \frac{2}{5}$, and so on and so forth.
The $\phi^H_{a_1\ldots a_s}$ contain the same information as the 
$\phi_{\mu_1\ldots\mu_s}$ in space-like gauge, but enjoy more powerful symmetries.
In particular, it follows that
\begin{align}
 -\{x^a,\{x_a,\phi^{(s)}\}_-\}_+
   &= \a_s  \{x^a,(\Box_H - 2 r^2) \phi^{H(s-1)}_a \}  \nn\\
  &= \a_s \left(\Box_H - 2 r^2(s+1) \right) \{x^a, \phi^{H(s-1)}_a \} \nn\\
  &= \a_s \left(\Box_H - 2 r^2(s+1) \right)\phi^{(s)}   
  \label{xa+xa-id} \,,\\
  -\{x^a,\{x_a,\phi^{(s)}\}_+\}_-
  &= \left((1-\a_s)\Box_H + 2(s+1)\a_s \right)\phi^{(s)} 
  \label{xa-xa+id}
\end{align}
using the intertwiner properties \eqref{x-BoxH-bracket}.
In particular, the operators on the right-hand-side of \eqref{xa+xa-id} and  
\eqref{xa-xa+id}  are positive, because the 
left-hand-side is positive\footnote{since $\int_{H^4} 
\{x^a,\phi\}_-\{x_a,\phi\}_- > 0$, because 
$\{x_a,.\}$  is tangential on the Euclidean space $H^4$, cf. \cite{Sperling:2018xrm}.} on $H^4$.
Reduced to $\cM^{3,1}$, this gives
\begin{align}
 -\{x^\mu,\{x_\mu,\phi^{(s)}\}_-\}_+
  &= \big(\a_s (\Box_H - 2 r^2(s+1)) + D^+ D^-\big) \phi^{(s)}  \,.
   \label{xmu+xmu-id}
\end{align}
Similarly, one finds
\begin{align}
\{x^\mu,\{x_\mu,\phi^{(s)}\}_-\}_- 
 &= - \Big((\Box_H + \{x^4,\{x^4,.\}\})\phi^{(s)}\Big)_{s-2}
 = -D^- D^-\phi^{(s)}  \nn\\
 \{x^\mu,\{x_\mu,\phi^{(s)}\}_+\}_+ &= -D^+ D^+\phi^{(s)}
\label{xpxp-id-4} 
\end{align}
 since $\{x^\mu,\cdot\}: \cC^s \to \cC^{s-1} \oplus \cC^{s+1}$ 
 while $\Box_H:  \cC^s \to \cC^{s}$.
Combining these, we  obtain 
\begin{align}
  -\{x^\mu,\{x_\mu,\phi^{(s)}\}_+\}_-
  &=  \left((1-\a_s)\Box_H + 2\a_s r^2(s+1) + D^-D^+ \right) \phi^{(s)} \ .
  \label{xppxm-id-4} 
\end{align}
 To summarize,
 $\phi^{(s)}$ corresponds to a 4-dimensional  spin $s$ field, 
 either represented as divergence-free symmetric traceless tensor on $H^4$ or as 
space-like symmetric trace-less 
 tensor on $\cM^{3,1}$ which is generically not divergence-free.
 The counting of degrees of freedom is the same in both interpretations, 
 and gives, for example, 5 in the case of $s=2$.
%
%
\section{Matrix model and  quantum space-time solutions}
\label{sec:matrix-model}
Let us move towards the matrix model, by considering the bosonic part of the  
action of the IKKT matrix model supplemented with a mass term
\begin{align}
 S[Y] &= \frac 1{g^2}\Tr \Big([Y^\mu,Y^\nu][Y^{\mu'},Y^{\nu'}] \eta_{\mu\mu'} \eta_{\nu\nu'} \, 
  - \mu^2 Y_\mu Y^\mu  \Big) \ . 
 \label{bosonic-action}
\end{align}
Here $\eta_{\mu\nu} = \diag(-1,1,1,1)$ 
is the flat Minkowski metric of the target space $\R^{1,3}$, and $\mu^2$ 
introduces a mass scale.
This  action leads to the classical equations of motion 
\begin{align}
 \Box_Y Y^\mu + \frac 12\mu^2 Y^\mu &= 0 
 \label{eom-lorentzian-M} \\
 \text{with} 
 \qquad 
 \Box_Y &= [Y^\mu,[Y_\mu,\cdot]] \ \sim - \{y^\mu,\{y_\mu,\cdot\}\} \,,
  \label{Box-Y}
\end{align}
where $\Box_Y$
plays the role of the Laplacian or rather d'Alembertian operator. 
We consider the following ansatz for solutions of \eqref{eom-lorentzian-M}  
\begin{align}
 Y^\mu = \cM^{\mu a}\a_a, \qquad \mu=0,1,2,3
\end{align}
where $\a_a$ is some constant $SO(4,2)$ vector. These describe
homogeneous and isotropic quantized cosmological space-times $\cM_n^{1,3}$ as 
introduced\footnote{We change notation from \cite{Steinacker:2017bhb}, 
where $Y^1$ was dropped instead of $Y^4$.} in \cite{Steinacker:2017bhb}.
Since  $\eta^{\mu\nu}$ is $SO(3,1)$-invariant, we have 
\begin{align}
  [Y_\r,[Y^\r,Y^\mu]] &= \im(\a\cdot\a)  [Y_\r,\cM^{\r\mu}]  = -\im (\a\cdot\a) [\cM^{\r\mu},Y_\r] \nn\\
    &= (\a\cdot\a)\left\{\begin{array}{ll}
               Y^\mu, & \mu\neq\r \\
               0, & \mu = \r 
              \end{array}\right.
\qquad\mbox{(no sum)} 
\end{align}
and therefore
\begin{align}
 \Box_{Y} Y^\mu 
 &= 3 (\a\cdot\a) \, Y^\mu \ .
 \label{Y-Box}
\end{align}
Consequently, we obtain three different types of 
quantized space-time solutions with Minkowski signature in the IKKT model with mass term,
\begin{align}
\begin{aligned}
 \Box_X X^\mu &= -3 r^2\, X^\mu, \qquad \text{for} \quad \frac 12 \mu^2 =  3 
r^2 \\
  \Box_T T^\mu &= 3 R^{-2}\,  T^\mu , \qquad  \text{for} \quad  \frac 12 \mu^2 
= - 3 R^{-2}  \\
  \Box_Z Z^\mu &= 0 \ . 
  \end{aligned}
  \label{MM-solutions}
\end{align}
The $Z^\mu$ solution describes a  light-cone and will not be considered  here.
The $X^\mu$ solution has been discussed to some extent in 
\cite{Steinacker:2017bhb}.
The  parameter $\mu^2$ sets the length scale $r^2$ of the space-time, while the quantum number
$n$ is undetermined.
In this paper, we will focus on the
\emph{momentum} solution
\begin{align}
  Y^\mu &= T^\mu \, .
 \label{T-solution-4}
\end{align}
The main reason for this choice is that the effective d'Alembertian $\Box \equiv \Box_T$
as well as  $[T_\mu,\cdot]$
respect the spin $\cS^2$ as discussed in section \ref{sec:spin-casimir}, 
such that the above decomposition of fluctuations into higher-spin modes is 
applicable.
Even though the embedding is realized via the momentum generator $T^\mu$
rather than $X^\mu$, we will see that the fluctuations  lead 
to a gauge theory on $\cM^{3,1}$.
%
%
\section{Fluctuations and higher-spin gauge theory}
\label{sec:fluctuations}
Now consider  tangential\footnote{In the 9+1-dimensional IKKT model, there are of course
also transversal fluctuation modes. Those are simply scalar fields as discussed 
in section \ref{sec:wavefunct-higherspin}. However, it is plausible that these scalar fields 
acquire a non-trivial vacuum structure, along the lines of 
\cite{Sperling:2018hys,Aoki:2014cya} and references therein. Here we focus on the 
tangential modes.} 
deformations of the above background solution given by 
$T^\mu$, i.e.
\begin{align}
 Y^\mu = T^\mu  + \cA^\mu, 
\end{align}
where $\cA \in \cC \otimes \C^4$ is an arbitrary (Hermitian) fluctuation.
We will give a complete fluctuation analysis in the following\footnote{
Note that the background admits $SO(3,1)$ as a symmetry, which is spontaneously broken, but
preserved modulo to a gauge transformation. Hence the corresponding Goldstone bosons are unphysical.}.
The full Yang-Mills action \eqref{bosonic-action}  can be expanded in the 
fluctuation modes as
\begin{align}
 S[Y] = S[T]  +  S_2[\cA] + O(\cA^3),  
 \end{align}
 where the quadratic fluctuations are  governed by  
 \begin{align}
S_2[\cA] = -\frac{2}{g^2}\int\limits_{} \dd \Omega\, \left( \cA_\mu 
\Big(\cD^2+\frac{1}{2}  \mu^2\Big) \cA^\mu + \cG\left(\cA\right)^2 \right) .
\label{eff-S-expand}
\end{align}
Here 
\begin{align}
\cD^2 \cA =  \left(\Box  - 2\cI \right)\cA  
\label{vector-Laplacian}
\end{align}
is the  vector Laplacian, which  is composed of   the scalar matrix Laplacian 
(or rather d'Alembertian)
\begin{align}
 \Box = [T^\mu,[T_\mu,\cdot]] \ \sim -\{t^\mu,\{t_\mu,.\}\} = \a^{-1} \Box_G
 \label{Box-T-def}
\end{align}
 on the $\cM^{3,1}$ background due to \eqref{G-Box-relation}, and the intertwiner
\begin{align}
 \cI (\cA)^\mu = - [[ Y^\mu, Y^\nu],\cA_\nu] =  \frac{\im}{r^2 R^2} 
[\Theta^{\mu\nu},\cA_\nu] 
 \eqqcolon -\frac{1}{r^2 R^2}\tilde\cI (\cA)^\mu \ ,
\end{align}
recalling $\Theta^{\mu\nu} = - r^2 \cM^{\mu\nu}$ as in 
\eqref{basic-CR-H4}.
As usual in Yang-Mills theories, the scalar mode
 \begin{align}
\cG(\cA) = -\im [T^\mu,\cA_\mu] \sim \{t^\mu,\cA_\mu\}  ,  
 \label{gaugefix-intertwiner}
 \end{align}
should be removed to get a meaningful theory.
This is achieved by adding a gauge-fixing term $-\cG(\cA)^2$ to the action
as well as the corresponding Faddeev-Popov (or BRST) ghost. Then the quadratic 
action becomes 
\begin{align}
 S_2[\cA] + S_{g.f} + S_{ghost} &= -\frac{2}{g^2}\int\limits_{} \dd \Omega\, 
\left( \cA_\mu \Big(\cD^2 + \frac{1}{2}  \mu^2\Big) \cA^\mu + 2 \obar{c} \Box 
c \right) \ 
\label{eff-S-gaugefixed}
\end{align}
where $c$ denotes the fermionic BRST ghost; see e.g.\ \cite{Blaschke:2011qu} 
for more details.
\paragraph{UV spectrum and no-ghost.}
For short wavelengths, the quadratic action can be simplified as follows
  \begin{align}
S_{2}^{\mathrm{UV}}[\cA] &\approx - \frac{2}{g^2}\int\limits_{} \dd \Omega\, 
\left( 
\cA_\mu \Box \cA^\mu + 2 \obar{c} \Box c \right) .
\end{align}
This has the same structure as a Yang-Mills action, which makes it very 
plausible that
the theory will be ghost-free, as the $c$ ghost cancels the unphysical 
polarizations of $\cA_\mu$. 
A simple general argument for having a ghost-free theory is as 
follows\footnote{we are grateful for H.\ Kawai for pointing this out.}:
Since the $\cA_0$ component can be diagonalized using the gauge invariance,
the physical, propagating degrees of freedom are carried by the remaining 
matrices, which are space-like such that their kinetic term 
has the standard sign. 
Indeed, we have seen that no internal higher-spin ghosts arise, due to the 
space-like gauge-fixing discussed above.
We will provide another argument  in section \ref{sec:gaugefix} that there are no ghosts, and verify
to some extent how these general arguments are borne out in the mode expansion below.

%
%
\subsection{Mode expansion and ansatz}
We consider the  mode expansion for vector modes similar to the case of  scalar fields 
\eqref{functions-expansion-momentum}
\begin{align}
 \cA^\mu &= A^{\mu}(x) + A^{\mu}_\a(x)\, t^\a +   A^{\mu}_{\a\b}(x)\, t^\a t^\b + \ldots \nn\\
  &\in \ \cC^0  \quad \oplus \quad \ \cC^1 \quad \oplus \quad \ \cC^2 \quad \oplus \ \ \ldots
 \label{A-M31-spins}
\end{align}
However these are neither irreducible nor eigenmodes of $\cD^2$. To find the 
spin $s$ eigenmodes $\cA_\mu \in \cC \otimes \C^4$, we choose
the following ansatz:
\begin{align}
\label{A2-mode-ansatz}
 \boxed{ \
\begin{aligned}
 \cA_\mu^{(g)}[\phi^{(s)}] &= \{t_\mu,\phi^{(s)}\}  \quad \in \cC^{s}\,,
\\
 \cA_\mu^{(+)}[\phi^{(s)}] &= \{x_\mu,\phi^{(s)}\}|_{\cC^{s+1}} \ \equiv  
\{x_\mu,\phi^{(s)}\}_+ \quad \in \cC^{s+1} \,, \\
 \cA_\mu^{(-)}[\phi^{(s)}] &= \{x_\mu,\phi^{(s)}\}|_{\cC^{s-1}} \  \equiv  
\{x_\mu,\phi^{(s)}\}_- \quad \in \cC^{s-1} \,.
 \end{aligned}
 }
\end{align}
These expressions should be viewed as $\mso(3,1)$ intertwiners 
\begin{align}
 \cA^{(i)}: \quad \cC^s \to \cC^s\otimes \C^4 , \qquad i\in\{g,+,-\} \ ,
\end{align}
where 
$\phi^{(s)} \in \cC^s$ is used to represent the vector mode.
Clearly $\cA_\mu^{(g)}$ is the pure gauge mode.
It is important to note that the $\cA^{(\pm)}$ intertwiners 
can be extended as $\mso(4,1)$ intertwiners via 
$\cA_a^{(\pm)}[\phi^{(s)}] = \{x_a,\phi^{(s)}\}_{\pm}$.
These are 2 linear independent modes\footnote{We note that $\cA_\mu^{(-)} = 
-\a_s (\Box_H-2r^2) \phi^{(s)}_\mu$
in the notation of \cite{Sperling:2018xrm}.
But the notation $\phi^{(s)}_a$ is inconsistent with the present, hence we avoid it here.}, which
will turn out to be exact eigenmodes of $\cD^2$. For completeness, we need one more such mode. 
We will use  the ansatz
\begin{align}
\boxed{
 \cA_\mu^{(\t)}[\phi^{(s)}] = x_\mu \phi^{(s)}
 }
 \label{A4-mode-ansatz}
\end{align}
as a starting point, because it is independent of the above modes and it also
extends as $\mso(4,1)$ intertwiner
$\cA_a^{(\t)}[\phi] = x_a \phi$.
This is a time-like mode, but we will argue that it is not part of 
the physical Hilbert space of the theory.
%
%
\subsection{Group-theoretical considerations}
\label{sec:group-th-results}
To evaluate $\cD^2$ on these modes, we need some group-theoretical preparations.
\paragraph{Intertwiners.}
Consider the following $SO(3,1)$ intertwiners 
\begin{align}
 \cC \otimes \C^4 \ & \to \ \cC \otimes \C^4\nn\\[1ex]
 \cA^\mu  &\mapsto \tilde\cI(\cA)^\mu = -\im [\Theta^{\mu\nu},\cA_\nu] \sim 
\{\theta^{\mu\nu},\cA_\nu\}  \nn\\
 \cA^\mu  &\mapsto \Box(\cA)_\mu = [T^\nu,[T_\nu,\cA_\mu]] \sim - 
\{t^\nu,\{t_\nu,\cA_\mu\}\}
\end{align}
which are Hermitian, in the sense
\begin{align}
 \int \dd\Omega\, \cA_\mu \tilde\cI(\cA')^\mu &= \int \dd\Omega\, 
\tilde\cI(\cA)_\mu \cA'^\mu 
 \label{intertwiner-A-hermitian}
 \end{align}
and similarly for $\Box$.
We recall that $\cS^2$ commutes with $\Box$, see \eqref{Box-T-SO41}, 
and it also commutes with $\tilde\cI$.
This will greatly facilitate the analysis.
Note that the gauge-fixing functional $\cG$ is also an  $SO(3,1)$ intertwiner.
\paragraph{Relation with Casimirs.}
Recall from \eqref{Box-T-SO41} the Laplacian
 \begin{align}
 \Box  &= [T_\mu,[T^\mu,\cdot]] 
 =  R^{-2}(C^2[\mso(4,1)]^{(\ad)} - C^2[\mso(3,1)]^{(\ad)})  \,,
 \label{Box-Casimirs}
\end{align}
where $(\ad)$ denotes the  representation of 
$\mso(4,1)$ or $\mso(3,1)$ acting on $\cC$ via 
$M_{\mu\nu}^{(\ad)} = [\cM_{\mu\nu},\cdot] \sim  \im \{\cM_{\mu\nu},\cdot \}$.
Then $\cA_a$ transforms in $(\rm full) = (ad) \otimes (5)$ of $SO(4,1)$. Now
 consider the intertwiners 
 \begin{align}
 \begin{aligned}
  \tilde\cI^{(5)}[\cA]^a &\coloneqq  -\im[\Theta^{ab},\cA_b] \sim 
\{\theta^{ab},\cA_b\}  \\ 
  \tilde\cI^{(4)}[\cA]^\mu &= -\im[\Theta^{\mu\nu},\cA_\nu] \sim 
\{\theta^{\mu\nu},\cA_\nu\} 
\end{aligned}
 \end{align}
 which arise in $\cD^2$.
 They can be related to the Casimir $C^2[\mso(4,1)]^{\rm (full)}$
 acting on vector modes 
 as follows:
\begin{align}
 C^2[\mso(4,1)]^{\rm (full)} \cA^a &= \frac{1}{2} \left( [\cM_{cd},\cdot] + 
  M^{(5)}_{cd} \right)^2\cA^a \nn\\
  &= \left( C^2[\mso(4,1)]^{(\ad)} - 2 r^{-2}\tilde\cI^{(5)} +4\right)\cA^a 
  \label{C2full-I-relation}
\end{align} 
using \eqref{Spin-casimir}, and $C^2[\mso(4,1)]^{(5)} = 4$ for the vector 
representation $\C^{5}$.
 This can be seen by expressing $\tilde\cI$ as follows:
\begin{align}
 - r^2(M_{cd}^{(\ad)} \otimes M_{cd}^{(5)}\cA)^a \sim
 -\big(M^{(5)}_{cd}\big)^a_b \,\im \{\theta^{cd},\cdot \}  \cA^b 
 = 2\{\theta_{ab},\cA^b\} = 2\, \tilde\cI^{(5)}(\cA)^a \, 
 \label{I-MM-id}
 \end{align}
where 
\begin{align}
\left(M_{ab}^{(5)} \right)^c_d &=  \im\left(\d^c_b \eta_{ad} - \d^c_a 
\eta_{bd}\right)\, 
\label{M-rep-5}
 \end{align}
is the $\mso(4,1)$ vector representation.
Thus
\begin{align}
 2r^{-2}\, \tilde\cI^{(5)} &= - C^2[\mso(4,1)]^{(5)\otimes(\ad)} + C^2[\mso(4,1)]^{(\ad)} + C^2[\mso(4,1)]^{(5)} \nn\\
 2r^{-2}\, \tilde\cI^{(4)} &=- C^2[\mso(3,1)]^{(4)\otimes(\ad)} + C^2[\mso(3,1)]^{(\ad)} + C^2[\mso(3,1)]^{(4)} 
 \label{box-Casimirs}
\end{align}
 cf.\ \cite{Sperling:2018xrm}, where 
 $C^2[\mso(3,1)]^{(4)}= 3$.
Using \eqref{Box-Casimirs}, we can rewrite \eqref{C2full-I-relation} 
and its $\mso(3,1)$ analog
as 
\begin{align} 
    C^2[\mso(4,1)]^{\rm (full)} \cA^a &= \big(C^2[\mso(4,1)]^{(\ad)} - 2 r^{-2}\tilde\cI^{(5)} +4\big)\cA^a \nn\\
&= (R^2\Box + C^2[\mso(3,1)]^{(\ad)} - 2 r^{-2}\tilde\cI^{(5)}  + 4) \cA^a   \nn\\
 &=  \cA^a[C^2[\mso(4,1)]\phi] \,, \label{C2-total-id-5} \\
C^2[\mso(3,1)]^{\rm (full)} \cA^\mu
&= \big(C^2[\mso(3,1)]^{(\ad)} - 2 r^{-2}\tilde\cI^{(4)} +3\big)\cA^\mu \nn\\
 &=  \cA^a[C^2[\mso(3,1)]\phi] \,,
\label{C2-total-id-4}
\end{align}
assuming that $\cA^a[\phi]$ is an $\mso(4,1)$ intertwiner such as \eqref{A2-mode-ansatz} and \eqref{A4-mode-ansatz}.
Subtracting the right-hand-side of \eqref{C2-total-id-4} from 
\eqref{C2-total-id-5} for $a=\mu$, we obtain
\begin{align}
 \left(R^2\Box  - 2 r^{-2}\left(\tilde\cI^{(5)}- \tilde\cI^{(4)}\right)  + 
1\right) \cA^\mu
  &=\cA^\mu\left[ \left(C^2[\mso(4,1)] - C^2[\mso(3,1)] \right)\phi\right]  
\nn\\
     &= R^2\cA^\mu[\Box\phi]
 \label{Box-I-A-SO41-id}
\end{align}
which gives 
\begin{align}
  \cD^2 \cA^\mu = \left(\Box  +\frac{2}{r^2R^2} \tilde\cI^{(4)} \right)\cA^\mu 
   &= \cA^\mu\left[\left(\Box-\frac{1}{R^2}\right)\phi\right] + \frac{2}{r^2 
R^2}\tilde\cI^{(5)}\cA^\mu \ .
 \label{Box-I-A-SO41-id-2}
\end{align}
This can now be evaluated using the results on $H^4$ in \cite{Sperling:2018xrm},
and we obtain 
an eigenmode of $\cD^2$ if $\cA^a[\phi]$ is an eigenmode of $\tilde\cI^{(5)}$.
Finally, we remark that all relations written in the semi-classical 
(Poisson) case generalize to the fully noncommutative case.
%
%
\subsection{\texorpdfstring{$\cD^2$}{D2} eigenvalues}
\label{sec:D2-eigenvalues}
Consider first the pure gauge mode
$\cA_\mu^{(g)}[\phi]  = \{t_\mu,\phi\}$ \eqref{A2-mode-ansatz},
 which define a flat direction for the gauge-invariant action 
\eqref{eff-S-expand}.
This means that
\begin{align}
 \left(\cD^2+\frac{1}{2} \mu^2 \right) \cA_\mu^{(g)} + 
\{t_\mu,\{t^\nu,\cA_\nu^{(g)}\}\} &= 0
 \label{Ag-D2-identity} \,
\end{align}
and therefore
\begin{align}
 \cD^2 \cA_\mu^{(g)}[\phi] &= 
\cA_\mu^{(g)}\left[\left(\Box+\frac{3}{R^2}\right) 
\phi\right] \,,
 \label{puregauge-D2}
\end{align}
using $\frac 12\mu^2 = -3R^{-2}$ \eqref{MM-solutions}. Alternatively, 
relation \eqref{puregauge-D2} can be computed directly as a consistency check.
Next, consider the  $\cA_\mu^{(\pm)}$ modes. Since they are part of the 
$SO(4,1)$-covariant modes $\{x_a,\phi\}_\pm$, we can use 
\eqref{Box-I-A-SO41-id-2} and \eqref{I5-X-phi} to obtain 
\begin{align}
 \cD^2 \cA_\mu^{(+)}[\phi^{(s)}]  &= \cA_\mu^{(+)}\left[\left(\Box - 
\frac{1}{R^2} \right)\phi\right] 
   +\frac{2}{R^2} (s+3) \{x^\mu,\phi^{(s)}\}_+ \nn\\
    &= \cA_\mu^{(+)}\left[\left(\Box + \frac{2s+5}{R^2} 
\right)\phi^{(s)}\right]
    \label{D2-A2p-eigenvalues} \,, \\
 \cD^2 \cA_\mu^{(-)}[\phi^{(s)}] 
   &= \cA_\mu^{(-)}\left[\left(\Box + 
\frac{-2s+3}{R^2}\right)\phi^{(s)}\right] \, .
   \label{D2-A2m-eigenvalues}
\end{align}
Defining $\End(\cH_n)$ to be the (Hilbert) space of Hilbert-Schmidt operators 
as in \eqref{inner-prod-End}, it decomposes into 
unitary irreps of $\mso(4,2)$, which can be decomposed further into eigenmodes 
of $\Box$ since the latter is expressed in terms of Casimirs 
\eqref{S^2-BoxH-commute}.
Thus diagonalizing $\Box$ on $\cC^s$  we obtain three series of 
eigenmodes of $\cD^2$, and in particular three series of on-shell modes  
which satisfy $\big(\cD^2  -\frac{3}{R^2} \big) \cA = 0$:
\begin{align}
 \boxed{ \
\begin{aligned}
 \cA^{(+)}[\phi^{(s)}]  \qquad & \text{for } \  \ \left(\Box + 
\frac{2s+2}{R^2}\right) \phi^{(s)} = 0 \,,\\
 \cA^{(-)}[\phi^{(s)}] \qquad& \text{for } \quad  \ \left(\Box + \frac{-2s}{R^2} 
\right)\phi^{(s)} = 0 \, \\
\cA^{(g)}[\phi^{(s)}] \qquad &\text{for } \qquad \qquad  \ \Box \phi^{(s)} = 0 \ .
  \label{on-shell-A}
\end{aligned}
}
\end{align}
Of course the  pure gauge mode $\cA^{(g)}$ is  unphysical.

Finally, the mode $\cA^{(\t)}_\mu$ is also part of an $SO(4,1)$ mode 
$\cA^{(\t)}_a[\phi] = x_a \phi$. Thus 
formula \eqref{Box-I-A-SO41-id-2} gives 
\begin{align}
  \cD^2 \cA_\mu^{(\t)}[\phi] = 
\cA^{(\t)}_\mu\left[\left(\Box-\frac{1}{R^2}\right)\phi\right]
  + \frac{2}{r^2 R^2}\tilde\cI^{(5)}\cA^{(\t)}_\mu[\phi] \,,
\end{align}
but now $\tilde\cI^{(5)}$ is no longer diagonal:
\begin{align}
 \tilde\cI^{(5)}\cA^{(\t)}_\mu[\phi] &= \{\theta^{\mu b},x_b\phi\} =  \{\theta^{\mu b},x_b\} \phi +  x_b\{\theta^{\mu b},\phi\} \nn\\
  &= 4 r^2 x_\mu \phi  - \theta^{\mu b}\{ x_b,\phi\} \nn\\
   &= 4 r^2 \cA^{(\t)}_\mu[\phi] + r^2 R^2  \eth^\mu \phi 
\end{align}
where 
\begin{align}
  \eth^a \phi^{(s)} \coloneqq -\frac{1}{r^2 R^2}\theta^{ab}\{x_b,\phi\}  \qquad \in\cC^{s} \; 
\end{align}
is the tangential derivative operator on $H^4$ introduced in \cite{Sperling:2018xrm}.
Thus, we arrive at 
\begin{align}
  \cD^2 \cA_\mu^{(\t)}[\phi] = 
\cA^{(\t)}_\mu\left[\left(\Box+\frac{7}{R^2}\right)\phi\right] + 2\eth_\mu \phi  
\, .
  \label{A4-eom}
\end{align}
\paragraph{Degeneracy.}
We can recognize a degeneracy of these modes by  
considering the  eigenmodes in the same $\cC^s$.
Then the above results give 
\begin{align}
\begin{aligned}
 \cD^2 \cA_\mu^{(+)}[\phi^{(s-1)}] &= \cA_\mu^{(+)}\left[\left(\Box + 
\frac{2s+3}{R^2} \right)\phi^{(s-1)}\right]   \qquad \in \cC^{s}\ , \\
 \cD^2 \cA_\mu^{(-)}[\phi^{(s+1)}] &= \cA_\mu^{(-)}\left[\left(\Box + 
\frac{-2s+1}{R^2} \right)\phi^{(s+1)}\right] \qquad \in \cC^{s} \ ,   \\
 \cD^2 \cA_\mu^{(g)}[\phi^{(s)}] &= \cA_\mu^{(g)}\left[\left(\Box + 
\frac{3}{R^2} \right)\phi^{(s)}\right]   \qquad \in \cC^s \ . 
\end{aligned}
\label{D2-A-modes}
\end{align}
Now recall that $D^\pm: \cC^s \to \cC^{s\pm 1}$ relates eigenfunctions and eigenvalues 
according to \eqref{D-def}--\eqref{Dpm-def}.
Therefore
\begin{align}
\begin{aligned}
 \left(\cD^2 + \frac{1}{2} \mu^2\right) \cA_\mu^{(+)}[D^-\phi^{(s)}] &= 
\cA_\mu^{(+)}\left[ \left(\Box + \frac{2s}{R^2} \right)D^-\phi^{(s)} \right]  
  = \cA_\mu^{(+)}[D^-\Box\phi^{(s)}] \,,   \\
 \left(\cD^2 + \frac{1}{2} \mu^2 \right) \cA_\mu^{(-)}[D^+\phi^{(s)}] &= 
\cA_\mu^{(-)}\left[\left(\Box + \frac{-2s-2}{R^2} \right)D^+\phi^{(s)}\right] 
  = \cA_\mu^{(-)}[D^+ \Box \phi^{(s)}]  \,,
  \end{aligned}
\end{align}
using \eqref{Box-x4-relations}.
Hence if $\Box\phi^{(s)} = m^2 \phi^{(s)}$ is an eigenmode, we obtain 3 degenerate eigenmodes of $\cD^2$ with 
the same eigenvalue 
\begin{align}
 \boxed{ \
\begin{aligned}
 \left(\cD^2 + \frac{1}{2} \mu^2 \right) \cA_\mu^{(+)}[D^-\phi^{(s)}] &= m^2\, 
\cA_\mu^{(+)}[D^-\phi^{(s)}]    \\
 \left(\cD^2 + \frac{1}{2} \mu^2 \right) \cA_\mu^{(-)}[D^+\phi^{(s)}] &= 
m^2\,\cA_\mu^{(-)}[D^+\phi^{(s)}] \\
 \left(\cD^2 + \frac{1}{2} \mu^2\right)\cA_\mu^{(g)}[\phi^{(s)}] &= m^2\, 
\cA_\mu^{(g)}[\phi^{(s)}] .
\end{aligned}
 }
  \label{Apmg-degeneracy}
\end{align}
Whether or not these modes are always linearly independent is not yet 
established.
This could be decided using the inner products \eqref{inner-1}.
A non-trivial consistency check is provided in appendix 
\ref{sec:app-consistency-I},
using the action of $\tilde \cI$ \eqref{tilde-I-Apm}: 
\begin{align}
\begin{aligned}
 \tilde \cI \cA^{(-)}_\mu[D^+\phi^{(s)}] 
 &= r^2 (-s + 1) \cA^{\mu(-)}[D^+\phi^{(s)}] + r^2 R 
\{t^{\mu},D^-D^+\phi^{(s)}\} \,, \\
 \tilde \cI \cA^{(+)}_\mu[D^-\phi^{(s)}] 
 &= r^2 (s + 2) \cA^{\mu(+)}[D^-\phi^{(s)}] + r^2 R 
\{t^{\mu},D^+D^-\phi^{(s)}\}  \, .
\end{aligned}
 \label{I-Apm-relation}
\end{align}
%
%
\subsection{Gauge fixing and physical Hilbert space}
\label{sec:gaugefix}
As always in Yang-Mills gauge theory, the fluctuations 
$\cA^\mu$ can be separated into  
gauge-fixed modes denoted as $\cB^\mu$  and an unphysical  mode $\chi$, 
which should be determined via
\begin{align}
  \cB^\mu &= \cA^\mu  + \cA^\mu_{(g)}[\chi]  \,,\\
  0 &= \cG(\cB)\equiv \{t^\mu,\cB_\mu \} =  \{t^\mu,\cA_\mu \} -\Box \chi \,.
 \label{on-shell-psi}
\end{align}
Then the equations of motion for $\cB^\m$ become   
\begin{align}
 \left(\cD^2 + \frac 12\mu^2\right)\cB_\mu
 &=  \left(\cD^2 + \frac 12\mu^2 \right)\cA_\mu + \left(\cD^2 + \frac 
12\mu^2 \right)\{t_\mu,\chi\} \nn\\
 &=  \left(\cD^2 + \frac 12\mu^2 \right)\cA_\mu + \{t_\mu,\Box\chi\} \nn\\
 &=  \left(\cD^2 + \frac 12\mu^2 \right)\cA_\mu + \{t_\mu, \{t^\nu,\cA_\nu \}\}
\end{align}
 using  \eqref{Ag-D2-identity}.
Hence $\cA_\mu$ is a solution 
of the non-gauge-fixed action \eqref{eff-S-expand} if and only if
$\cB_\mu$ is a  solution  
$(\cD^2 + \frac 12\mu^2)\cB_\mu = 0$ of the gauge-fixed action\footnote{In the presence of matter,
this generalizes as $(\cD^2 + \frac 12 \mu^2) \cB_\mu = \cJ_\mu$, where $\cJ_\mu$ is the conserved current. }. 
 $\chi$ is determined only up to the kernel of $\Box$,
which must be factored out in the physical Hilbert space.

Let us make this explicit for the above modes.
The following relations \eqref{A2-gaugefix-ids} are  shown in 
appendix \ref{sec:checks-group-th-results}
\begin{align}
\begin{aligned}
\{t^{\mu},\cA_\mu^{(+)}[\phi^{(s)}]\} &= \frac{s+3}{R} D^+\phi^{(s)} \,,\\
\{t^{\mu},\cA_\mu^{(-)}[\phi^{(s)}]\} &= \frac{-s + 2}{R} D^-\phi^{(s)} \, .
\end{aligned}
 \label{A2-gaugefix}
\end{align}
Hence in the divergence-free sector and for $s=2$, $\cA_\mu^{(-)}$ is 
already gauge fixed.
For the time-like mode $\cA_\mu^{(\t)} = x_\mu \phi^{(s)}$, we obtain 
\begin{align}
 \{t^\mu,\cA_\mu^{(\t)}[\phi^{(s)}]\} &= \{t^\mu,x_\mu \, \phi^{(s)}\} \
 =  \sinh(\eta)\big(4 + s + \t \big) \phi^{(s)} \ ,
 \label{gaugefix-timelike}
\end{align}
using \eqref{t-x-poisson-id}.
We will argue below that the $\cA_\mu^{(\t)}$ modes do not contribute to the 
physical Hilbert space.
Finally, the pure gauge mode satisfies
\begin{align}
 \{t^\mu,\cA_\mu^{(g)}[\phi]\} &= - \Box \phi \ ,
 \label{gaugefix-g}
\end{align}
which vanishes for on-shell pure gauge fields.
Together with \eqref{Box-x4-relations}, this allows 
to determine the $\cB_\mu^{(\pm)}$  explicitly:
\begin{align}
\begin{aligned}
 \chi_{(+)} &=  \frac{s+3}{R}\Box^{-1} D^+\phi^{(s)} = 
\frac{s+3}{R}D^+\left(\left(\Box+\frac{2s+2}{R^2} \right)^{-1}\phi^{(s)}\right) 
\,, \\
 \chi_{(-)} &=  \frac{2-s}{R}\Box^{-1} D^-\phi^{(s)} = 
\frac{2-s}{R}D^-\left(\left(\Box-\frac{2s}{R^2}\right)^{-1}\phi^{(s)}\right) \, 
.
\end{aligned}
 \label{psi-box-relations}
\end{align}
Assuming  that these expressions make sense,
it follows that the gauge-fixed modes satisfy the intertwiner 
relations
\begin{align}
\begin{aligned}
 \cD^2\cB^{(+)}_\mu[\phi^{(s)}] &= \cA^{(+)}_\mu\left[ \left(\Box + 
\frac{2s+5}{R^2}\right)\phi^{(s)}\right]
 + \frac{s+3}R\left\{t_\mu,\left(\Box+\frac{3}{R^2}\right)\Box^{-1} 
D^+\phi^{(s)}\right\}  \\
  &= \cB^{(+)}_\mu\left[\left(\Box + \frac{2s+5}{R^2}\right)\phi^{(s)}\right]  
\,, \\
 \cD^2\cB^{(-)}_\mu[\phi^{(s)}] 
  &= \cB^{(-)}_\mu\left[\left(\Box + \frac{-2s+3}{R^2}\right)\phi^{(s)}\right]
  \end{aligned}
  \label{D2-B-modes}
\end{align}
using \eqref{puregauge-D2} and \eqref{psi-box-relations}. 
Hence they are eigenmodes of $\cD^2$ if the underlying modes $ \phi^{(s)}$ are eigenmodes of $\Box$.
In particular,
\begin{align}
 \boxed{ \
\begin{aligned}
 \left(\cD^2 + \frac 12\mu^2\right)\cB^{(+)}_\mu[D^-\phi^{(s)}] &=  
\cB^{(+)}_\mu[D^-\Box\phi^{(s)}] \,, \\
 \left(\cD^2 + \frac 12\mu^2\right)\cB^{(-)}_\mu[D^+\phi^{(s)}] &=  
\cB^{(-)}_\mu[D^+\Box \phi^{(s)}] \,, \\
 \left(\cD^2 + \frac 12\mu^2\right) \cA_\mu^{(g)}[\phi^{(s)}] &= 
\cA_\mu^{(g)}[\Box \phi^{(s)}]\,,
 \end{aligned}
 }
 \label{D2-B-modes-degenerate}
\end{align}
cf.\ \eqref{Apmg-degeneracy}. We observe again the triple degeneracy of 
$\cD^2$, unless some  modes coincide or vanish.
\paragraph{Physical Hilbert space.}

Now consider \eqref{psi-box-relations} in more detail.
Comparing with \eqref{on-shell-A}, we see that these expressions are well-defined 
for off-shell modes, 
but not for on-shell modes. 
Indeed if $\cA$ is on-shell, the added pure gauge term must also be 
on-shell. But this  means  by \eqref{gaugefix-g} that it is  gauge-fixed, so that
it cannot change the gauge of $\cA$.
Hence on-shell modes cannot simply be gauge-fixed, and the physical Hilbert space 
contains only those on-shell modes which are gauge-fixed (modulo pure gauge modes).
Due to the  degeneracy \eqref{D2-B-modes-degenerate}, we can always find a linear combination 
\begin{align}
 \cA^{(ph)} &\coloneqq \cA^{(-)}[\phi_+^{(s+1)}] + \cA^{(+)}[\phi_-^{(s-1)}]  \ \ \in \cC^s,
 \qquad \Big(\cD^2 +\frac{1}{2}\mu^2\Big)\cA^{(ph)} = 0
  \label{A-phys-comb}
\end{align}
of on-shell modes with
\begin{align}
 \Big(\Box+\frac{2s}{R^2}\Big)\phi_-^{(s-1)} = 0 = \Big(\Box-\frac{2s+2}{R^2}\Big)\phi_+^{(s+1)} 
\end{align}
which 
is gauge-fixed, i.e.
\begin{align}
0 &=  \{t^\mu,\cA^{(ph)}_\mu\}
   = \frac{-s + 1}{R} D^-\phi_+^{(s+1)} + \frac{s+2}{R} D^+\phi_-^{(s-1)} \ .  
\end{align}
Here
we have to distingish generic $s$ and $s=1$. For $s=1$, the physical solutions are
\begin{align}
 \cA^{(ph-)}[\phi^{(2)}] &\coloneqq \cA^{(-)}[\phi^{(2)}] , \qquad \Big(\Box - \frac{4}{R^2}\Big)\phi^{(2)} = 0
\end{align}
while none of the $\cA^{(+)}[\phi_-^{(0)}]$ is physical.
For $s\neq 1$,
 $\phi_+^{(s+1)}$ is uniquely determined by the above equation as 
\begin{align}
 \phi^{(s+1)}_+ = \frac{s+3}{s-1}
 (D^+D^-)^{-1} D^+D^+\phi_-^{(s-1)} , \qquad \Big(\Box + \frac{2s}{R^2}\Big) \phi_-^{(s-1)} = 0
 \label{phi--p-relation}
\end{align}
Note that $D^+D^- = D^+ (D^+)^\dagger$ commutes with $\Box$ due to \eqref{Box-x4-relations}
and is positive semi-definite, and vanishes only on $\phi^{(s,0)}$.
Therefore  the inverse in \eqref{phi--p-relation} exists, and the relation is compatible with the 
on-shell conditions.
Explicitly, this gives the following physical solutions
\begin{align}
 \cA^{(ph-)}[\phi^{(2)}] &\coloneqq \cA^{(-)}[\phi^{(2)}]    \nn\\
 \cA^{(ph-)}[\phi^{(s+1,0)}] &\coloneqq \cA^{(-)}[\phi^{(s+1,0)}] ,    \nn\\
 \cA^{(ph)}[\phi^{(s-1)}] &\coloneqq  \cA^{(+)}[\phi^{(s-1)}] 
 + \frac{s+3}{s-1}\cA^{(-)}[(D^+D^-)^{-1} D^+D^+\phi^{(s-1)}]  , \ s > 1 \ .
 \label{phys-modes-explicit}
\end{align}
The lowest physical modes are 
\begin{align}
&\{ \cA^{(-)}[\phi^{(0)}], 
 \cA^{(-)}[\phi^{(1,0)}], \cA^{(-)}[\phi^{(2)}],\cA^{(+)}[\phi^{(1)}] + 5\cA^{(-)}[(D^+D^-)^{-1}D^+D^+\phi^{(1)}], ...\}
\end{align}
The physical Hilbert space $\cH_{\rm phys}$ of the linearized theory, 
therefore, 
consists of the on-shell modes 
\begin{alignat}{3}
\begin{aligned}
 &\cA^{(ph)}[\phi^{(s-1)}] &\qquad &\text{for } \ & \ \Big(\Box + \frac{2s}{R^2}\Big) \phi^{(s-1)} &= 0 \,,  \quad s \geq 2 \\
 &\cA^{(ph-)}[\phi^{(s+1,0)}] &\qquad &\text{for } \ & \ \Big(\Box - \frac{2s+2}{R^2} 
\Big)\phi^{(s+1,0)} &= 0 \, ,  \quad s \geq 0   \\
 &\cA^{(ph-)}[\phi^{(2)}] &\qquad &\text{for } \ & \ \Big(\Box - \frac{4}{R^2} 
\Big)\phi^{(2)} &= 0 \, 
\end{aligned}
 \label{on-shell-B2pm}
\end{alignat}
which satisfy $\big(\cD^2 +\frac{1}{2}\mu^2\big)\cA^{(ph)} = 0$.
The on-shell pure gauge solutions 
\begin{align}
 \cA^{(g)}[\phi^{(s)}] &\qquad \text{for } \  \ \Box\phi^{(s)} = 0 \,
\end{align}
are null (see section \ref{sec:inner}) and must be factored out
from $\cH_{\rm phys}$. 
Finally, we  argue in appendix \ref{sec:app-time-like-modes}
that there are no on-shell gauge-fixed solutions involving
$\cA^{(\t)}_\mu[\phi]$.
To summarize,
\begin{align}
\boxed{
\;
 \cH_{\rm phys} = \left\{\cA^{(ph)}[\phi^{(s-1)}]\,, \ s=2,3,\ldots 
\right\} \cup \left\{\cA^{(ph-)}[\phi^{(s+1,0)}]\,, \ s=0,1,\ldots 
\right\} \cup \left\{\cA^{(ph-)}[\phi^{(2)}]\right\}
 \ }
\end{align}
 subject to the on-shell conditions \eqref{on-shell-B2pm}. 
 Here we indicate also the cutoff $n$, which disappears in the semi-classical limit $n\to\infty$.
 Since $\Box$ 
encodes the effective metric $G_{\mu\nu}$, the propagation of these modes respects
local Lorentz-invariance, and the extra structures 
$\t$ and $\kappa^{\mu\nu}$ do not enter.
We conjecture that this list is complete, although this has not been shown.
However off-shell, a $4^{th}$ series of modes 
is clearly missing, which is needed e.g. for the propagator.

In view of the discussion on space-like gauge in section \ref{sec:D-substructure},
the $\phi^{(s,0)}$ modes should be interpreted as massless spin $s$ fields, 
while the full $\phi^{(s)}$ modes are interpreted as (would-be) massive spin $s$ fields.
Thus we have found one massless and one massive tower of spin $s$ fields,
as well as a would-be massive spin $2$ field. 
The term ``would-be massive`` fields indicates the degrees of freedom
of massive  fields but without explicit mass term.
The physical significance of these modes needs further investigation.
For example, the pure gauge modes $\cA^{(g)}$ may mix or even coincide  with some of the 
would-be massive gauge fields, which could make them massless 
or partially massless. 
Off-shell, the $\cB^{\pm}$  provide  the 
degrees of freedom of two massive spin $s$ fields.

%

%
%
\subsection{Inner products and quadratic action}
\label{sec:inner}
We have argued following the quadratic action \eqref{eff-S-expand} that the 
matrix model should lead to a ghost-free theory,
based on rather general arguments. To see this explicitly and to clarify the on-shell Hilbert space structure,
we elaborate the inner product matrix for the fluctuation modes $\cA$ introduced 
above. This inner products is also needed to 
compute the propagator.
To simplify the notation we set $\int \equiv \int \dd\Omega$. Then
using \eqref{t-x-poisson-id}, \eqref{xmu+xmu-id}, \eqref{xpxp-id-4}, 
\eqref{xppxm-id-4} and \eqref{A2-gaugefix-ids}, we obtain
\begin{subequations}
\label{eq:inner_products}
\begin{align}
\int \cA_\mu^{(g)}[\phi'^{}] \cA^{(g)\mu}[\phi^{}] &=  \int \phi'^{} \Box \phi^{}  
\label{puregauge-inner}\\
 \int \cA_\mu^{(g)}[\phi'^{(s')}] \cA^{(+)\mu}[\phi^{(s)}] 
 &= - \frac{s+3}R\int_{\cM^{3,1}}  \phi'^{(s')} D^+ \phi^{(s)}   
\\
\int \cA_\mu^{(g)}[\phi'^{(s')}]  \cA^{(-)\mu}[\phi^{(s)}] 
 &=   \frac{s-2}R \int\,  \phi'^{(s')} D^-\phi^{(s)}  \quad 
 \\ 
\int \cA_\mu^{(-)}[\phi'^{}] \cA^{(+)\mu}[\phi^{}] 
&= \int\, D^- D^-\phi'^{}\, \phi^{}  \\
\int \cA_\mu^{(+)}[\phi'^{(s')}] \cA^{(+)\mu}[\phi^{(s)}]  
 &=   \int\,\phi'^{(s')}\big((1-\a_s)\Box_H + 2\a_s r^2(s+1) + D^-D^+  \big) 
\phi^{(s)}  \\
\int \cA_\mu^{(-)}[\phi'^{(s')}] \cA^{(-)\mu}[\phi^{(s)}]  
  &=   \int\, \phi'^{(s')}\big(\a_s(\Box_H - 2 r^2(s+1)) + D^+ D^-\big) 
\phi^{(s)}  \\
\int \cA_\mu^{(\t)}[\phi'^{}] \cA^{(\pm)\mu}[\phi^{}]  
 &= - \int \phi'^{} x_4 D^\pm\phi^{}   \\
%
 \int \cA_\mu^{(\t)}[\phi'^{}] \cA^{(g)\mu}[\phi^{}] 
  &= \int  \frac{x_4}{R}\, \phi'^{(s')} (s+\t)\phi^{(s)} \\
%
 \int \cA_\mu^{(\t)}[\phi'^{}]  \cA^{(\t)\mu}[\phi^{}]  
  &=  -R^2 \int \cosh^2(\eta) \phi'^{} \phi^{} \ .
  \label{inner-1}
\end{align}
\end{subequations}
Note that the on-shell pure gauge modes $\cA^{(g)}[\phi^{}]$ with $\Box\phi=0$ are null, but these
are  factored out from $\cH_{\rm phys}$. 
Observe also the negative sign of the time-like $\cA^{(\t)}$ mode,
which couples to  $\cA^{(g)}$. 

Now consider the physical sector $\cH_{\rm phys}$.
Since the null modes $\cA^{(g)}$ are orthogonal to all gauge-fixed modes, 
these must all be positive, because 
the 4-vectors $\cA^\mu$ have only one negative and three positive components 
at each point of $\cM^{3,1}$. Here we use the fact that $\cC^s$
has a positive definite inner product \eqref{inner-prod-End}, which is related 
to the space-like gauge as discussed in section \ref{sec:space-like gauge}.
Hence the theory is ghost-free, as expected\footnote{It is quite remarkable that 
this is possible, 
in spite of having only 3 rather than 4 diffeomorphism 
degrees of freedom.
The underlying reason is the reduced Lorentz invariance, which enables the 
internal space-like gauge \eqref{inner-gauge-fixing-tau}.}.

We can also check this explicitly for some modes.
The positivity of $(\Box_H - 2 r^2(s+1))$, see \eqref{xa+xa-id}, implies 
that the inner product of the $\cA^{(-)}[\phi^{(s,0)}]$
is indeed positive.  This follows also from $x^\mu\cA_\mu^{(-)}[\phi^{(s,0)}] = 0$, which 
means that the time-like component vanishes.

%
%
\paragraph{Discussion.}

The physical significance of these modes is most transparent in  space-like gauge.
This suggests to interpret the divergence-free  modes 
$\cA^{(-)}[\phi^{(s,0)}]$ as massless spin $s$ gauge fields,
while the generic modes  $\cA^{(ph)}[\phi^{(s)}]$  describe would-be massive 
spin $s$ fields, 
according to the discussion in section \ref{sec:so41-gauge}. Indeed
we will see in the case of spin 2  that the  $\cA^{(-)}[\phi^{(2,0)}]$ 
correspond
to  massless gravitons with 2 degrees of freedom, while the 
remaining $\cA^{(-)}[\phi^{(2,1)}]$ and $\cA^{(-)}[\phi^{(2,2)}]$ 
contribute the remaining 3 degrees of freedom
for a generic spin 2 mode.
\paragraph{Quadratic form for $\cD^2$.}
Using the above results, we obtain the quadratic form for $\cD^2$ as follows:
\begin{subequations}
\label{eq:quad_form}
\begin{align}
 \int \cA_\mu^{(g)}[\phi'^{}] \Big(\cD^2+\frac 12\mu^2\Big) 
\cA^{(g)\mu}[\phi^{}] 
 &= \int \Box \phi'^{} \Box\phi^{} \\
 \int \cA_\mu^{(+)}[\phi'^{(s)}] \Big(\cD^2+\frac 12\mu^2 \Big) 
\cA^{(g)\mu}[\phi^{(s+1)}] 
&=  - \frac{s+3}R\int  D^+\phi'^{(s)}\, \Box\phi^{(s+1)} \\
\int \cA_\mu^{(-)}[\phi'^{(s)}] \Big(\cD^2+\frac 12\mu^2\Big) 
\cA^{(g)\mu}[\phi^{(s-1)}]
&=  - \frac{-s+2}R\int  D^-\phi'^{(s)} \Box\phi^{(s-1)} \\
\int \cA_\mu^{(+)}[\phi'^{(s-1)}] \Big(\cD^2+\frac 12\mu^2\Big) 
\cA^{(-)\mu}[\phi^{(s+1)}] 
 &= -\int  D^+\phi'^{(s-1)} D^- \Big(\Box+\frac{-2s}{R^2}\Big)\phi^{(s+1)} \\
\int  \cA_\mu^{(\t)}[\phi'] \Big(\cD^2+\frac 12\mu^2\Big)\cA^{(4)\mu}[\phi] 
    &= - \int R^2 \cosh^2(\eta) \phi' \Big(\Box+\frac{4}{R^2}\Big)\phi \nn \\
    &\qquad -\int 2 \phi'\sinh^2(\eta) (s+\t)\phi 
\end{align}
\end{subequations}
using \eqref{geometry-H-M}, \eqref{puregauge-D2},
\eqref{inner-1}, \eqref{A2-gaugefix-ids}, and \eqref{Box-x4-relations}.
As a consistency check, we also compute
\begin{subequations}
\label{}
\begin{align}
 \int \Big(\cD^2+\frac 12\mu^2\Big)  \cA_\mu^{(+)}[\phi'^{(s)}]  
\cA^{(g)\mu}[\phi^{(s+1)}] 
  &=  -\frac{s+3}R\int  \Box D^+\phi'^{(s)} \phi^{(s+1)} \,, \\
 \int  \Big(\cD^2+\frac 12\mu^2\Big) \cA_\mu^{(-)}[\phi'^{(s)}]  
\cA^{(g)\mu}[\phi^{(s-1)}] 
&=- \frac{-s+2}R\int  D^-\phi'^{(s)} \Box\phi^{(s-1)} \,,
\\
 \int \Big(\cD^2+\frac 12\mu^2\Big)  
\cA_\mu^{(+)}[\phi'^{}]\cA^{(-)\mu}[\phi^{}] 
 &= -\int \Box D^+\phi'^{}  D^-\phi^{} \, .
\end{align}
\end{subequations}
in agreement with the above.
The terms $\int \cA_\mu^{(\t)}(\cD^2+\frac 12\mu^2)\cA^{(i)\mu}$ for 
$i\in\{\pm,g\}$
can be evaluated using the above results for $\cD^2\cA^{(i)}$, and we  skip the verification of Hermiticity for $\cD^2$.
As expected, the kinetic term for time-like 
$\cA_\mu^{(\t)}$ has a negative sign, but this mode is not part of the 
physical Hilbert space. Nevertheless,
these modes are needed to compute the propagator.
%
%
\section{Metric and gravitons on \texorpdfstring{$\cM^{3,1}$}{M31}}
\label{sec:graviton}
In this paper, we consider only the basic metric aspects of linearized gravity. Further  developments  and more formal aspects
will be studied elsewhere.
The effective metric on a perturbed background is extracted from the kinetic 
term as in \eqref{scalar-action-G}, and is formally obtained from the 
bi-derivation
\begin{align}
\begin{aligned}
 \g:\quad \cC\times \cC  \ &\to  \quad \cC  \\
  (\phi,\phi') &\mapsto \{Y^\a,\phi\}\{Y_\a,\phi'\}
  \end{aligned}
\end{align}
up to a conformal factor as discussed in section \ref{sec:metric}.
Specializing to $\phi=x^\mu, \phi' = x^\nu$ we obtain  the  form
 $\g^{\mu\nu} =  \obar\g^{\mu\nu} + \d_\cA \g^{\mu\nu} + O(\cA^2)$
 in Cartesian coordinates, with 
 metric fluctuation given by
\begin{align}
\begin{aligned}
 \d_\cA \g^{\mu\nu} &\coloneqq 
  \{t^\a,x^\mu\}\{\cA_\a,x^\nu\} + (\mu \leftrightarrow \nu)  \\
  &= \sinh(\eta) \{\cA_\mu,x^\nu\} + (\mu \leftrightarrow \nu) \ .
  \label{gravitons-H1}
  \end{aligned}
\end{align}
To evaluate this for the above $\cA^{(i)}$ modes,
it is convenient to consider the following rescaled graviton mode: 
\begin{align}
  H^{\mu\nu}[\cA] &\coloneqq 
 \frac{1}{\sinh(\eta)} \d_\cA \g^{\mu\nu} =  \{\cA^\mu,x^\nu\} + (\mu 
\leftrightarrow \nu) \nn\\
  h^{\mu\nu}[\cA] &\coloneqq [ H^{\mu\nu}[\cA]]_0
 =  \{\cA^\mu,x^\nu\}_- + (\mu \leftrightarrow \nu) \ .
 \label{tilde-H-def}
\end{align}
Clearly only $\cA \in \cC^1$ can contribute to $ h^{\mu\nu}[\cA]$.
The trace contributions are 
\begin{align}
  H[\cA] &= 2\{\cA^\mu,x_\nu\}, \qquad \quad
  h[\cA] = 2\{\cA^\mu,x_\mu\}_0 \ .
 \label{h-trace}
\end{align}
We observe 
\begin{align}
 \{t_\mu, h^{\mu\nu}[\cA]\} 
 &= \{\{t_\mu,\cA^\mu\},x^\nu\}_- - \frac 2R\, D^-\cA^\nu  \nn\\
 &= - \frac 2R\, D^-\cA^\nu 
 \label{t-hmunu-id}
\end{align}
using \eqref{A2-gaugefix-ids} for $\cA^\nu\in \cC^1$, and
assuming the gauge-fixing $\{t_\mu,\cA^\mu\} = 0$ in the last line.
Then 
\begin{align}
 \{t_\nu,\{t_\mu, h^{\mu\nu}[\cA]\} \}
 &= - \{t_\nu,\{x^\nu,\{t_\mu,\cA^\mu\}\}_-\} 
 - \frac 2R\,\{t_\nu,D^-\cA^\nu \}  \nn\\
 &= - \frac 1R D^-\{t_\mu,\cA^\mu\}
 - \frac 2R\,\{t_\nu,D^-\cA^\nu \}   \nn\\
 &= - \frac 3R D^-\{t_\mu,\cA^\mu\}
 - \frac 1{R^2}\, h \ .
 \label{tth-contract-id}
\end{align}
However, recall that the effective metric, as identified in section \ref{sec:metric}, 
differs from the above by a conformal factor: hence, \eqref{eff-metric-G} becomes  
\begin{align}
 G^{\mu\nu} = \obar G^{\mu\nu} + \d G^{\mu\nu}  \; , \qquad 
\text{with} \qquad 
{\d G}^{\mu\nu} \coloneqq \a\left[ \d_\cA \g^{\mu\nu} 
  - \frac{1}{2} \eta^{\mu\nu}\,\left(\eta_{\a\b}\ \d_\cA \g^{\a\b}\right) 
\right]_0 \; ,
\label{eq:def_phys_graviton}
\end{align}
where $\obar G^{\mu\nu} = \a\g^{\mu\nu} = \sinh^{-1}(\eta)\eta^{\mu\nu}$, see \eqref{G-effective},
is the effective background metric, and
$\a = \sinh^{-3}(\eta)$ is the conformal factor arising from the fixed symplectic measure on $\C P^{1,2}$.
Note that $\eta^{\mu\nu}\,\eta_{\a\b}  = \g^{\mu\nu}\,\g_{\a\b}$, 
i.e.\ the conformal factor drops out here.
We therefore proceed with the 
auxiliary metric fluctuation $ h^{\mu\nu}$.
\paragraph{Pure gauge modes.}
Suppose $\phi = \phi^{(1)}$  is a  spin 1 field. Then the auxiliary metric 
fluctuation of $\cA^{(g)}$ has the following properties:
\begin{subequations}
\label{eq:prop_pure_gauge}
\begin{align}
  h^{\mu\nu}_{(g)}[\phi] &\coloneqq
  h^{\mu\nu}[\cA^{(g)}]  
  = -\{t^\mu,\cA^{\nu(-)}[\phi]\} + (\mu \leftrightarrow \nu)  
  + \frac{2}{R} D^-\phi\, \eta^{\mu\nu}\nn\\
  &= -\{t^\mu,\cA^{\nu(-)}[\phi]\} + (\mu \leftrightarrow \nu)  
  + \frac{1}{3}  h^{(g)} \eta^{\mu\nu} \,,
  \label{pure-gauge-metric} \\
h^{(g)}[\phi] &\coloneqq
 h[\cA^{(g)}] =  -2\{t_\mu,\cA^{\mu(-)}[\phi^{(1)}]\}  
  + \frac 8R D^-\phi\, 
  = \frac 6R D^-\phi\, \nn\\
 &= 6 \{t_\mu,\cA^{\mu(-)}[\phi^{(1)}]\} \,,
 \label{pure-gauge-H}\\
  \{t_\mu, h^{\mu\nu}[\cA^{(g)}]\} 
 &=-\{\Box \phi,x^\nu\}_- - \frac 2R\, D^-\{t^\nu,\phi\} \,,
 \label{t-h-g-relation}  \\
\{t_\mu,\{t^\a, h^{(g)}_{\a\nu}\}\} 
 &=  -\frac 13 \{t_\mu,\{t_\nu, h^{(g)} \}\}
 + \{t_\mu,\{x_\nu,(\Box - \frac{2}{R^2} )\phi\}_-\} \,,
\end{align}
\end{subequations}
using \eqref{A2-gaugefix-ids}.
Hence the $ h^{\mu\nu}_{(g)}$ are traceless for divergence-free modes 
$\phi^{(1,0)}$.
Even for the $\phi^{(1,1)}=D\phi^{(0)}$ modes, the trace contribution $h$ is 
subleading and can be dropped at scales much shorter than the cosmological 
scales. 
To see this, consider for $\phi=\phi^{(0)}$
\begin{align}
 h^{(g)}_{\mu\nu}[D\phi] 
  &= -\{x_\mu,\{t_\nu,D\phi\}\} + (\mu \leftrightarrow \nu)  \nn\\
&=  O(r^2 R^2 \cosh^3(\eta))\del\del \del\phi + (\mu \leftrightarrow \nu)  
\nn\\ 
 h^{(g)}[D\phi] &=  \frac{6}{R}\  D^-D\phi = O\left(r^2 R \cosh^2(\eta)\right) 
\Delta_3 
\phi
\end{align}
noting that $D(\phi) = O(\theta\del\phi)$ due to \eqref{D-ids-1}. 
Therefore 
\begin{align}
 h^{(g)}_{\mu\nu} = O\left(x \del\right) h^{(g)} \ \gg \ h^{(g)} \ .
 \label{h-g-traceless}
\end{align}
Therefore the pure gauge modes
are effectively traceless for sufficiently short scales, 
and $h^{(g)}_{\mu\nu}$ has the usual form $\sim \del_\mu \cA_\nu + 
\del_\nu \cA_\mu $ 
of volume-preserving diffeomorphisms due to \eqref{pure-gauge-metric}.
Finally, we note  that the gauge transformation of the combination
\begin{align}
  \d_\phi\left( h_{\mu\nu} - \frac 13 \eta_{\mu\nu}  h\right) 
  =  h^{(g)}_{\mu\nu} - \frac 13 \eta_{\mu\nu}  h^{(g)}
   = -\{t^\mu,\cA^{\nu(-)}[\phi]\} + (\mu \leftrightarrow \nu)  
 \label{special-gaugetrafo-metric}
\end{align}
is close to the usual form of a pure gauge diffeomorphism contribution,  
determined by  $\phi \in \cC^{(1)}$. 
This will be a useful starting point for the construction of a linearized 
Einstein-Hilbert-like action. 
In particular, since the trace-contribution is subleading, 
the different factors $\tfrac{1}{3}$ versus $\tfrac{1}{2}$ in 
\eqref{eq:def_phys_graviton} is not significant. 
\paragraph{Physical $\cA^{(-)}$ modes.}
Among the $\cA^{(-)}[\phi^{(s)}]$ modes, only  the ones with spin $s=2$ can 
contribute to the metric, as
\begin{subequations}
\label{eq:prop_A2-mode}
\begin{align}
  h_{(-)}^{\mu\nu}[\phi^{(2)}] &\coloneqq
  h^{\mu\nu}[\cA^{(-)}[\phi^{(2)}]] = - \{x^\mu,\{x^\nu,\phi^{(2)}\}_-\}_- + 
(\mu\leftrightarrow\nu)  \, , \\
h_{(-)} &\coloneqq
 \eta_{\mu\nu}  h_{(-)}^{\mu\nu}
  = -2\{x^\mu,\{x_\mu,\phi^{(2)}\}_-\}_-  \
  = 2 D_- D_-\phi^{(2)}  \,,
    \label{Hmunu-tr} \\
\{t_\mu, h_{(-)}^{\mu\nu}[\phi^{(2)}]\} 
  &= -\frac{2}{R} \{x^\nu, D^-\phi^{(2)}\}_- 
  =
  -\frac{2}{R}\cA_\nu^{(-)}[D^-\phi^{(2)}] \ ,
  \label{Hmunu-div}  \\
\{t_\mu,\{t^\a, h^{(-)}_{\a\nu}\}\}  
 + (\mu\leftrightarrow \nu)
   &=\frac{2}{R^2} \left(h^{(g)}_{\mu\nu}  - \frac 
13\eta_{\mu\nu}h^{(g)} \right)[D^-\phi^{(2)}]
\end{align}
\end{subequations}
using \eqref{xpxp-id-4} and \eqref{t-hmunu-id}, 
cf.\ \eqref{phi-tensorfield-H}. 
Recall from \eqref{A2-gaugefix} that $\cA^{(-)}[\phi^{(2)}]$ is automatically gauge-fixed. 
Therefore the physical 
Hilbert space $\cH_{\rm phys}$ contains
not only  $\cA^{(-)}[\phi^{(2,0)}]$, but also 
 $\cA^{(-)}[\phi^{(2,1)}]$ and  $\cA^{(-)}[\phi^{(2,2)}]$, unless the last two are 
equivalent to pure gauge modes.
Moreover, \eqref{Hmunu-tr} shows that $ h_{(-)}$ vanishes for $\phi^{(2,0)}$ 
and $\phi^{(2,1)}$, 
but it is non-vanishing  for $\phi^{(2,2)}$. 
The  $h^{(-)}_{\mu\nu}[\phi^{(2,0)}]$  are divergence-free and traceless 
gravitons, which
realize the usual 2 propagating physical degrees of freedom of GR.
The $h^{(-)}_{\mu\nu}[\phi^{(2,1)}]$  lead to graviton modes 
which turn out to be Ricci-flat on-shell in section \ref{sec:curvature-eval}, 
and therefore may be equivalent\footnote{in the present approximation which 
is restricted to scales shorter than the cosmic curvature scale, we cannot decide 
whether the $h^{(-)}_{\mu\nu}[\phi^{(2,1)}]$ are exactly or only approximately equal 
to pure gauge modes. } 
to unphysical pure gauge modes.
Finally,  $h^{(-)}_{\mu\nu}[D^+D^+\phi^{(0)}]$ provides 
an extra propagating scalar metric mode which has no counterpart in GR;
its significance  should be clarified elsewhere.
Off-shell, the modes in $h^{(-)}_{\mu\nu}[\phi^{(2)}]$ provide the degrees of 
freedom of a massive spin 2 multiplet. 
These modes are approximately in 
de Donder gauge, since
\eqref{Hmunu-div} is suppressed at scales  
shorter than the cosmological curvature scale.

%
\paragraph{Unphysical $\cA^{(+)}$ modes.}
None of the  $\cA^{(+)}[\phi^{(s)}]$ modes with $s\geq 1$  can contribute
to $ h^{\mu\nu}$, because 
\begin{align}
 H^{\mu\nu}[\cA^{(+)}[\phi^{(s)}]] &= - \{x^\mu,\{x^\nu,\phi^{(s)}\}_+\} + 
(\mu\leftrightarrow\nu)  \qquad \in \cC^s \oplus \cC^{s+2} \  \ .
\end{align}
However the scalar $\cA^{(+)}[\phi^{(0)}]$ mode contributes, which gives 
\begin{subequations}
\label{eq:prop_A2+mode}
\begin{align}
 h_{(+)}^{\mu\nu}[\phi^{(0)}] &= - \{x^\mu,\{x^\nu,\phi^{(0)}\}_+\}_- + 
(\mu\leftrightarrow\nu) \ \nn\\
  &= -2[\theta^{\mu\a}\theta^{\nu\b}]_0\del_\a\del_\b\phi - 
\left(\{x^\mu,\theta^{\nu\b}\}\del_\b +  \{x^\nu,\theta^{\mu\b}\}\del_\b\right) 
\phi^{(0)} \nn\\
  &= -\frac{2r^2R^2}3\left(P^{\mu\nu}P^{\a\b} - P^{\mu\b}P^{\nu\a}\right)\del_\a\del_\b\phi^{(0)} 
  - (\{x^\mu,\theta^{\nu\b}\}\del_\b +  \{x^\nu,\theta^{\mu\b}\}\del_\b) \phi^{(0)}\nn\\
 &= \frac{2r^2R^2}3\left(\del^\mu \del^\nu 
   - (\eta^{\mu\nu} + R^{-2}x^\mu x^\nu)\del^\a\del_\a\right)\phi^{(0)} \nn\\
 &\qquad  
  - \frac 23 r^2 \eta^{\mu\nu}(\t  + 2)\t\phi^{(0)}
  + \frac 13 r^2(x^\nu \del^\mu + x^\mu \del^\nu)(1+ 2\t) \phi^{(0)} \,  \\
h_{(+)} &= -2\{x_\mu,\{x^\mu,\phi^{(0)}\}\}_- 
   = -2 \{x_\mu,\{x^\mu,\phi^{(0)}\}\}_0 \,,\\
 \{t_\mu, h_{(+)}^{\mu\nu}\} 
 &= \{\{t_\mu,\cA^{(+)\mu}\},x^\nu\}_- - \frac 2R\, D^-\cA^{(+)\nu}  \nn\\
 &=  \frac{3}R\{D^+\phi^{(0)},x^\nu\}_- - \frac 2R\, D^-\cA^{(+)\nu}  
  =  -2r^2 \{t^\nu,\phi^{(0)}\} - \frac{5}{R}\{x^\nu,D\phi^{(0)} \}_- 
 \label{th-rel-A2+} 
\end{align}
\end{subequations}
recalling $\t=x\del$ and
using \eqref{average-H}, \eqref{A2-gaugefix}, and \eqref{t-hmunu-id}.
Even though $\cA^{(+)}[\phi^{(0)}]$ is unphysical, this 
provides the  missing scalar degree of freedom for the 
off-shell metric fluctuations.
Finally, $\cA^{(\t)}$  
is not expected to provide an additional physical mode,
and will not be considered any further here.
\paragraph{Discussion.}

To summarize, we have found 9 independent off-shell metric fluctuations:
5 from the $\cA^{(-)}[\phi^{(2)}]$ modes, one from the scalar 
$\cA^{(+)}[\phi^{(0)}]$ mode, and the 3 pure gauge modes.
All these modes are governed by the appropriate wave equations 
\eqref{on-shell-B2pm},
which are massless up to cosmological scales. There may be an extra 
off-shell (unphysical) mode based on $\cA^{(\t)}$, which should be clarified elsewhere.
These metric fluctuations provide  all $(5{+}1)$  degrees of 
freedom required for gravity in the presence of matter.
However, in contrast to GR there are only 3 instead of 4 pure gauge modes 
$\cA^{(g)}$, 
corresponding essentially to volume-preserving diffeomorphisms\footnote{They  preserve
the invariant symplectic volume form on $\C P^{1,2}$.}
\eqref{h-g-traceless}.
We expect that this is also 
related to the absence of a cosmological constant term as discussed
in section \ref{sec:lin-EH-sction}.

On-shell, the physical Hilbert space $\cH_{\rm phys}$ of vacuum modes  
certainly contains the 2 standard Ricci-flat graviton modes, while the 
significance of the other modes remains to be understood. 

These gravitational modes will certainly be sourced by matter.
However to 
obtain the linearized Einstein equations to a sufficiently good approximation 
may require an induced Einstein-Hilbert-like action, as discussed 
below.
In fact, induced gravity actions are known to arise quite generically upon 
taking quantum effects into account \cite{Sakharov:1967pk,Visser:2002ew}. 
The large gauge 
invariance of the present framework will allow to determine these actions,
as shown in the remaining sections.
%
%
\subsection{Linearized curvature-like tensors}
\label{sec:lin-curv}
To understand the significance of the above metric modes, we consider their 
linearized Ricci tensor.
Recall that the linearized Ricci tensor $\cR_{(\rm lin)}^{\mu\nu}$ 
for a metric fluctuation $\d g_{\mu\nu}$ is given by
\begin{align}
R_{(\rm lin)}[\d g] 
 &= -\nabla^\a \nabla_\a \d  g^{\mu \nu}  
 + \nabla^{\mu} \nabla_\r \d  g^{\nu\r} + \nabla^{\nu} \nabla_\r \d  g^{\mu\r} 
  - \nabla_\mu\nabla_\nu \d g  \nn\\
 &\quad +\eta^{\mu\nu}(\nabla^\a\nabla_\a \d g -  \nabla_{\s} \nabla_\r \d  g^{\s\r}) \ .
 \label{gmunu-lin}
\end{align}
We need to find analogous expressions that are suitable for the matrix 
model setting, which should  be defined solely in terms 
of  (commutators or) Poisson brackets.
We tentatively define a linearized \emph{Poisson-Ricci} tensor  $\cR_{(\rm 
lin)}^{\mu\nu}$ and scalar $\cR_{(\rm lin)}$, as well as a linearized 
\emph{Poisson-Einstein} tensor $\cG_{(\rm lin)}^{\mu\nu}$ as follows:
\begin{align}
 2 \cR_{(\rm lin)}^{\mu\nu}[\d g] 
 &= \Box\d   g^{\mu\nu}
  + \big(\{t_\mu, \{t_\r,\d   g^{\r\nu}\}\} - \frac 12\{t_\mu,\{t_\nu,\d g\}\}
  +  (\mu \leftrightarrow \nu)\big) \,, \nn\\
 \cR_{(\rm lin)} &=  \Box\d g + \{t_\mu, \{t_\r,\d   g^{\r\mu}\}\}    \,, 
  \label{ricci-poisson-def} \\[1ex]
 2\cG_{(\rm lin)}^{\mu\nu}[\d   g] &= 2R_{(\rm lin)}^{\mu\nu}[\d  g] - \eta^{\mu\nu} R_{(\rm lin)}[\d  g] \nn\\
  &= \Box\d   g^{\mu\nu}
  + \big(\{t_\mu, \{t_\r,\d   g^{\r\nu}\}\}  -  \frac 12\{t_\mu,\{t_\nu,\d g\}\}
  +  (\mu \leftrightarrow \nu)\big) \nn\\
   &\quad - \eta^{\mu\nu}(\Box \d g + \{t_\s, \{t_\r, \d   g^{\s\r}\}\})\,, \nn\\
 \cG_{(\rm lin)} &= -(\Box\d g   + \{t_\mu, \{t_\r,\d   g^{\r\mu}\}\}) \, .
 \label{einstein-poisson-def}
\end{align}
To understand these expressions in terms of usual tensor calculus, we recall 
$\{t_\mu,\cdot\} \sim \sinh(\eta)\del_\mu$ in Cartesian (background) 
coordinates,
and similarly $\Box \sim \sinh^2(\eta) \big(\eta^{\mu\nu}\del_\mu\del_\nu + 
O\big(\frac{1}{x_4}\del\big) \big)$ from
\eqref{G-Box-relation}, because
\begin{align}
 \frac{1}{\sinh(\eta)}\del\sinh(\eta)= O\left(\frac{1}{x_4}\right) 
 \label{estimate-del}
\end{align}
and $x_4= R\sinh(\eta)$ measures the current size of the universe. 
Then the linearized Poisson-Ricci tensor \eqref{ricci-poisson-def} reduces to  
\begin{align}
 \cR_{(\rm lin)}^{\mu\nu}[\d g^{\a\b}]
 &= \frac 12\Box \d g^{\mu\nu}
  + \left(\{t^\mu,\{t_\a,\d g^{\a\nu}\}\}
  - \frac 12 \{t^\mu,\{t^\nu,\d g\}\}
   + (\mu \leftrightarrow \nu)\right) \nn\\
  &=  \frac 12 \sinh^2(\eta) \left(-\del^\a\del_\a \d g^{\mu \nu}  
  - \del^\mu\del^\mu \d g
 + \del^{\mu}\del_\r \d g^{\nu\r} + \del^{\nu}\del_\r \d g^{\mu\r} \   + 
O\left(\frac{1}{x_4}\del \d g\right)  \right) \nn\\
  &= \sinh^2(\eta) \left( R_{(\rm lin)}^{\mu\nu}[\d g^{\a\b}] \ 
  + O\left(\frac{1}{x_4}\del \d g\right)\right) \ .
  \label{box-tilde-h-R}
\end{align}
Up to the sub-leading corrections,
this is  the classical linearized Ricci tensor for
a metric fluctuation  $\d g^{\mu\nu}$ 
around a background $\obar G^{\mu\nu} = \sinh^{2}(\obar\eta)\eta^{\mu\nu}$, or 
some suitable conformal rescaling thereof such as \eqref{G-effective}.
Note that the indices in the second line are contracted with $\eta_{\\a\b}$.

%
\subsection{Curvature and gravitational waves}
\label{sec:curvature-eval}
%

Now we relate the linearized Ricci tensor to the equations of 
 motion of the physical modes $\cA^{(\pm)}$ underlying the metric fluctuations,
 and  show that the linearized gravitational waves of GR are recovered.
 
To evaluate the Ricci tensor, we 
need a relation for $\Box h^{\mu\nu}$. This can be achieved using the intertwiner relation \eqref{D2-h-intertwiner}, 
\begin{align}
 \cD^2  h^{\mu\nu}[\cA]
  &= \left(\Box + \frac{2}{R^2r^2}\tilde\cI\right) h^{\mu\nu}[\cA]
  =  h_{\mu\nu}[\cD^2\cA]   +  \frac{2}{R^2}\left(3 h^{\mu\nu}[\cA] - 
\eta^{\mu \nu} h[\cA]\right) \, 
   \label{D2-h-approx}
\end{align}
and noting that  $\cD^2 = \Box + O(\frac{1}{x_4}\del) \approx \Box$ 
up to corrections of  the order of the cosmological curvature.
Let us discuss this for the different modes.

 \paragraph{Pure gauge modes.}
 Consider first the pure gauge mode  $\cA^{(g)}[\phi]$. 
 From the properties 
derived in \eqref{eq:prop_pure_gauge} we obtain for the Poisson-Ricci tensor
\begin{align}
 2\cR_{(\rm lin)}^{\mu\nu}[ h^{(g)}[\phi]]
  &= \Box h^{(g)}_{\mu\nu} 
   + \frac{2}{R^2} \left(h^{(g)}_{\mu\nu}  - \frac 13\eta_{\mu\nu}h^{(g)} 
\right) 
[\phi]
    - \left(h^{(g)}_{\mu\nu}  - \frac 13\eta_{\mu\nu}h^{(g)} \right)[\Box\phi] 
\nn\\
  &\quad  -\left(\frac{1}{2} + \frac{1}{3} \right)
  \left( \{t_\mu,\{t_\nu,h^{(g)}\}\} 
+ (\mu\leftrightarrow \nu) \right) \nn\\
  &\approx \Box h^{(g)}_{\mu\nu}  - h^{(g)}_{\mu\nu}[\Box\phi] 
 \label{labelRlin-puregauge}
\end{align}
at sufficiently short scales, using 
\eqref{h-g-traceless} and
dropping $\frac 1{R^2}$ terms. 
Together with \eqref{D2-h-approx}
to this approximation\footnote{Of course these modes 
cannot have any physical effect whatsoever. This is just a consistency check for the approximation 
which is used for the other, physical modes.},
the linearized Ricci tensor \eqref{labelRlin-puregauge} vanishes both on and off shell
as it should, 
\begin{align}
 R_{(\rm lin)}^{\mu\nu}[ h] \ = 0 \   + 
O\left(\frac{1}{x_4}\del  h\right) \ .
\end{align}

\paragraph{Physical $\cA^{(-)}$ modes.}
Now consider  $\cA^{(-)}[\phi^{(2)}]$, for which the properties listed in 
\eqref{eq:prop_A2-mode} imply 
\begin{align}
 2\cR_{(\rm lin)}^{\mu\nu}[ h^{(-)}[\phi]]
  &= \Box h^{(-)}_{\mu\nu} 
   + \frac{2}{R^2} \left(h^{(g)}_{\mu\nu}  - \frac 13\eta_{\mu\nu}h^{(g)} 
\right)[D^-\phi]
    +\left(\{t_\mu,\{t_\nu,D^-D^-\phi\}\} + (\mu\leftrightarrow\nu) \right) 
\nn\\
  &\approx \Box h^{(-)}_{\mu\nu} 
  +\left(\{t_\mu,\{t_\nu,D^-D^-\phi\}\} + (\mu\leftrightarrow\nu) \right) \nn\\
  &= \Box h^{(-)}_{\mu\nu} 
  +\frac 12\left(\{t_\mu,\{t_\nu,h^{(-)}\}\} + (\mu\leftrightarrow\nu) \right) 
\end{align}
dropping $\frac 1{R^2}$ terms. 
We will see  that
this  vanishes on-shell but not off-shell for $\cA^{(-)}[\phi^{(2,0)}]$
and $\cA^{(-)}[\phi^{(2,1)}]$. 

Consider first the 2 physical  modes $\cA^{(-)}[\phi^{(2,0)}]$. The on-shell 
condition \eqref{on-shell-B2pm} for the bare action is 
$(\Box-\frac{4}{R^2}) \phi^{(2,0)} = 0$, and
the corresponding metric fluctuation $ h_{\mu\nu}$ is 
traceless and satisfies $\{t^\mu, h_{\mu\nu}\} = 0$ due to \eqref{Hmunu-div}.
It follows with \eqref{D2-h-approx} that
the linearized Ricci tensor vanishes on-shell up to corrections of  the order of the cosmological curvature,
\begin{align}
\boxed{
 \ \  \sinh^2(\eta)\, R_{(\rm lin)}^{\mu\nu}[ h] 
 \ \approx  \ \cR_{(\rm lin)}^{\mu\nu}[ h] 
 \  \approx \ \frac 12  h^{\mu\nu}[\cD^2\cA^{(-)}]  \ = 0 \   + 
O\left(\frac{1}{x_4}\del  h\right)  \ .
 }
\end{align}
These  are essentially the standard Ricci-flat gravitational wave solutions of 
GR, which are thus recovered in the bare matrix model.
This is one of the main results of 
this paper.
In the presence of an induced Einstein-Hilbert term $S_{\rm EH}$ \eqref{S-EH}  as discussed below, 
these  will of course remain to be 
solutions, up to small corrections.

Now consider the  physical modes 
$\cA^{(-)}[\phi^{(2,1)}]$. Since they are trace-free \eqref{Hmunu-tr}, 
it follows with
 \eqref{D2-h-approx} that they are also Ricci-flat up to subleading corrections,
\begin{align}
 R_{(\rm lin)}^{\mu\nu}[ h] \ = 0 \   + 
O\left(\frac{1}{x_4}\del  h\right) \ .
\end{align}
In general, there exist only 2 independent Ricci-flat metric modes (apart from
trivial diffeomorphism modes). Therefore in this approximation, the $\cA^{(-)}[\phi^{(2,1)}]$ should not be independent modes.
This would suggest that the present theory contains only
the 2 propagating metric modes of massless spin 2 gravitons as in GR, 
rather than the 5 propagating 
modes of massive gravitons. 
However, is is conceivable that the latter are masked by 
pure gauge modes, and are not visible in the present approximation.
To settle this question, a more refined analysis is required.

Now consider the scalar metric mode arising from $\cA^{(-)}[\phi^{(2,2)}]$. This is not expected to be Ricci-flat in general, 
and provides an extra physical mode 
which has no counterpart in GR. 
This is not surprising due to the reduced gauge invariance, corresponding 
to a 3-parameter volume-preserving diffeomorphism group which 
preserves the invariant volume form.
The existence of extra scalar modes in modified theories of 
gravity is not uncommon and may be of interest e.g. in the context of cosmology 
or possibly dark matter.
However, we leave a more detailed investigation to future work.
Finally,
the $\cA^{(+)}[\phi^{(0)}]$ mode is not physical and hence discarded.

In the presence of an induced Einstein-Hilbert action $S_{\rm EH}$ \eqref{S-EH}
discussed below,
the 2 propagating Ricci-flat modes will clearly survive, while the 
fate of the scalar gravity mode remains to be understood.
Of course off-shell, all 5+1 non-trivial metric modes are present, which 
are needed to describe  gravity in the presence of matter.
We  therefore expect to recover linearized GR to a 
 good approximation, presumably with an extra scalar mode, and a 
non-linear completion defined by the matrix model.

\subsection{Linearized Einstein-Hilbert-like action}
\label{sec:lin-EH-sction}

In this final section, we  derive an action 
for $h^{\mu\nu}$ which plays the role of the linearized 
Einstein-Hilbert (E-H) action. The strategy for the construction 
is based on two requirements: first, the action should reduce to the usual 
linearized E-H action in the  appropriate limit and, second,
 the action should be gauge invariant.
 We  obtain such an action which is expressed solely in terms of Poisson brackets,
as appropriate for the present framework.
Much of the following considerations would generalize to 
higher-spin modes. However to keeps things within bounds we restrict ourselves 
to the spin 2 sector, postponing the general case.

In the classical case, gauge invariance of the Einstein-Hilbert action 
is guaranteed by the 
Bianchi identity $\nabla_\mu \cG^{\mu\nu} = 0$.
To find a gauge-invariant action, we recall from 
\eqref{special-gaugetrafo-metric} that the combination 
$ h_{\mu\nu} - \frac 13 h\, \eta_{\mu\nu}$ has a particularly simple 
gauge transformation.
Therefore, consider the action
\begin{align}
 S_1 &= \int  \left( h_{\mu\nu} - \frac 13 h \,\eta_{\mu\nu}\right) 
 \cG_{(\rm lin)}^{\mu\nu}\left[ h_{\a\b} - \frac 13 h\, \eta_{\a\b}\right] \ .
 \label{S-1-curvature}
\end{align}
The gauge transformation of the metric fluctuation is given by
$\d_\phi  h^{\mu\nu} =  h_{(g)}^{\mu\nu}[\phi]$ in \eqref{pure-gauge-metric}, 
which is non-vanishing only for
spin 1 gauge transformations generated by $\phi\in\cC^1$.
Since the bracket structure of $\cG_{(\rm lin)}^{\mu\nu}$ \eqref{einstein-poisson-def}
amounts to a self-adjoint operator, 
the gauge variation of this action is given by 
\begin{align}
 \d_\phi S_1 &= 2\int  \d_\phi\left( h_{\mu\nu} - \frac 13 h\, \eta_{\mu\nu} 
\right) 
 \cG_{(\rm lin)}^{\mu\nu}\left[ h_{\a\b} - \frac 13 h \,\eta_{\a\b} 
\right]   \nn\\
 &= -4 \int  \{t_\mu,\cA^{(-)}_{\nu}[\phi]\}\,
 \cG_{(\rm lin)}^{\mu\nu}[ h_{\a\b} - \frac 13 h \, \eta_{\a\b}]  \nn\\
 &= 4 \int \cA^{(-)}_{\nu}[\phi]\,
  \left\{t_\mu,\cG_{(\rm lin)}^{\mu\nu}\left[ h_{\a\b} - \frac 13 h \, 
\eta_{\a\b}\right] \right\} 
  \label{del-S1-1}
\end{align}
using \eqref{pure-gauge-metric}.
Although $\{t_\mu,\cG_{(\rm lin)}^{\mu\nu}\}$ is reminiscent of 
the Bianchi identity, it does not vanish  here, but reduces to 
\eqref{conservation-law-ricci}. 
This will determine the required modifications to obtain a gauge-invariant action.
To simplify the evaluation, 
we define the traceless metric fluctuation
\begin{align}
 \d  g_0^{\mu\nu} &=  h^{\mu\nu} - \frac{1}{4} \eta^{\mu\nu}\, h, 
\qquad \quad
  \d  g_0 = 0 \ .
  \label{traceless-tildeh}
\end{align}
In order to compensate the gauge variation of $S_1$, we introduce the following 
quadratic actions:
\begin{subequations}
\label{eq:add_actions}
\begin{align}
S_{h} &= \int  h\Box   h \,,\\
 S_{3}  &=\int  \d  g^0_{\mu\nu} \d g_0^{\mu\nu} \,, \\
 S_{4}  &= \int  \d  g^0_{\mu\nu} \{\cM^{\mu\r},\d g_0^{\r\nu}\}
  = \frac 1{2r^2} \int  \d  g^0_{\mu\nu} \tilde \cI(\d g_0^{\r\nu}) 
\,, \\
 S_{M3}  &= \int f_{\mu\nu} \{\cM^{\nu\r}, h^{\r\mu}\}
  \ =  -\int  h_{\mu\nu} \{\cM^{\nu\r},f^{\r\mu}\} \,,
  \\
 S_{gf2} &=  R \int \{x_\mu, A^\mu\}_-\, D^-\{t_\r,A^\r\} \,,\\
S_{gf3} &= \int (D^-\{t_\nu,A^\nu\})\Box^{-1} (D^-\{t_\nu,A^\nu\})\,,
 \label{S-gf3}
\end{align}
\end{subequations}
assuming that $\Box^{-1}$ makes sense. For the additional action terms, we 
employ some short-hand notation
\begin{align}
  f_{\mu\nu} \coloneqq \{x^\mu,\cA^\nu\}_- - \{x^\nu,\cA^\mu\}_- 
  \,, \qquad
  F_{\mu\nu}[\cA] \coloneqq \{t_\mu,\cA_\nu\} - \{t_\nu,\cA_\mu\} 
\end{align}
which satisfy
\begin{align}
 F_{\mu\nu}[\cA^{(-)}[\phi]] 
   = f_{\nu\mu}[\cA^{(g)}[\phi]] 
  = \d_\phi f_{\nu\mu} \ .
   \label{F-A-phi-id}
\end{align}
The  gauge variations are computed  to be (cf. appendix \ref{app:derivation_action})
\begin{subequations}
\label{eq:del_add_actions}
\begin{align}
\d_\phi S_{h}  &=  2\int  h^{(g)} \Box    h \,,\\
 \d_\phi S_{3}  &= 2 \int \d g_0^{\m\nu} h^{(g)}_{\mu\nu} \,,\\
 \d_\phi S_{4}  
 &= 2\int \d  g_0^{\mu\nu}\{\cM^{\nu\r}, h^{(g)}_{\r\mu} \} \,, \\
\d_\phi S_{M3} 
  &=  -2\int \left(
\{x^\mu,A^\nu\}_-\{\cM^{\nu\r},\{x^\mu,A_{(g)}^\r\}_-\}
    - \{x^\nu,A^\mu\}_- \{\cM^{\nu\r},\{x^\r,A_{(g)}^\mu\}_- \}  \right)
  \,, \\
 \d_\phi  S_{gf2} &= R\int \{x_\mu, A^\mu\}_- \,D^- \{t_\r,A_{(g)}^\r\}
 +  \{x_\mu, A^\mu_{(g)}\}_- D^-\{t_\r,A^\r\}   \,,\\
 \d_\phi S_{gf3}  &=  2\int (D^- \phi)  D^-(\{t_\nu,A^\nu\})  \,,
 \label{S-gf3-del}
\end{align}
\end{subequations}
using tracelessness of $\d g_0^{\m\nu}$.
These additional action terms allow to compensate the gauge variance of $S_1$ 
and, as detailed in appendix \ref{app:derivation_action}, one explicitly finds
\begin{align}
\d_\phi S_1 
  &=-\frac{3}{2 R^2} \d S_{M3} + \frac{6}{R^2}\d_\phi S_{gf2}
  -\frac{9}{2R^2} \d_\phi S_{gf3}
+\frac{1}{2R^2} \d_\phi S_4 + \frac{1}{R^2} \d_\phi S_3 
  + \frac 1{24} \d_\phi S_{h} \,.
\end{align}
We have thus found the analog of the linearized Einstein-Hilbert 
action
\begin{align}
\boxed{
 S_{\rm EH} \coloneqq S_1 +\frac{1}{16} S_h 
 - \frac{1}{R^2} S_3
 - \frac{1}{2R^2}  S_4
 +\frac{3}{R^2} \left( \frac{1}{2}S_{M3} - 2 S_{gf2} + \frac 32 S_{gf3} \right)
 }
 \label{S-EH}
\end{align}
 which is gauge-invariant,
\begin{align}
 \d_\phi S_{\rm EH} = 0 \ .
\end{align}
It is therefore very plausible that this action  is induced 
by quantum corrections at one loop (and beyond), 
upon integrating out fields coupled to the  background with metric fluctuation $ 
h^{\mu\nu}$, 
in the spirit of Sakharov's induced gravity \cite{Sakharov:1967pk,Visser:2002ew}
or the Seeley-de~Wit expansion \cite{Gilkey:1995mj}.
Upon gauge-fixing $\cG(\cA)\equiv\{t_\mu,\cA^\mu\} = 0$, this action simplifies 
to
\begin{align}
 S_{\rm EH,gf} = S_1 + \frac{1}{16} S_h
 - \frac{1}{R^2}\Big(S_3 +\frac{1}{2}  S_4 -\frac{3}{2} S_{M3}\Big) \ .
\end{align}
Although we do not have a formal proof, it seems that $S_{\rm EH}$ of 
\eqref{S-EH} is the only gauge-invariant term 
at second order of derivatives of $ h_{\mu\nu}$;
however, this needs to be studied in more detail elsewhere.
In particular,
there is apparently no gauge-invariant counterpart of
the cosmological constant 
 $\int d^4 x\sqrt{g} \Lambda$.
The natural analog is actually the original Yang-Mills action 
\eqref{bosonic-action}, which 
stabilizes the background.
This suggests that the cosmological constant problem may not arise here,
because the FLRW background is obtained  from the bare 
Yang-Mills-type matrix model action,
without fine-tuning and even without requiring any matter.
 Adding matter would  modify the background at shorter scales 
 in a way which should be similar to GR, provided 
the above linearized Einstein-Hilbert-like action dominates in the quantum 
effective action.

To understand and interpret the above expressions in terms of usual tensor calculus, we recall 
the results and approximations in section \ref{sec:lin-curv}, and in particular \eqref{estimate-del}.
Then the action term $S_1$ of
\eqref{S-1-curvature} clearly reduces to
the Einstein-Hilbert action for the linearized 
metric $ h_{\mu\nu} - \frac 13\eta_{\mu\nu}  h$ of
\eqref{special-gaugetrafo-metric},
up to corrections suppressed by the cosmological curvature scale. 
Moreover,
recall that the trace  $h$ of the physical spin 2 metric modes can be neglected to leading order.
However,
the trace contribution $S_h$ for the scalar metric fluctuations cannot be 
neglected, and remains to be understood.
The terms $S_4, \ S_{M3}$ and $S_3$ in brackets contain only one or no derivative of the metric, and
they are  suppressed by the cosmological curvature scale just like
 the subleading corrections in \eqref{box-tilde-h-R}.
This follows from 
\begin{align}
\frac{1}{R^2}\{\cM^{\mu\nu'},\d g^{\a\b}\} \sim \frac{1}{R^2} (x^\mu\del^\nu - x^\nu\del^\mu) \d g^{\a\b}
 \ll \cR_{(\rm lin)}^{\mu\nu}[\d g^{\a\b}]
 \label{M-estimate}
\end{align}
as long as  $\del \gg \frac 1{x_4}$. 
Hence,  $S_4$, $S_{M3}$, and $S_3$ 
should be negligible for short and intermediate scales, but they may be significant for very long
(cosmological) length scales.
The remaining terms $S_{gf2}$ and  $S_{gf2}$ vanish upon gauge-fixing and are not considered 
any further. 

Therefore $S_{\rm EH}$
reduces to the linearized E-H action for the physical spin 2 modes on short and 
intermediate scales, but not 
for the trace contribution.
The significance of this deviation is not obvious.
Eventually one should also work out the sub-leading contributions 
$O\big(\frac{1}{x_4}\del \d g\big)$, and try to interpret them in terms of the 
cosmological background geometry. These  points are postponed to future 
work.

It is interesting to observe that gauge-invariance apparently requires the non-local contribution 
$S_{gf3}$, see \eqref{S-gf3}. Of course, this term vanishes for gauge-fixed 
fields, leaving a local 
action whose leading  (2nd derivative) contributions are
given by $S_1 -\frac{1}{24} S_h$ in \eqref{S-1-curvature}. Nevertheless, such a 
non-local term is not unreasonable 
in a quantum effective action.
%
%

%
%
\subsection{Discussion}
\label{sec:discussion}
The precise interplay between the  Yang-Mills action \eqref{eff-S-expand} and 
the above 
Einstein-Hilbert-like action $S_{\rm EH}$ \eqref{S-EH} is not yet clear. 
In principle, it should be possible to rewrite the quadratic Yang-Mills action 
$S_2[\cA]$ \eqref{eff-S-expand} for the gravitational modes
in terms of the metric fluctuations $ h^{\mu\nu}$, 
as in \cite{Sperling:2018xrm}. This would  lead to a non-local 
action\footnote{This simply reflects the fact that the 
$\cA_\mu$ are the fundamental fields here, while the metric fluctuations are derived 
fields. Observe that in contrast to $H^4_n$ \cite{Sperling:2018xrm}, the two box operators do not 
cancel here, leading to the vacuum Einstein equations even from the bare 
action.}, which for $\cA^{(-)}$ has the structure
\begin{align}
 S_2[\cA^{(-)}[\phi^{(2,0)}]] \ \propto \ 
 -  \int  \,  h^{\mu\nu}[\phi^{(2,0)}] \Big(\Box-\frac{2}{R^2}\Big) 
   (\Box_H-2 r^2)^{-1}  h_{\mu\nu}[\phi^{(2,0)}] \,,
   \label{bare-action-h}
\end{align}
but we have not yet found an appealing general form. Note that 
$(\Box_H - 2r^2)$ is the positive-definite
Euclidean Laplacian on $H^4$, so that the inverse is perfectly well-defined and introduces no zeros or poles.
However, we can surely say that the full quadratic action $S_{2} + S_{\rm EH}$ will admit 
the usual two Ricci-flat gravitational waves of GR as solutions, 
since they are solutions of both actions. 
In contrast, the extra non-Ricci-flat  mode arising from the scalar
mode in $\cA^{(-)}[\phi^{(2,0}]$  and possibly  $\cA^{(-)}[\phi^{(2,1}]$
are presumably \emph{not} solutions of the full action.

In the presence of matter with standard metric coupling
\begin{align}
 \d_\cA S_{\rm matter}  &= -\frac 12 \int d^4 x \, \d G^{\mu\nu}[\cA]  T_{\mu\nu} \ ,
\end{align}
the bare matrix model action  \eqref{bare-action-h} would  not lead to the linearized Einstein equations but rather to equations 
of the form 
\begin{align}
 \Big(\Box -\frac{2}{R^2}\Big)  h_{\mu\nu} \sim -(\Box_H-2 r^2) T_{\mu\nu} \ .
 \label{box-h-Box-T}
\end{align}
The source term $\Box_H T_{\mu\nu}$  would lead to a short-range metric perturbation, but
a small 
long-distance contribution might survive due to the constant shift. However in the presence of an induced
$S_{\rm EH}$, we 
expect that the linearized Einstein equations in the presence of matter  
are recovered to a good approximation
 at intermediate scales, provided that
induced $S_{\rm EH}$ dominates. At shorter scales, 
corrections due to the right-hand-side of \eqref{box-h-Box-T} are expected.

At the non-linear level, it is not entirely clear how to proceed.
The problem is that the derivations $\{t^\mu + \cA^\mu,\cdot \}$
defined by  a deformed background no longer respect the 
spin operator $\cS^2$.  It remains to be seen how this can be handled, 
possibly by defining a deformed spin operator.
However this is just a technical rather 
than a conceptual problem, and it is clear that the matrix model provides a 
fully non-linear completion of the linearized gravity discussed 
in this paper. It is also clear that this will not be identical to 
GR, and significant differences are bound to arise at cosmological scales.
This is manifest by the fact that the cosmological background solution 
is obtained even without any matter. 
But this is also the regime where our present understanding of gravity is very 
limited, which is manifest in the big puzzles around dark energy, dark matter, and 
the related fine-tuning issues. The gravity theory outlined in this paper  clearly
has the potential to address these problems, and it remains to be seen where it 
leads to.
%
%
\section{Conclusion and outlook}
\label{sec:conclusion}
We briefly summarize some of the most significant points of this paper:
\begin{itemize}
\item 
 Fuzzy $\cM^{3,1}_n$ is defined in terms of the minimal discrete series of 
 unitary representations of $SO(4,2)$.
$\cM^{3,1}_n$ respects a global $SO(3,1)$ isometry, but  (local) 
Lorentz invariance is not manifest. It describes 2-fold cover of a FLRW space-time, linked by a 
Big Bounce.
\item 
The space of functions on $\cM^{3,1}_n$ decomposes into different sectors 
$\oplus \cC^s$, each of which contains the degrees of freedom of an irreducible
massive spin $s-k$ 
tensor field. The constraints allow some internal gauge freedom, and 
they can be represented either as irreducible tangential tensor 
fields on $H^4$, or in space-like gauge eliminating the time-like components.
\item  
Due to the lack of Lorentz invariance, these multiplets decompose further into 
$\cC^s = \oplus_{k} \cC^{(s,k)}$. 
$\cC^{(s,k)} = (D^+)^k\, \cC^{(s-k,0)}$.
The ``lowest'' sector $\cC^{(s-k,0)}$ 
has the degrees of freedom of massless spin $s$ fields.
\item The matrix model defines a gauge theory of tangential fluctuation modes 
on $\cM^{3,1}_n$, with a higher-spin-type gauge invariance. 
We argue that it is ghost-free even though there are only 3 rather than 4 diffeomorphism modes, 
due to the internal space-like gauge and the restricted Lorentz invariance. 
In particular, the (maximally supersymmetric) IKKT matrix model is
expected to define a good quantum theory around this background solution.
\item The bosonic on-shell fluctuation modes are classified, and found to respect  
effectively local Lorentz invariance, corresponding to an effective universal metric.
In contrast to the Euclidean case \cite{Sperling:2018xrm} they are propagating and are
either exactly massless, or massless up to small corrections.
\item 
The effective metric is not fundamental, but a derived 
quantity determined be the matrix model background and its fluctuations.
In particular, the cosmological background solution is not determined by the Friedmann 
equations.
The on-shell fluctuation modes of the effective metric  
contain those of massless gravitons, plus a scalar mode.
The off-shell metric fluctuations provide all degrees of freedom required for 
gravity in the presence of matter.
\item 
Gauge invariance allows to determine an effective Einstein-Hilbert-type 
action $S_{\rm EH}$, which is expected to be induced upon quantization.
It appears that no cosmological constant term can exist.
This leads to the scenario where the cosmological background is directly 
determined by 
the matrix model quite independently of the matter content, 
while gravity arises only as an effective theory which 
describes the fluctuations on the background. Deviations from GR at cosmological
scales are expected.
\end{itemize}

This provides a promising basis for a quantum theory of gravity.
However there are many open issues and loose ends, which need to be tied up in following work.  
One  task is to find and clarify the $4^{th}$ off-shell mode, presumably based on 
$\cA^{(\t)}$. 
Also, the inner product matrix
of the triplet of degenerate modes \eqref{Apmg-degeneracy} should be diagonalized.
This would allow to compute the propagator, and to
verify explicitly the no-ghost statement
in section \ref{sec:inner}. 
Furthermore, it is important to determine whether or not  
the extra degrees of freedom of massive gravitons arising from $\cA^{(-)}[\phi^{(2)}]$ 
are physical, 
and to clarify their physical significance.

 Since we argued that the Einstein-Hilbert-type action $S_{\rm EH}$ is induced upon
 integrating out matter, determining the effective Newton constant requires at least 
 a one-loop computation. This is also required in the computation of quantum corrections 
 to the scale parameter $r$ of the background.
 Since  $H^4_n$ has a similar structure as $S^4_N$, one 
 should be able to repeat the 1-loop computation along the lines of 
 \cite{Steinacker:2015dra,Steinacker:2016vgf}.
In particular, the techniques developed in \cite{Steinacker:2016nsc} should be 
very helpful.
Eventually, the quantum effective action should also allow to justify the choice 
of a background solution in the matrix model.

Another task is to compute the sub-leading corrections to the curvature and the 
linearized Einstein equations. 
Since we have the full expressions, it should be 
possible to clarify the geometrical meaning of the linearized Ricci-tensors 
on the cosmological background. 
This would then allow to give a meaningful comparison with the standard picture of a cosmological constant, and to 
address possible relations with apparent dark energy.

Furthermore, the significance of the bare action for the gravitational sector should be 
clarified, and the full equations for gravity in the presence of matter
should be derived and studied. Further topics are finding an analog of the 
Schwarzschild solution, investigating the deviations from GR in detail, etc.

 In any case, having a mechanism for gravity on a $(3{+}1)$-dimensional solution of 
the IKKT matrix model is very significant. 
For example, this may allow to avoid the issue of a landscape in string theory. 
That problem arises in the standard approach to string theory, because gravity is assumed to 
originate from the $10$-dimensional bulk gravity,
which must thus be compactified to 4 dimensions. 
In the present mechanism this is no longer required; extra structure for 
low-energy gauge theory
can arise in different ways, see e.g.\ \cite{Aoki:2014cya,Sperling:2018hys} and 
references therein.
Ultimately, one should try to relate these analytic studies with non-perturbative simulations 
as in \cite{Kim:2011cr,Kim:2012mw,Ito:2015mxa,Nishimura:2019qal}.

Looking forward, a particularly interesting perspective is to elaborate the early 
universe near the 
Big Bounce. While the late-time $k=-1$  cosmology  is not quite in line with 
the present concordance model, the resulting 
picture is very reasonable, 
given the simplicity of the model and the crudeness of our analysis.
Ultimately, the present framework could be powerful enough to address also
the deep questions around black holes and the resolution of singularities.

\paragraph{Acknowledgements.}
Useful conversations with David Berenstein, Stefan Fredenhagen, 
Karapet Mkrtchyan,  Hikaru Kawai, Jun Nishimura, Jan Rosseel, Mark van Raamsdonk,
Peter Schupp,  Evgeny Skvortsov, and Asato Tsuchiya
are gratefully acknowledged, as well as a related collaboration with Joanna Karczmarek.
This work was supported by the Austrian Science Fund (FWF) grant
 P28590. 
 Related support by the Action MP1405 QSPACE from the European 
 Cooperation in Science and Technology (COST) is also acknowledged.

\appendix
\section{Mathematical supplements and details}
\label{app:proofs}
%

%
%
\subsection{Derivation of the Einstein-Hilbert-like action}
\label{app:derivation_action}
Here, we provide the details for the derivation of the 
Einstein-Hilbert-like action \eqref{S-EH}.
To begin with, we split the variation \eqref{del-S1-1} of $S_1$ according to 
the traceless perturbation \eqref{traceless-tildeh} and obtain 
\begin{align}
 \d_\phi S_1 
 &=  \underbrace{4 \int \cA^{(-)}_{\nu}[\phi]\,
  \{t_\mu,\cG_{(\rm lin)}^{\mu\nu}[\d  g_0^{\a\b}]\} }_{\equiv \Upsilon_1}
\underbrace{- \frac{1}{3} \int \cA^{(-)}_{\nu}[\phi]\,
  \{t_\mu,\cG_{(\rm lin)}^{\mu\nu}[  h\, \eta_{\a\b} ]\} }_{\equiv 
\Upsilon_2}
\equiv \Upsilon_1+\Upsilon_2 \,.
\label{eq:del_S1_new}
\end{align}
Next, we note that first term in \eqref{eq:del_S1_new} can be 
simplified using \eqref{conservation-law-ricci} and \eqref{F-A-phi-id} as
\begin{align}
\Upsilon_1 &= 4\int  \cA^{(-)}_{\nu}[\phi] \{t_\mu,\cG_{(\rm 
lin)}^{\mu\nu}[\d g_0]\} \nn\\
 &= 2\int \cA^{(-)}_{\nu}[\phi] 
    \Big( \frac{1}{r^2R^2}\Big(
 2\{t_{\r},\{\theta^{\r'\r},\d g_0^{\r'\nu}\}\} 
 - \{t_\r,\{\theta^{\mu\nu},\d  g_0^{\r\mu}\}\} 
 \Big) + \frac 2{R^2} \{t_\r,\d g_0^{\r\nu}\}  \Big) \nn\\
 &= -\frac 2{R^2}\int \{t_{\r},\cA^{(-)}_\nu[\phi]\}
    \Big(-2\{\cM^{\r'\r},\d g_0^{\r'\nu}\}
 + \{\cM^{\mu\nu},\d  g_0^{\r\mu}\}  + 2 \d 
g_0^{\r\nu}  \Big) \nn\\
  &= -\frac 2{R^2}\int \{t_{\r},\cA^{(-)}_{\nu}[\phi]\}
    \Big(-\frac 12(\{\cM^{\r'\r},\d g_0^{\r'\nu}\} + 
\{\cM^{\r'\nu},\d g_0^{\r'\r}\}) \nn\\
 &\qquad \qquad -\frac 32 (\{\cM^{\r'\r},\d g_0^{\r'\nu}\} - 
\{\cM^{\r'\nu},\d g_0^{\r'\r}\})
 + 2 \d g_0^{\r\nu}  \Big) \nn\\
  &=  \frac 1{R^2}\int 3 
  F_{\r\nu}[\cA^{(-)}]\{\cM^{\mu\r},\d  g_0^{\mu\nu}\} 
  - h_{\r\nu}^{(g)} \{\cM^{\mu\nu},\d  g_0^{\r\mu}\} 
   + 2 h_{\mu\nu}^{(g)}\d g_0^{\mu\nu}\nn\\
%
%
%
  &=  \frac 1{R^2}\int   
  3\d  g_0^{\mu \nu} \{\cM^{\mu\r},\d_\phi f_{\r\nu} \}
   + \d  g_0^{\r\mu}\{\cM^{\mu\nu}, 
h_{\nu\r}^{(g)}[\phi] \} 
   + 2\d g_0^{\mu\nu} h_{\mu\nu}^{(g)}[\phi] \ .
   \label{del-S12-div}
\end{align}
With the additional action terms \eqref{eq:add_actions} and their variations 
\eqref{eq:del_add_actions},  $\Upsilon_1$ can be written as
\begin{align}
 \Upsilon_1 &=  \frac 1{R^2}\int \left(  
   3\d  g_0^{\mu \nu} \{\cM^{\mu\r},\d_\phi f_{\r\nu} \}
   + \d  g_0^{\r\mu}\{\cM^{\mu\nu}, 
h_{\nu\r}^{(g)}[\phi] \} 
   +  2\d g_0^{\mu\nu} h_{\mu\nu}^{(g)}[\phi]  
\right) \notag\\
&= \frac 1{R^2}\int  3\d  g_0^{\mu \nu} \{\cM^{\mu\r},\d_\phi f_{\r\nu} \}
  +\frac{1}{2R^2} \d_\phi S_4 + \frac{1}{R^2} \d_\phi S_3  \ .
  \label{upsilon-1-2}
\end{align}
The first term in \eqref{upsilon-1-2} can be re-written by noting that
 \begin{align}
 -\int   h^{\mu\nu}\{\cM^{\nu\r},f_{\r\mu}[\cA^{(g)}]\}
%
  &= -  \int \{x^\mu,A^\nu\}_-\{\cM^{\nu\r},\{x^\mu,A_{(g)}^\r\}_- 
\}
    - \{x^\nu,A^\mu\}_- \{\cM^{\nu\r},\{x^\r,A_{(g)}^\mu\}_- \}  \nn\\
  &\qquad + \{x^\nu,A^\mu\}_- \{\cM^{\nu\r},\{x^\mu,A_{(g)}^\r\}_-\}
   - \{x^\mu,A^\nu\}_-\{\cM^{\nu\r},\{x^\r,A_{(g)}^\mu\}_-\} \nn\\
%
%
 &= \frac 12 \d_\phi S_{M3} +  \int \{\cM^{\nu\r},\{x^\nu,A^\mu\}_- 
\}\{x^\mu,A_{(g)}^\r\}_- \nn \\
   &\qquad \qquad \qquad 
+\int\{x^\mu,A^\nu\}_-\{\cM^{\nu\r},\{x^\r,A_{(g)}^\mu\}_-\} 
\ . 
 \end{align} 
Applying \eqref{tilde-I-Apm} twice yields
\begin{align}
 -\int   h^{\mu\nu}\{\cM^{\nu\r},f_{\r\mu}[\cA^{(g)}]\} 
 &= \frac 12 \d_\phi S_{M3} + \int (\{x^\r,A^\mu\}_-  
+R\{t^\r,D^- A^\mu\})\{x^\mu,A_{(g)}^\r\}_- \nn\\
  & \qquad  - \int\{x^\mu,A^\nu\}_-(\{x^\nu,A_{(g)}^\mu\}_-  
+R\{t^\nu,D^- A_{(g)}^\mu\}) \nn\\
 &= \frac 12 \d_\phi S_{M3} + R \int\Big( 
\{t^\r,D^-A^\mu\}\{x^\mu,A_{(g)}^\r\}_-
   -\{x^\mu,A^\nu\}_- \{t^\nu,D^- A_{(g)}^\mu\} \Big) \nn\\
%
%
 &= \frac 12 \d_\phi S_{M3}  - R \int \Big(
 (D^- A^\mu) \{x^\mu,\{t_\r,A_{(g)}^\r\}\}_-
   - \{x^\mu,\{t_\nu,A^\nu\}\}_- (D^- A_{(g)}^\mu) \Big) \nn\\  
  &= \frac 12 \d_\phi S_{M3}  + R \int \Big(\{x^\mu,D^- 
A^\mu\}_+\{t_\r,A_{(g)}^\r\}
   - \{t_\nu,A^\nu\} \{x^\mu,D^- A_{(g)}^\mu\}_+  \Big) \nn\\
  &= \frac 12 \d S_{M3} + R \int \Big(D^+( \{x^\mu, 
A^\mu\}_-)\{t_\r,A_{(g)}^\r\}
   -D^+\{x^\mu,A_{(g)}^\mu\}_- \, \{t_\nu,A^\nu\} \Big) \nn\\
  &= \frac 12 \d S_{M3}   - R \int \Big( \{x^\mu, 
A^\mu\}_- D^-(\{t_\r,A_{(g)}^\r\})
   -\{x^\mu,A_{(g)}^\mu\}_- \, D^-(\{t_\nu,A^\nu\}) \Big)  \nn\\
   &= \frac 12 \d S_{M3}  -\d_\phi S_{gf2}  + 2 R \int \Big( 
\{x^\mu,A_{(g)}^\mu\}_- \, D^-(\{t_\nu,A^\nu\}) \Big)  
\end{align}
where we used
\begin{align}
  \{x^\mu,D^- A^\mu\}_+
   &= D^+ (\{x^\mu,A^\mu\}_-)  -  \{D^+ x^\mu,A^\mu\}_- \ .
\end{align}
Now  $\Upsilon_1$ needs 
\begin{align}
  \d  g_0^{\r\mu} \{\cM^{\mu\nu},\d_\phi f_{\nu\r} \} 
   =  h^{\r\mu} \{\cM^{\mu\nu},\d_\phi f_{\nu\r} \} 
 - \frac 14   h  \{\cM^{\mu\nu},\d_\phi f_{\nu\mu} \} .\nn
\end{align}
The first term was just evaluated above, and the last term is 
\begin{align}
 \int  h \{\cM^{\mu\nu},\d_\phi f_{\nu\mu} \} 
  &= \int  h \{\cM^{\mu\nu},\{x_\nu,\cA^{(g)}_\mu[\phi] \}_- - 
\{x_\mu,\cA^{(g)}_\nu[\phi] \}_-\}  \nn\\
  &= 2\int  h \{\cM^{\mu\nu},\{x_\nu,\cA^{(g)}_\mu[\phi] \}_- \}  \nn\\
  &= 2\int  h (\{x^\mu,\cA^{(g)}_\mu[\phi] \}_- + R 
\{t^\mu,D^-\cA^{(g)}_\mu[\phi] \}) \nn\\
  &= -2R \int \tilde{h} D^-
(\{t^\mu,\cA^{(g)}_\mu[\phi] \})  \nn \\
  &= -4R\int \{x^\nu,\cA_\nu\}_-   D^- \{t^\mu,\cA^{(g)}_\mu[\phi] \}  
\end{align}
using again \eqref{tilde-I-Apm} and 
\begin{align}
 \{t^\mu,D^- \cA_\mu\} &= D^- \{t^\mu,\cA_\mu\} - \{D^- t^\mu,\cA_\mu\}  \nn\\
 &= D^- \{t^\mu,\cA_\mu\} - \frac 1R\{x^\mu,\cA_\mu\}_-  \ .
\end{align}
Therefore
\begin{align}
\int \d  g_0^{\r\mu} \{\cM^{\mu\nu},\d_\phi f_{\nu\r} \} 
   &= \int  h^{\r\mu} \{\cM^{\mu\nu},\d_\phi f_{\nu\r} \} 
- \int\frac 14   h  \{\cM^{\mu\nu},\d_\phi f_{\nu\mu} 
\} \nn\\
&=  -\frac 12 \d S_{M3} +\d_\phi S_{gf2}  -2 R \int \Big( 
\{x^\mu,A_{(g)}^\mu\}_- \, D^-(\{t_\nu,A^\nu\}) \Big) \nn\\
 &\qquad  +R\int \{x_\nu,\cA^\nu\}_-   D^- \{t_\mu,\cA_{(g)}^\mu[\phi] \} \nn \\
&= -\frac 12 \d S_{M3} + 2\d_\phi S_{gf2}  -3 R \int \
\{x_\mu,A_{(g)}^\mu\}_- \, D^-(\{t_\nu,A^\nu\})  \nn\\
&= -\frac 12 \d S_{M3} + 2\d_\phi S_{gf2} 
 + 3  \int  D^-\phi \, D^-(\{t_\nu,A^\nu\})  \nn . 
\end{align}
The last term is non-vanishing only for 
$\phi = D^+\phi^0$, and  only  the spin 0 mode 
$\cA^\mu = \cA^{(+)}[\psi^0] = \{x^\mu,\psi^0\}$ for $\psi^0 \in \cC^0$ 
can contribute (since $\{t^\nu,A^{(-)}_\nu[\phi^{(2)}]\} = 0)$.
Then 
\begin{align}
D^-(\{t^\nu,\cA^{(g)}_\nu[\phi^{0}]\}) = - D^-(\Box\phi^{0}) = -\Box D^-\phi^{0}
\end{align}
so that the last term can be written as gauge variation \eqref{S-gf3-del} of 
the action \eqref{S-gf3}. Consequently, we are lead to
\begin{align} 
\Upsilon_1
&=  \frac 1{R^2}\int   
  3\d  g_0^{\mu \nu} \{\cM^{\mu\r},\d_\phi f_{\r\nu} \}
  +\frac{1}{2R^2} \d_\phi S_4 + \frac{1}{R^2} \d_\phi S_3 
  \nn \\
&= \frac{3}{R^2}
\left( -\frac 12 \d S_{M3} + 2\d_\phi S_{gf2}  -3 R \int \Big( 
\{x_\mu,A_{(g)}^\mu\}_- \, D^-(\{t_\nu,A^\nu\}) \Big) 
\right) 
\nn \\
&\qquad 
+\frac{1}{2R^2} \d_\phi S_4 + \frac{1}{R^2} \d_\phi S_3 
   \nn \\
   &=-\frac{3}{2 R^2} \d S_{M3} + \frac{6}{R^2}\d_\phi S_{gf2}
  -\frac{9}{2R^2} \d_\phi S_{gf3}
+\frac{1}{2R^2} \d_\phi S_4 + \frac{1}{R^2} \d_\phi S_3 \,.
\end{align}
Finally, the conformal metric fluctuations contribute using 
\eqref{conservation-law-ricci-conf} as follows:
\begin{align}
 \Upsilon_2 &= -\frac 13\int  \cA^{(-)}_{\nu}[\phi^{(1)}] 
\{t_\mu,\cG_{(\rm lin)}^{\mu\nu}[ h\, \eta_{\a\b}]\} \nn\\
 &= -\frac 1{2R^2} \int  \cA^{(-)}_{\nu}[\phi]\big(\{t_\nu, h \}
  - \frac 1{r^2} \tilde\cI\{t_\nu, h \}  \big) \nn\\
 &= -\frac 1{2R^2} \int \left(
 \cA^{(-)}_{\nu}[\phi]\{t_\nu, h \}
  - \frac 1{r^2} \tilde\cI \left(  \cA^{(-)}_{\nu}[\phi] \right)\{t^\nu, 
h \}  \right) \nn\\
 &= -\frac 1{2R^2} \int \left(  
\cA^{(-)}_{\nu}[\phi]\{t_\nu, h \}
  - \frac 1{r^2} ( r^2 \cA^{\mu(-)}[\phi] + r^2 R \{t^{\mu},D^-\phi\} ) 
\{t_\mu, h \} \right) \nn\\
&= \frac 1{2R^2} \int  R \{t^{\mu},D^-\phi\}  \{t_\mu, h \} 
\nn\\
 &= \frac 1{2R^2} \int R D^-\phi^{(1)}  \Box h 
 =  \frac 1{12}\int  h^{(g)} \Box    h      \nn\\
 &=  \frac 1{24} \d_\phi S_{h} \,,
\end{align}
using \eqref{tilde-I-Apm} and \eqref{pure-gauge-H}.
%
%
\subsection{Conservation law}
Using the definitions \eqref{ricci-poisson-def}, we compute first for any 
trace-less metric $\d g_0^{\mu\nu}$
\begin{align}
 2 \{t_\mu,\cR_{(\rm lin)}^{\mu\nu}[\d g_0]\} 
 &= \{t_\mu,\Box\d  g_0^{\mu\nu}\}
 -\Box\{t_\r,\d  g_0^{\r\nu}\}
 + \{t_\mu,\{t^\nu, \{t_\r,\d  g_0^{\r\mu}\}\}\}  \nn\\
 &= \{t_\mu,\Box\d  g_0^{\mu\nu}\}
 -\Box\{t_\r,\d  g_0^{\r\nu}\}
 -\frac{1}{r^2 R^2} \{\theta^{\mu\nu}, \{t_\r,\d  g_0^{\r\mu}\}\}
 + \{t^\nu,\{t_\mu, \{t_\r,\d  g_0^{\r\mu}\}\}\} \ . \nn 
\end{align}
Now we use 
 \begin{align}
 \Box (\{t_\r,\d g_0^{\r\nu}\})
 &= \{t_\r,(\Box+\frac{3}{R^2})\d g_0^{\r\nu}\}
 - \frac{2}{r^2R^2}\{\theta^{\r\r'},\{t_{\r'},\d g_0^{\r\nu}\}\} \nn\\
 &=  \{t_\r,(\Box+\frac{3}{R^2})\d g_0^{\r\nu}\}
 - \frac{2}{r^2R^2}\Big(
 \{t_{\r'},\{\theta^{\r\r'},\d g_0^{\r\nu}\}\} 
 + \{\{\theta^{\r\r'},t_{\r'}\},\d g_0^{\r\nu}\} \Big) \nn\\
 &=  \{t_\r,(\Box+\frac{3}{R^2})\d g_0^{\r\nu}\}
 - \frac{2}{r^2R^2}\Big(
 \{t_{\r'},\{\theta^{\r\r'},\d g_0^{\r\nu}\}\} 
 + 3r^2\{t^\r,\d g_0^{\r\nu}\} \Big)\nn\\
 &=  \{t_\r,\Box\d g_0^{\r\nu}\}  - \frac{2}{r^2R^2}\Big(
 \{t_{\r'},\{\theta^{\r\r'},\d g_0^{\r\nu}\}\} 
 + \frac 32 r^2\{t^\r,\d g_0^{\r\nu}\} \Big)
\end{align}
due to \eqref{puregauge-D2}.
Hence
\begin{align}
  \{t_\r,\Box\d  g_0^{\r\nu}\}
   -  \Box (\{t_\r,\d g_0^{\r\nu}\})
  &= \frac{2}{r^2R^2}
 \{t_{\r'},\{\theta^{\r\r'},\d g_0^{\r\nu}\}\} 
 + \frac{3}{R^2} \{t_\r,\d g_0^{\r\nu}\}
\end{align}
Therefore 
\begin{align}
 2 \{t_\mu,\cR_{(\rm lin)}^{\mu\nu}[\d g_0]\} 
 &= \frac{1}{r^2R^2}\Big(
 2\{t_{\r'},\{\theta^{\r\r'},\d g_0^{\r\nu}\}\} 
 - \{\theta^{\mu\nu}, \{t_\r,\d  g_0^{\r\mu}\}\}\Big)
  + \frac  3{R^2} \{t^\r,\d g_0^{\r\nu}\}  \nn\\
&\qquad  + \{t^\nu,\{t_\mu, \{t_\r,\d  g_0^{\r\mu}\}\}\}
\end{align}
The second term can be written as
\begin{align}
 \{\theta^{\mu\nu}, \{t_\r,\d  g_0^{\r\mu}\}\}
  &=  \{\{\theta^{\mu\nu},t_\r\},\d  g_0^{\r\mu}\}
   + \{t_\r,\{\theta^{\mu\nu},\d  g_0^{\r\mu}\}\} \nn\\
 &=  -r^2\{\eta^{\mu\r} t^{\nu} - \eta^{\nu\r} t^\mu ,\d  g_0^{\r\mu}\}
   + \{t_\r,\{\theta^{\mu\nu},\d  g_0^{\r\mu}\}\} \nn\\
 &= r^2\{t^\mu ,\d  g_0^{\nu\mu}\}
      + \{t_\r,\{\theta^{\mu\nu},\d  g_0^{\r\mu}\}\} 
\end{align}
using $\d g_0 = 0$.
Hence 
\begin{align}
 2 \{t_\mu,\cR_{(\rm lin)}^{\mu\nu}[\d g_0]\} 
 &= \frac{1}{r^2R^2}\Big(
 2\{t_{\r'},\{\theta^{\r\r'},\d g_0^{\r\nu}\}\} 
 - \{t_\r,\{\theta^{\mu\nu},\d  g_0^{\r\mu}\}\} 
 \Big)
  + \frac 2{R^2} \{t_\r,\d g_0^{\r\nu}\}  \nn\\
&\qquad  + \{t^\nu,\{t_\mu, \{t_\r,\d  g_0^{\r\mu}\}\}\}
\end{align}
Similarly
 $\cR_{(\rm lin)}[\d g_0^{\mu\nu}] =  \{t_\mu, \{t_\r,\d  g_0^{\r\mu}\}\}$, and 
\eqref{ricci-poisson-def} gives
\begin{align}
2 \{t_\mu,\cG_{(\rm lin)}^{\mu\nu}[\d g_0]\}
 &= 2 \{t_\mu,\cR_{(\rm lin)}^{\mu\nu}\} -  \{t_\nu,\cR_{(\rm lin)}\}  \nn\\
 &= \frac{1}{r^2R^2}\Big(
 2\{t_{\r'},\{\theta^{\r\r'},\d g_0^{\r\nu}\}\} 
 - \{t_\r,\{\theta^{\mu\nu},\d  g_0^{\r\mu}\}\} 
 \Big) + \frac 2{R^2} \{t_\r,\d g_0^{\r\nu}\} \ .
 \label{conservation-law-ricci}
\end{align}
Now  consider  trace contributions 
to the metric
\begin{align}
 \cG_{(\rm lin)}^{\mu\nu}[h \eta_{\a\b}] &= - \Box h \eta^{\mu\nu}
   - \frac 12(\{t_\mu, \{t_\nu,\ h \}\}   +  (\mu \leftrightarrow \nu))\ .
\end{align}
These contribute as follows
\begin{align}
  \{t_\mu,\cG_{(\rm lin)}^{\mu\nu}[h \eta]\}
 &= -\{t_\nu,\Box h\}
  - \frac 12 \{t_\mu,\{t_\mu, \{t_\nu, h \}\}\} 
  - \frac 12 \{t_\mu,\{t_\nu, \{t_\mu, h \}\}\} \nn\\
 &= -\{t_\nu,\Box h\}
  + \frac 12 \Box \{t_\nu, h \}
  + \frac 12 \{t_\nu,\Box h \}
  - \frac 1{2r^2R^2} \{\theta^{\nu\mu}, \{t_\mu, h \}\} \nn\\
 &= -\{t_\nu,\Box h\}
  + \frac 12 (\cD^2 -\frac 2{r^2R^2}\tilde \cI) \{t_\nu, h \}
  + \frac 12 \{t_\nu,\Box h \}
  - \frac 1{2r^2R^2} \tilde\cI\{t_\mu, h \} \nn\\
 &= -\{t_\nu,\Box h\}
  - \frac 1{r^2R^2}\tilde \cI \{t_\nu, h \}
  +\frac 12 \{t_\nu,(\Box+\frac{3}{R^2}) h \}
  + \frac 12 \{t_\nu,\Box h \}
  - \frac 1{2r^2R^2} \tilde\cI\{t_\mu, h \} \nn\\
 &= \frac{3}{2R^2} \{t_\nu,h \}
  - \frac 3{2r^2R^2} \tilde\cI\{t_\mu, h \}
  \label{conservation-law-ricci-conf}
\end{align}
%
%
\subsection{Time-like mode}
\label{sec:app-time-like-modes}
Suppose there exists an extra on-shell gauge-fixed solution to $(\cD^2 + \frac 
12\mu^2) \cB^{(\t)} = 0$ of the form
\begin{align}
 \cB^{(\t)}[\phi^{(s)}] \coloneqq \cA^{(\t)}[\phi^{(s)}] + \{t^\mu,\chi^{(s)}\} 
+ \sum_\pm \cB^{(\pm)}[\phi_\pm] \qquad \in \cC^s
\end{align}
where $\chi$ is determined by $0 = \cG(\cB^{(\t)})\equiv 
\{t^\mu,\cB_\mu^{(\t)}\}$; hence, by  using 
\eqref{t-x-phi-id}, this reduces to
\begin{align}
 \sinh(\eta)(4+s+\t)\phi^{(s)} = \Box\chi^{(s)} .
 \label{phi-chi-4}
\end{align}
The on-shell condition gives 
\begin{align}
 0 &= \left(\cD^2 + \frac 12 \mu^2\right) \cB^{(\t)}_\mu[\phi]  = 
 x_\mu\left(\Box+\frac{4}{R^2}\right)\phi +2\eth_\mu\phi  +  \{t^\mu,\Box\chi\}
  + \sum_\pm \cB^{(\pm)}[(\Box + \ldots)\phi_\pm]   \nn\\
  &= 
 x_\mu\left(\Box+\frac{4}{R^2}\right)\phi +2\eth_\mu\phi  +  \{t^\mu, 
\sinh(\eta)(4+s+\t)\phi \}
  + \sum_\pm \cB^{(\pm)}[(\Box + \ldots)\phi_\pm]  
  \label{onshell-A4}
\end{align}
Taking $\{t^\mu,\cdot\}$ of that expression yields 
\begin{align}
\ 0  &= \{t^\mu,x_\mu\left(\Box+\frac{1}{R^2}\right)\phi^{(s)}\} 
   + 2 \{t^\mu,\eth_\mu\phi^{(s)}\} 
  -\Box(\sinh(\eta)(4+s+\t)\phi^{(s)})   \nn\\
  &= \frac{x_4}{R} (4+s+\t)\left(\left(\Box+\frac{1}{R^2}\right)\phi\right)
   + 2 \{t^\mu,\eth_\mu\phi^{(s)}\} 
  - \frac 1R\Box\left(x_4\left(4+s+\t\right)\phi\right)  \nn\\
 &= \frac{x_4}{R^3} (4+s+\t)\phi
   + 2 \{t^\mu,\eth_\mu\phi^{(s)}\} 
 - \frac 4{R^3} x_4(4+s+\t)\phi 
  + \frac 2R \{t_\mu, x_4\}\{t^\mu,(4+s+\t)\phi\} \nn\\
 &=  2 \{t^\mu,\eth_\mu\phi^{(s)}\} 
 - \frac {3x_4}{R^3} (4+s+\t)\phi - \frac {2x_4}{R^3} (s+\t)((4+s+\t)\phi)
\label{onshell-A4-id}
\end{align}
since the $\cB^{(\pm)}$ are gauge-fixed, 
and we used \eqref{phi-chi-4} and $\Box x_4 = \frac{4}{R^2} x^4$.
This provides some (Lorentz-violating) on-shell condition for $\phi$.
However, \eqref{onshell-A4} states  that $x_\mu(\Box+\frac{4}{R^2})\phi +2\eth_\mu\phi +  \{t^\mu, \sinh(\eta)(4+s+\t)\phi \}$  
is a linear combination of the two $\cB_\mu^{(\pm)}$ modes.
This  implies that there is another, independent equation for $\phi$, which strongly suggests that 
there is no generic solution. Unfortunately we cannot provide  
a formal proof here.
\subsection{Group theory and  useful identities}
\label{sec:checks-group-th-results}%
Consider 
\begin{align}
 [C^2[\mso(4,1)],\cM^{45}] 
  &= -\im \left(\cM_{b4}\cM^{b5} +\cM^{b5}\cM_{b4}\right)  \nn\\
  &= -2\im\cM_{b4}\cM^{b5} -\im[\cM^{b5},\cM_{b4}] \nn\\
  &= -2\im\cM_{b4}\cM^{b5} + 4\cM^{54} =  -2\im\frac{R}{r} T_\mu X^{\mu} + 
4\cM^{54} \,,
\end{align}
such that
\begin{align}
 C^2[\mso(4,1)]\cM^{45} &= \cM^{45} \left(C^2[\mso(4,1)]-4\right) -2\im 
\frac{R}{r} T_\mu X^{\mu}
\end{align}
This holds in any representation, in particular in the adjoint acting on 
functions.
Thus
\begin{align}
 -\im \frac{2R}{r}\Tr\left( \phi' [T_\mu, [X^{\mu}, \phi]]\right)
 &= \Tr  \phi' \left(C^{2 \ad}[\mso(4,1)][\cM^{45},\phi] - [\cM^{45}, (C^{2 
\ad}[\mso(4,1)]-4) \phi] \right) \nn\\
 &= \left(C^2[\phi'] - C^2[\phi]+4\right)\, \Tr \phi'[\cM^{45},\phi] \ .
 \label{T-X-Casimir-relation}
\end{align}
Similarly,
\begin{align}
 [T_\mu T^\mu,\cM^{45}] &= T^\mu[T_\mu ,\cM^{45}] + [T_\mu,\cM^{45}]  T^\mu 
\nn\\
 &= - \im\frac{1}{rR} ( T^\mu X^\mu + X^\mu T^\mu)  \nn\\
 &= - \im\frac{2}{rR} T^\mu X^\mu - \frac 4{rR^2} X^4  
\end{align}
in any representation\footnote{Note that \eqref{X-X-orthogonal} only holds on 
$\cH_n$, but not e.g.\ in the
adjoint representation.}. 
This is consistent with \eqref{Box-T-SO41}. In particular,
from 
\begin{align}
 [C^2[\mso(4,1)],\cM^{45}] &=\frac{1}{2} [ \cS^2,\cM^{45}] = R^2[T_\mu 
T^\mu,\cM^{45}]\,
\end{align}
(since $[C^2[\mso(3,1)],\cM^{45}] =0$) 
we  obtain in the adjoint representation 
\begin{align}
  -\im\frac{2R}{r} \Tr \phi' [T_\mu, [X^{\mu}, \phi]]
 &= \Tr  \phi' \left(R^2 \Box[\cM^{45},\phi] - [\cM^{45}, (R^2 \Box -4) 
\phi]\right) \,.
 \label{TT-X4-relation}
\end{align}
\paragraph{Identities for $\cI(\{x^\mu,\phi\})$.}
Next, recall the following result of \cite{Sperling:2018xrm} 
\begin{align}
2 \tilde\cI^{(5)}(\{x_a,\phi\}) =
 2\{\theta^{ab},\{x_b,\phi\}\} =  r^2\left(\frac 12 \cS^2 - s(s+1) + 
4\right)(\{x_a,\phi\}) , \qquad a=0,\ldots,4
\label{I-x-id}
 \end{align}
and in particular
\begin{align}
 \label{I5-X-phi}
\begin{aligned}
\tilde\cI^{(5)}\left(\{x_a,\phi^{(s)}\}_+ \right) &= r^2(s + 
3)\{x_a,\phi^{(2)}\}_+    \\
 \tilde\cI^{(5)}\left(\{x_a,\phi^{(s)}\}_- \right) &= r^2(-s + 
2)\{x_a,\phi^{(2)}\}_- 
 \end{aligned}
\end{align}
which then implies
\begin{align}
 \tilde\cI(\{x^\mu,\phi^{(s)}\}) &= 
  \frac{r^2}2\Big(\frac 12 \cS^2 - s(s+1) + 4\Big) \{x^\mu,\phi^{(s)}\}   +  r^2 
R \{t^{\mu},\{x^4,\phi^{(s)}\}\} \,.
 \label{I-A2}
\end{align}
For the $\cA^{(\pm)}$ modes, this gives 
\begin{align}
 \label{tilde-I-Apm}
\begin{aligned}
\tilde\cI(\cA^{\mu(+)}) &= r^2 (s + 3) \cA^{\mu(+)} +  r^2 R 
\{t^{\mu},D^+\phi^{(s)}\} \\
\tilde\cI(\cA^{\mu(-)}) &=  r^2 (-s + 2) \cA^{\mu(-)} + r^2 R 
\{t^{\mu},D^-\phi^{(s)}\}  \ .
\end{aligned}
\end{align}
Furthermore, evaluating the $4$ component of \eqref{I-x-id} yields 
\begin{align}
 R \{t^{\mu},\{x_\mu,\phi^{(s)}\}\} = \frac 12\Big(\frac 12 \cS^2 - s(s+1) + 
4\Big) \{x^4,\phi^{(s)}\} \,,
\end{align}
which is useful to understand the orthogonality.
Using $\cS^2 = 2 C^2[\mso(4,1)] - C^2[\mso(4,2)]$ as in \eqref{Spin-casimir},
one can see that 
this relation is nothing but \eqref{T-X-Casimir-relation} in the adjoint representation.
For the $\cA^{(\pm)}$ modes, this gives 
\begin{align}
 \label{A2-gaugefix-ids}
\begin{aligned}
 R \{t^{\mu},\cA_\mu^{(+)}[\phi^{(s)}]\} &= (s+3) \{x^4,\phi^{(s)}\}_+  \,,\\
 R \{t^{\mu},\cA_\mu^{(-)}[\phi^{(s)}]\} &= (-s + 2) \{x^4,\phi^{(s)}\}_-  \,.
 \end{aligned}
\end{align}
Similarly, \eqref{TT-X4-relation} gives along the same lines 
\begin{align}
 R \{t^{\mu},\{x_\mu,\phi^{(s)}\}\} = \frac 12\Big((R^2\Box+4) \{x^4,\phi^{(s)}\} - \{x^4,R^2\Box\phi^{(s)}\}\Big)
 \label{t-c-comm-id}
\end{align}
which immediately implies 
\begin{align}
 R \{t^{\mu},\cA_\mu^{(\pm)}[\phi^{(s)}]\} = \frac 12\Big((R^2\Box+4) 
\{x^4,\phi^{(s)}\}_\pm - \{x^4,R^2\Box\phi^{(s)}\}_\pm\Big)
\end{align}
and together with \eqref{A2-gaugefix-ids}, we find
\begin{align}
 \label{Box-x4-relations}
\begin{aligned}
 \Box D^+\phi^{(s)}   &= D^+\left(\Box+\frac{2s+2}{R^2}\right)\phi^{(s)}  \,,\\
 \Box D^-\phi^{(s)}   &= D^-\left(\Box-\frac{2s}{R^2}\right)\phi^{(s)}  \,.
 \end{aligned}
\end{align}
This implies
\begin{align}
 \Box D^+D^-\phi^{(s)}   &=  D^+D^-\Box\phi^{(s)} , \qquad 
 \Box D^-D^+\phi^{(s)}   =  D^-D^+\Box\phi^{(s)} 
\end{align}
and
\begin{align}
\label{Boxinv-x4-phi}
\begin{aligned}
\Box^{-1} D^+ \phi &= D^+\left(\Box+\frac{2s+2}{R^2} \right)^{-1}\phi  \,,\\
    \Box^{-1} D^- \phi &= D^-\left(\Box-\frac{2s}{R^2}\right)^{-1}\phi \,,
    \end{aligned}
\end{align}
provided the inverses on the right-hand-side exist.
We finally note
\begin{align}
 D^-\{x^\nu,\phi^{(2)}\}_- &= \{x^4,\{x_\nu,\phi^{(2)}\}_-\}_- 
 = \{x^\nu,\{x_4,\phi^{(2)}\}_-\}_- =\{x^\nu, D^-\phi^{(2)}\}_- 
 \label{D-x-id}
\end{align}
since $\{t^\nu,\cdot\}$ respects the spin.
\paragraph{$\cD^2$ intertwiner for $h_{\mu\nu}$.} 
To evaluate $\cR_{(\rm lin)}^{\mu\nu}[ h^{\mu\nu}]$, 
it is useful to extend the definition \eqref{tilde-H-def} of $ h^{\mu\nu}$ as 
follows:
\begin{align}
 \cA \mapsto  h^{\mu\nu}_{(4)}[\cA] &= \{\cA^\mu,x^\nu\}_- + 
(\mu\leftrightarrow 
\nu) \nn\\
 \cA \mapsto  h^{ab}_{(5)}[\cA] &= \{\cA^a,x^b\}_- + (a \leftrightarrow b) 
\end{align}
viewed as $SO(3,1)$ and $SO(4,1)$  intertwiners, respectively.
Here we assume that $\cA^\mu$ arises from $\cA^b$ in $(5)\otimes (\ad)$;
this is the case for all physical fluctuation modes as discussed in section \ref{sec:D2-eigenvalues}. 
Then the following the intertwiner property holds:
\begin{align}
 C^2[\mso(4,1)]^{\rm (full)}   h^{ab}_{(5)}[\cA] 
  &=  h^{ab}_{(5)}\left[C^2[\mso(4,1)]^{\rm (full)}\cA\right]
  \label{C2full-I-relation-H}
\end{align} 
and similarly for $SO(3,1)$. The full Casimirs can be expressed in terms of 
$\tilde\cI$ as explained in section \ref{sec:group-th-results}, and we obtain
\begin{align}
 \left(C^2[\mso(4,1)]^{(\ad)} - 2 r^{-2}\tilde\cI^{(5)} + C^2_{(5)}\right)  
h^{ab}_{(5)}[\cA]
  &=  h^{ab}_{(5)}\left[ \left( C^2[\mso(4,1)]^{(\ad)} - 2 
r^{-2}\tilde\cI^{(5)} +4\right)\cA\right]  \nn\\
  \left(C^2[\mso(3,1)]^{(\ad)} - 2 r^{-2}\tilde\cI^{(4)} +C^2_{(4)}\right)  
h^{\mu\nu}_{(4)}[\cA]
  &=  h^{\mu\nu}_{(4)}\left[ \left( C^2[\mso(3,1)]^{(\ad)} - 2 
r^{-2}\tilde\cI^{(4)} +3\right)\cA\right] 
\end{align}
using \eqref{Spin-casimir}, and $C^2_{(5)}\coloneqq C^2[\mso(4,1)]^{(5)\otimes(5)}$ and  
$C^2_{(4)}\coloneqq C^2[\mso(3,1)]^{(4)\otimes(4)}$ denotes the Casimirs acting on the indices.
Here 
\begin{align}
 \tilde\cI^{(5)} h^{\mu\nu}_{(5)}[\cA] &= 
 \{\theta^{\mu a}, h^{a\nu}_{(5)}\} +  \{\theta^{\nu a}, h^{\mu a}_{(5)}\} 
.
 \end{align}
Subtracting these for
$ h^{\mu\nu}_{(5)} =  h^{\mu\nu}_{(4)} =  h^{\mu\nu}$, we obtain 
\begin{align}
 \left({R}^2\Box - 2 r^{-2}\left(\tilde\cI^{(5)} - \tilde\cI^{(4)} \right)  + 
C^2_{(5)} - C^2_{(4)} \right) 
  h^{\mu\nu}[\cA]
  &=  h^{\mu\nu}\left[ \left({R}^2\Box - 2 r^{-2}\left(\tilde\cI^{(5)} - 
\tilde\cI^{(4)}\right) +1 \right)\cA \right]  \nn\\
 \left(R^2\cD^2 -2 r^{-2}\tilde\cI^{(5)}+ C^2_{(5)} - C^2_{(4)}  \right)  
h^{\mu\nu}[\cA]
  &= h^{\mu\nu}\left[ \left( {R}^2\cD^2- 2 
r^{-2}\tilde\cI^{(5)}+1 \right)\cA \right] 
\end{align}
where $\cD^2 = \Box + \frac{2}{r^2 R^2} \tilde\cI^{(4)}$. To proceed,
we need
\begin{align}
 \cI^{(5)} \{\cA^\mu,x^\nu\}_- -  \{\cI^{(5)}\cA^\mu, x^\nu\}_-
&=\{\theta^{\mu b}, \{\cA_b, x^\nu\}_- \} + \{\theta^{\nu b} ,\{\cA^\mu, x_b\}_- \} - \{\{\theta^{\mu b} ,\cA_b\} , x^\nu\}_- \nn\\
 &=r^2 \{\cA^\mu,x^\nu\}_- + \{\theta^{\mu b},\{\cA_b,x^\nu\}_-\} +\{x^\nu,\{\theta^{\mu b},\cA_b\}\}_-  \nn\\
 &=r^2 \{\cA^\mu,x^\nu\}_- + \{\cA_b,\{\theta^{\mu b},x^\nu\}_-\}  \nn\\
 &=r^2 \{\cA^\mu,x^\nu\}_- - r^2 \{\cA_b,x^b\}_- \eta_{\mu \nu} 
 + r^2 \{\cA^\nu, x^\mu\}_-  \nn
\end{align}
using \eqref{M-X-relations}, \eqref{I5-X-phi}, and the Jacobi identity.
This gives
\begin{align}
 \cI^{(5)}  h^{\mu\nu}[\cA]
-  h^{\mu\nu}[\tilde\cI^{(5)}\cA]
&= 2 r^2  h^{\mu\nu}[\cA] - 2r^2 \{\cA_b,x^b\}_- \eta_{\mu \nu} 
\end{align}
and therefore 
\begin{align}
 \left(\cD^2 + \frac 1{R^2}\left(C^2_{(5)} - C^2_{(4)}\right)\right)  
h^{\mu\nu}[\cA]
  &=  h^{\mu\nu}\left[\left(\cD^2+\frac{4}{R^2} \right)\cA\right] + 
\frac{4}{R^2} \{\cA_b,x^b\}_- \eta_{\mu \nu} 
   \end{align} 
hence 
\begin{align}
\cD^2   h^{\mu\nu}[\cA]
  &=  h^{\mu\nu}\left[\left(\cD^2+\frac{6}{R^2}\right)\cA\right]  
     + \frac 2{R^2}\eta_{\mu \nu}\left(- h_{(5)}[\cA] +   
h_{(5)}^{44}[\cA]\right) \nn\\
      &=  h^{\mu\nu}[\cD^2\cA]  
      +  \frac{2}{R^2}\left(3 h^{\mu\nu}[\cA] - \eta^{\mu \nu} h[\cA]\right)
  \label{D2-h-intertwiner}
\end{align} 
 (where $ h \equiv  h_{(4)}$), using
 $2\{\cA_b,x^b\}_- = -  h_{(5)} $  and
\begin{align}
 (C^2_{(5)} - C^2_{(4)})  h^{\mu\nu}[\cA] 
  &= -2  h^{\mu\nu}[\cA] -2 \eta^{\mu\nu} h^{44}[\cA]
\end{align}
since $M_{(5)}^{\a 4} v_\mu = -i\d^\a_\mu \, v^4$ using  \eqref{M-rep-5}.

\subsection{Consistency checks}
\label{sec:app-consistency-I}
Here, we check the implications of \eqref{I-Apm-relation} for consistency. To 
begin with, \eqref{I-Apm-relation} implies 
 \begin{align}
  \int &\cA^{(+)}_\mu[D^-\phi'^{(s)}] \tilde \cI 
\cA^{(-)}_\mu[D^+\phi^{(s)}] \\
   &= \int \cA^{(+)}_\mu[D^-\phi'^{(s)}] 
   \Big( r^2 (-s + 1) \cA^{\mu(-)}[D^+\phi^{(s)}]  + r^2 R 
\{t^{\mu},D^-D^+\phi^{(s)}\}  \Big)  \nn\\
   &= \int  \tilde \cI\cA^{(+)}_\mu[D^-\phi'^{(s)}] 
\cA^{(-)}_\mu[D^+\phi^{(s)}] \nn\\
   &=  \int  \Big( r^2 (s + 2) \cA^{\mu(+)}[D^-\phi'^{(s)}] + r^2 R 
\{t^{\mu},D^+D^-\phi'^{(s)}\}   \Big)    \cA^{(-)}_\mu[D^+\phi^{(s)}] \nn
 \end{align}
 using the fact that $\tilde\cI$ is self-adjoint, see  
\eqref{intertwiner-A-hermitian}. From this equality, we learn that 
\begin{align}
 \int &(-s + 1)  \cA^{(+)}_\mu[D^-\phi'^{(s)}] 
   \cA^{\mu(-)}[D^+\phi^{(s)}]
   +  R \cA^{(+)}_\mu[D^-\phi'^{(s)}]  \{t^{\mu},D^-D^+\phi^{(s)}\}   \nn\\
&=  \int (s + 2) \cA^{\mu(+)}[D^-\phi'^{(s)}]\cA^{(-)}_\mu[D^+\phi^{(s)}]
    + R \{t^{\mu},D^+D^-\phi'^{(s)}\}\cA^{(-)}_\mu[D^+\phi^{(s)}]  \nn\\
&\quad - (2s +1) \int \cA^{(+)}_\mu[D^-\phi'^{(s)}] 
   \cA^{\mu(-)}[D^+\phi^{(s)}]
   -  R \int \{t^{\mu}, \cA^{(+)}_\mu[D^-\phi'^{(s)}]\} D^-D^+\phi^{(s)} \nn\\
&= -R\int D^+D^-\phi'^{(s)} \{t^{\mu},\cA^{(-)}_\mu[D^+\phi^{(s)}] \} \nn\\
&\quad -(2s +1) \int \cA^{(+)}_\mu[D^-\phi'^{(s)}] \cA^{\mu(-)}[D^+\phi^{(s)}]
   - (s+2)  \int D^+ D^-\phi'^{(s)} D^-D^+\phi^{(s)} \nn\\
&= - (-s+1)\int D^+D^-\phi'^{(s)} D^-D^+\phi^{(s)} 
\end{align}
such that
\begin{align}
- \int \cA^{(+)}_\mu[D^-\phi'^{(s)}] \cA^{\mu(-)}[D^+\phi^{(s)}]
   =  \int D^+ D^-\phi'^{(s)} D^-D^+\phi^{(s)}  \,.
\end{align}
This is the same as \eqref{inner-1}, which provides a consistency test.
%
%
\subsection{Relations for \texorpdfstring{$\Box_H$}{Box}}
\label{sec:relations-box-H}
Consider the following intertwiner relation discussed 
in\cite{Sperling:2018xrm}: 
 \begin{align}
 r^2 C^2[\mso(4,1)]^{\rm (full)} \cA^{(-)}_a[\phi^{(s)}] &=
 -\left(\Box_H  + 2 \cI^{(5)} - r^2(\cS^2 +4)\right) \cA^{(-)}_a[\phi^{(s)}] 
\nn\\
  &= \cA_a^{(-)}[r^2 C^2[\mso(4,1)]\phi^{(s)}]
  \label{C2-A-phi-general}
\end{align}
for $a=0,\ldots,4$. More explicitly, this reads
\begin{align}
 -(\Box_H  + 2 r^2(2-s) - r^2(2s(s-1) +4)) \cA^{(-)}_a[\phi^{(s)}] 
  &= \cA_a^{(-)}[(-\Box_H + r^2\cS^2)\phi^{(s)}]  \nn\\
(-\Box_H  + 2 r^2 s^2) \cA^{(-)}_a[\phi^{(s)}] 
  &= \cA_a^{(-)}[(-\Box_H + 2r^2s(s+1))\phi^{(s)}]  \nn\\
 \Box_H \cA^{(-)}_a[\phi^{(s)}] 
  &= \cA_a^{(-)}[(\Box_H - 2r^2s)\phi^{(s)}] 
  \label{BoxH-A-relation}
\end{align}
 using $ \cI^{(5)}\cA^{(-)}_a[\phi^{(s)}] = r^2(2-s)\cA^{(-)}_a[\phi^{(s)}] 
$, see  
\eqref{I5-X-phi}.
Similarly, the  intertwiner property 
 \begin{align}
 C^2[\mso(4,1)]^{\rm(ad)}\{x^a, \phi^{H(s-1)}_a \} 
 &= \{x^a, C^2[\mso(4,1)]^{\rm(full)}\phi^{H(s-1)}_a \} 
\end{align}
provides us with
 \begin{align}
 (\Box_H - 2r^2s(s+1))\{x^a, \phi^{H(s-1)}_a \} 
   &= \{x^a,(\Box_H  + 2 r^2(2-s) - r^2(2s(s-1) +4)) \phi^{H(s-1)}_a \} \nn\\
 (\Box_H -2r^2 s)\{x^a, \phi^{H(s-1)}_a \} 
  &= \{x^a,\Box_H\phi^{H(s-1)}_a \} \, .
 \label{x-BoxH-bracket}
\end{align}

\bibliographystyle{JHEP}
\bibliography{papers}

\end{document}